\documentclass[english,aps,preprint,nofootinbib]{revtex4-1}
\usepackage[T1]{fontenc}
\usepackage[latin9]{inputenc}
\usepackage{color}
\usepackage{float}
\usepackage{amsmath}
\usepackage{amssymb}
\usepackage{stmaryrd}
\usepackage{graphicx}
\usepackage{wasysym}
\usepackage{esint}
\usepackage[colorlinks]{hyperref}

\makeatletter

\providecommand{\tabularnewline}{\\}

\@ifundefined{textcolor}{}
{%
 \definecolor{BLACK}{gray}{0}
 \definecolor{WHITE}{gray}{1}
 \definecolor{RED}{rgb}{1,0,0}
 \definecolor{GREEN}{rgb}{0,1,0}
 \definecolor{BLUE}{rgb}{0,0,1}
 \definecolor{CYAN}{cmyk}{1,0,0,0}
 \definecolor{MAGENTA}{cmyk}{0,1,0,0}
 \definecolor{YELLOW}{cmyk}{0,0,1,0}
}

\makeatother

\usepackage{babel}

\begin{document}

\title{Gravitational waves in higher-order $R^{2}$ gravity}

\author{S. G. Vilhena}
\email{silas.vilhena@ga.ita.br}

\selectlanguage{english}%

\address{Departamento de F\'\i sica, Instituto Tecnol\'ogico da Aeron\'autica, Pra\c ca
Mal. Eduardo Gomes, 50, CEP 12228-900, S\~ao Jos\'e dos Campos, S\~ao Paulo,
Brazil}

\author{L. G. Medeiros}
\email{leo.medeiros@ufrn.br }

\selectlanguage{english}%

\address{Escola de Ci\^encia e Tecnologia, Universidade Federal do Rio Grande
do Norte, Campus Universit\'ario, s/n\textendash Lagoa Nova, CEP 59078-970,
Natal, Rio Grande do Norte, Brazil}

\author{R. R. Cuzinatto}
\email{rodrigo.cuzinatto@unifal-mg.edu.br}

\selectlanguage{english}%

\address{Instituto de Ci\^encia e Tecnologia, Universidade Federal de Alfenas,
Rodovia Jos\'e Aur\'elio Vilela, 11999, Cidade Universit\'aria, CEP 37715-400,
Po\c cos de Caldas, Minas Gerais, Brazil}

\begin{abstract}
We perform a comprehensive study of gravitational waves in the context
of the higher-order quadratic scalar curvature gravity, which encompasses
the ordinary Einstein-Hilbert term in the action plus an $R^{2}$ 
contribution and a term of the type $R\square R$. The main focus
is on gravitational waves emitted by binary systems such as binary
black holes and binary pulsars in the approximation of circular orbits
and nonrelativistic motion. The waveform of higher-order gravitational
waves from binary black holes is constructed and compared with the
waveform predicted by standard general relativity; we conclude that
the merger occurs earlier in our model than what would be expected
from GR. The decreasing rate of the orbital period in binary pulsars
is used to constrain the coupling parameters of our higher-order
$R^{2}$ gravity; this is done with Hulse-Taylor binary pulsar data
leading to $\kappa_{0}^{-1}\lesssim1.1\times10^{16}\,\text{m}^{2}$
, where $\kappa_{0}^{-1}$ is the coupling constant for the $R^{2}$ contribution.
\end{abstract}
\maketitle

\section{INTRODUCTION}

It is hardly necessary to motivate the appropriateness of studying
gravitational waves (GWs) in the light of LIGO-VIRGO Collaboration
observations, from the discovery event \cite{Abbott2016}, to the
neutron-star merger detection \cite{Abbott2017} inaugurating the
multimessenger astronomy, up to the most recent observational results
\cite{Abbott2019,Abbott2021}.

As a matter of fact, the direct detection of GW events confirmed the
indirect predictions stemming from the Hulse and Taylor binary pulsar
\cite{Hulse1975,Taylor1982,Weisberg2004}: The emission of gravitational
waves carries energy and momentum away from the parent binary system,
which in turn reduces its orbital period and mass separation, leading
to an inspiral motion ending with a coalescence. The importance of
the papers mentioned above includes the confirmations of all the predictions
of general relativity (GR) on this regard. On the one hand, this fact
is rather satisfying for Einstein's theory of gravity; on the other
hand, it imposes serious constraints onto modified theories for gravitational
interaction---so much so, that modified gravity frameworks that predict
propagation velocities for detectable spin-2 gravitational waves different
from the speed of light in vacuum are in difficulty \cite{Abbott2019,Kase2018}.

Even though gravitational wave observations definitely favor GR as
the correct theory of gravity, the physics community knows it cannot
be the final theory. The reason for this consensus comes from high-energy
regimes, where quantization is specially required. In effect, theoretical
contributions going back to the works, e.g., by Stelle \cite{Stelle1977,Stelle1978}
propose the modification of the Einstein-Hilbert action by the inclusion
of terms like the square of the curvature scalar, $R^{2}$, and other
invariants built from the curvature scalar and its derivatives \cite{LeeWick1970,Donoghue2019}.
In a similar direction, the Starobinsky inflation model uses the $R^{2}$ term
\cite{Starobinsky1979,Gurovich1979,Starobinsky1980} to produce one
of the strongest realizations of inflationary scenarios---in fact,
it predicts a tensor-to-scalar ratio $r$ per scalar tilt $n_{s}$
deep within the confidence contours produced with Plack satellite
observations \cite{PlanckCollab2018} and BICEP2/Keck plus BAO data
\cite{BICEP2Collab2018}.

That GR is not the ultimate theory for gravitational interaction is
also hinted at from the infrared regime in cosmology. The description
of late-time Universe dynamics hinges on the ubiquitous dark energy
\cite{Huterer2017,Motta2021,Slozar2019}. It is well known that the
simple cosmological constant could serve as a good candidate for describing
dark energy and the present-day acceleration of the cosmos. But then
again, the very concept of a cosmological constant faces problems, either
regarding its physical interpretation \cite{Weinberg1989,Carroll2001}
or related to the challenges against the concordance $\Lambda$CDM
model \cite{Skara2021}. Indeed, the $H_{0}$ tension, the $\sigma_{8}$ problem
and the lensing amplitude $A_{L}$ anomaly \cite{Verde2019,Valentino2021,Efstathiou2021,Handley2021,Valentino2019,Valentino2017,Gannouji2018,Vagnozzi2019,Vagnozzi2021a,Vagnozzi2021b}
have galvanized the ongoing ``cosmology crisis.\footnote{For a different perspective advocating against the crisis in cosmology,
please see Refs. \cite{Freedman2021,Nunes2021}.}'' This makes room for alternative theories of gravity \cite{Saridakis2021}.

One of the possible avenues of research in the context of modified
gravity is to consider higher-order quadratic-curvature scalar models.
By higher-order quadratic-curvature scalar gravity, we mean models
containing the $R^{2}$ term in the action\footnote{Recent developments in pure $R^{2}$ gravity can be found, e.g., in Ref. \cite{Edery2019} and references therein.} and also invariants built from derivatives of the curvature scalar,
viz. $R\square R$. In actuality, some papers from the 1990s explored
such terms \cite{Wands1994} in the inflationary context \cite{Berkin1990,Gottlober1990,Gottlober1991,Amendola1993}.
References \cite{JCAP2019,Castellanos2018,Iihoshi2011} explore an
equivalent inflationary model on a deeper level and constrain the
coupling constant associated with the higher-order contribution via
observational data. Higher-order gravity models of the type $f\left(R,\nabla R,\dots,\nabla^{n}R\right)$
may be cast in their scalar-multitensor equivalent form with some
interesting consequences \cite{PRD2016,PRD2019}.

Usually, the business of constraining modified gravity models is done
in the realm of cosmological data \cite{Sotiriou2010,DeFelice2010,Nojiri2011,Nojiri2017,GERG2015,Capozziello2017},
but gravitational wave observations could also contribute to the task
on a different scale \cite{Berti2015,Naf2011,DeLaurentis2013,Capozziello2009,Kuntz2017}.
It is with this in mind that we seek to investigate the GW predicted
for binary inspirals in the context of a higher-order quadratic-curvature scalar
model. Our model contains the terms $R^{2}$ and $R\square R$ alongside
the Einstein-Hilbert term in the action integral of the model. We
will show that this model, when linearized on a flat background, gives
rise to a spin-2 mode of the gravitational waves (consonant to the
$h_{\mu\nu}$ predicted within GR) plus two spin-0 modes, $\Phi_{-}$
and $\Phi_{+}$. These two scalar modes are coupled in a nontrivial
way, one serving as source for the other. The scalar mode $\Phi_{+}$
associated with the $R\square R$ contribution is treated perturbatively.\footnote{It is worth emphasizing that although we refer to ``two spin-0 modes
$\Phi_{-}$ and $\Phi_{+}$,'' there exists only one scalar degree
of freedom. }

Great care is devoted to building the solutions to the linearized
field equations in a self-contained way. Accordingly, our paper shows
explicitly how to apply the Green's method to solve the differential
equations for the modified gravity model, and how the functional forms
of $\Phi_{-}$ in terms of integrals involving Bessel functions are
born. In the same spirit, we detail the calculation of the energy-momentum
pseudotensor in the higher-order $R^{2}$ model, which involves many
more terms than the treatment given by GR. This is potentially useful
for other works dealing with models containing higher-order derivatives
of the curvature-based scalars. We should point out that other works
have approached gravitational waves for models involving the $R^{2}$ term---see,
e.g., Refs. \cite{Cano2019,Berry2011,Naf2010,Capozziello2011,Aguiar2003}.
Our contribution is the thorough step-by-step derivation of the results,
from the linearized field equation until the final constraining of
the associated coupling, besides the inclusion of the term $R\square R$
and the detailed study of its corrections to gravitational wave
physics.

Our modified gravitational wave solutions are specified for binary
systems, aiming for contact with observations later in the paper. We take
the approximation of the nonrelativistic motion of the pair of astrophysical
objects in the binary system; a circular orbit is also assumed. The
spin-0 mode $\Phi_{-}$ related to the $R^{2}$ contribution is carefully
computed in this context; it exhibits two different regimes according
to a well-defined scale. In fact, there is a propagating oscillatory
regime for $\Phi_{-}$ and a damped regime.

The modified GWs emitted by binary black holes are calculated. Their
waveform is compared to the one predicted by GR. It is shown that
the rise in the amplitude and frequency of the strain before the chirp
is more pronounced in our higher-order model than in GR, although the
difference is subtle. This result is qualitatively interesting, but
the real constraint to our model comes from its application to the
description of gravitational waves emitted by binary pulsars. Incidentally,
there is a difference in the treatment of black-hole binaries with
respect to neutron-star binaries in our higher-order quadratic-curvature scalar
model: the orbital dynamics of the binary system is different in each
instance. In any case, we use the binary pulsar investigated by Hulse,
Taylor, and Weisberg \cite{Hulse1975,Taylor1982,Weisberg2004} to constrain
the coupling constants $\kappa_{0}$ and $\beta_{0}$ associated with
the extra terms $R^{2}$ and $R\square R$ in the action. These constraints
contribute to the existing ones in the literature \cite{Berry2011,Naf2011}.

Our conventions follow the ones in Ref. \cite{Carroll2004}: i.e.,
the Minkowski metric is written as $\left(\eta_{\mu\nu}\right)=\text{diag}\left(-1,+1,+1,+1\right)$
in Cartesian coordinates; the Riemann tensor is defined as $R_{\,\sigma\mu\nu}^{\rho}=\partial_{\mu}\Gamma_{\,\nu\sigma}^{\rho}-\partial_{\nu}\Gamma_{\,\mu\sigma}^{\rho}+\Gamma_{\,\mu\lambda}^{\rho}\Gamma_{\,\nu\sigma}^{\lambda}-\Gamma_{\,\nu\lambda}^{\rho}\Gamma_{\,\mu\sigma}^{\lambda}$
; and, the Ricci tensor is given by the contraction $R_{\mu\nu}=R_{\,\mu\rho\nu}^{\rho}$.

The structure of the paper is as follows: In Sec. \ref{sec:HOSG-WeakField},
the action integral for the higher-order $R^{2}$ model is presented,
the equations of motion are written, and the weak-field regime of
those equations is carried out. Even in the linearized form, the field
equations are complex enough to demand a method of order deduction
for enabling their solution; this method is laid down in Sec. \ref{sec:Scalar-tensor-decomposition}
and is based on a scalar-tensor decomposition of the metric tensor.
Section \ref{sec:Solutions} presents the solutions to the field equations
and their multipole expansion; the expressions are given in terms
of the energy-momentum tensor of the source. In Sec. \ref{sec:GW-BS},
the form of the energy-momentum tensor is particularized to the case
of a binary system, the closed functional forms for the gravitational
waves are determined, and the expression for the power radiated in
the form of GWs is computed under the assumption of a circular orbit
approximation. Section \ref{sec:BH-Inspirals} is devoted to the study
of binary black holes: the focus is to determine how the additional
terms in our modified gravity model alter the waveform of the gravitational
wave. Binary pulsar inspirals are studied in Sec. \ref{sec:BP-Inspirals},
where the main objective is to constrain the extra couplings of our
higher-order quadratic-curvature scalar model through the observational
data coming from the Hulse-Taylor binary pulsar. Our final comments
are given in Sec. \ref{sec:Final-comments}.

\section{HIGHER-ORDER $R^{2}$ GRAVITY IN THE WEAK-FIELD REGIME \label{sec:HOSG-WeakField}}

We begin by stating the action of our model: 
\begin{equation}
S=\frac{1}{c}\frac{1}{2\chi}\int d^{4}x\sqrt{-g}\left[R+\frac{1}{2\kappa_{0}}R^{2}-\frac{\beta_{0}}{2\kappa_{0}^{2}}R\square R+\mathcal{L}_{m}\right]\qquad\left(\chi=\frac{8\pi G}{c^{4}}=\frac{1}{M_{P}^{2}}\right)\,.\label{eq:Action}
\end{equation}
We call quadratic-curvature scalar the contribution from the term
scaling with $R^{2}$. The higher-order contribution corresponds to
the $R\square R$ term. (The D'Alembertian operator $\square\equiv g^{\mu\nu}\nabla_{\mu}\nabla_{\nu}$
is given in terms of the covariant derivative $\nabla_{\mu}$ contraction.)
The coupling constant of the higher-order term is written as $\left(\beta_{0}/2\kappa_{0}^{2}\right)$
so that $\beta_{0}$ is a dimensionless constant. The coupling $\kappa_{0}$
has the dimension of $\left(\text{length}\right){}^{-2}$. Notice that
the Einstein-Hilbert action is recovered in the limit of a large $\kappa_{0}$
and a small $\beta_{0}$.

The equation of motion is obtained from Eq. (\ref{eq:Action}) by varying
with respect to $g^{\mu\nu}$: 
\begin{equation}
G_{\mu\nu}+\frac{1}{2\kappa_{0}}H_{\mu\nu}+\frac{\beta_{0}}{2\kappa_{0}^{2}}I_{\mu\nu}=\chi T_{\mu\nu}\,,\label{eq:FieldEq}
\end{equation}
where 
\begin{equation}
G_{\mu\nu}=R_{\mu\nu}-\frac{1}{2}g_{\mu\nu}R\label{eq:EinsteinTensor}
\end{equation}
is the standard Einstein tensor \cite{Aguiar2003}, 
\begin{equation}
H_{\mu\nu}\equiv R\left(R_{\mu\nu}+G_{\mu\nu}\right)-2\left(\nabla_{\nu}\nabla_{\mu}R-g_{\mu\nu}\square R\right)\label{eq:Hmunu}
\end{equation}
is the contribution coming from the $R^{2}$ term, and \cite{AstroSpScien2011}
\begin{equation}
I_{\mu\nu}\equiv\nabla_{\mu}R\nabla_{\nu}R-2R_{\mu\nu}\square R-\frac{1}{2}g_{\mu\nu}\nabla_{\rho}R\nabla^{\rho}R+2\left(\nabla_{\nu}\nabla_{\mu}\square R-2g_{\mu\nu}\square^{2}R\right)\label{eq:Imunu}
\end{equation}
stems from the higher-order term.

In the following, we work out the weak-field approximation of the
above field equation in a Minkowski background. The metric tensor will
be written as 
\begin{equation}
g_{\mu\nu}=\eta_{\mu\nu}+h_{\mu\nu}\,,\label{eq:g_WF}
\end{equation}
with $\left|h_{\mu\nu}\right|\ll1$ so that the linear approximations
hold. Then, $g^{\mu\nu}=\eta^{\mu\nu}-h^{\mu\nu}$ and $h=\eta^{\mu\nu}h_{\mu\nu}$.
It pays off to introduce the definition of the the trace-reverse tensor:
\begin{equation}
\bar{h}_{\mu\nu}=h_{\mu\nu}-\frac{1}{2}\eta_{\mu\nu}h\,,\label{eq:hbar}
\end{equation}
in terms of which Eq. (\ref{eq:EinsteinTensor}) reads 
\begin{equation}
G_{\mu\nu}=-\frac{1}{2}\left\{ \eta^{\rho\sigma}\partial_{\rho}\partial_{\sigma}\bar{h}_{\mu\nu}+\left[\eta_{\mu\nu}\partial_{\rho}\left(\partial_{\sigma}\bar{h}^{\rho\sigma}\right)-\partial_{\mu}\left(\partial^{\rho}\bar{h}_{\nu\rho}\right)-\partial_{\nu}\left(\partial^{\rho}\bar{h}_{\mu\rho}\right)\right]\right\} \,.\label{eq:EinsteinTensor_WF(hbar)}
\end{equation}

The $R^{2}$-term contribution $H_{\mu\nu}$ in the weak-field regime
is 
\begin{equation}
H_{\mu\nu}=-2\left[\partial_{\nu}\partial_{\mu}-\eta_{\mu\nu}\eta^{\rho\sigma}\partial_{\rho}\partial_{\sigma}\right]R\,,\label{eq:Hmunu_WF(R)}
\end{equation}
where $R$ is now expressed in terms of the trace-reverse tensor:
\begin{equation}
R=\partial_{\alpha}\partial_{\beta}\bar{h}^{\alpha\beta}+\frac{1}{2}\eta^{\alpha\beta}\partial_{\alpha}\partial_{\beta}\bar{h}\,.\label{eq:CurvScalar_WF(hbar)}
\end{equation}
This quantity also appears in the contribution $I_{\mu\nu}$ from
the $R\square R$ term in the weak-field regime: 
\begin{equation}
I_{\mu\nu}=2\left[\partial_{\nu}\partial_{\mu}-2\eta_{\mu\nu}\left(\eta^{\lambda\tau}\partial_{\lambda}\partial_{\tau}\right)\right]\left(\eta^{\rho\sigma}\partial_{\rho}\partial_{\sigma}R\right)\,.\label{eq:Imunu_WF(R)}
\end{equation}

We are now in a position to write down the entire field equation (\ref{eq:FieldEq})
in the weak-field approximation by using the results (\ref{eq:EinsteinTensor_WF(hbar)}),
(\ref{eq:Hmunu_WF(R)}), and (\ref{eq:Imunu_WF(R)}) for $G_{\mu\nu}$,
$H_{\mu\nu}$, and $I_{\mu\nu}$, respectively. In fact, 
\begin{align}
\square\bar{h}_{\mu\nu}+\left[\eta_{\mu\nu}\partial_{\rho}\partial_{\sigma}\bar{h}^{\rho\sigma}-\partial_{\mu}\partial^{\rho}\bar{h}_{\nu\rho}-\partial_{\nu}\partial^{\rho}\bar{h}_{\mu\rho}\right]\nonumber \\
+\frac{2}{\kappa_{0}}\left(\partial_{\mu}\partial_{\nu}-\eta_{\mu\nu}\square\right)\left[\partial^{\rho}\partial^{\sigma}\left(\bar{h}_{\rho\sigma}+\frac{1}{2}\eta_{\rho\sigma}\bar{h}\right)\right]\nonumber \\
-\frac{2\beta_{0}}{\kappa_{0}^{2}}\left(\partial_{\mu}\partial_{\nu}-\eta_{\mu\nu}\square\right)\left[\square\partial^{\rho}\partial^{\sigma}\left(\bar{h}_{\rho\sigma}+\frac{1}{2}\eta_{\rho\sigma}\bar{h}\right)\right] & =-2\chi T_{\mu\nu}\,,\label{eq:FieldEq_WF_Box}
\end{align}
where the first and second terms on the lhs are those obtained
in standard GR, the third term comes from the $R^{2}$ contribution
to the action, and the fourth term is born from the higher-order term
$R\square R$. (In the weak-field regime, it turns out that $\square=\eta^{\rho\sigma}\partial_{\rho}\partial_{\sigma}$.)

The field equation in the weak-field approximation is considerably
simplified in GR when one uses the harmonic gauge fixing condition,
$\partial_{\nu}\bar{h}^{\mu\nu}=0$. This is very satisfying, because
it eliminates the last three terms in the first line of the field
equation (\ref{eq:FieldEq_WF_Box}), and we are left with $\boxempty\bar{h}_{\mu\nu}=-2\chi T_{\mu\nu}$.
However, the harmonic gauge on $\bar{h}^{\mu\nu}$ might not be useful
or even attainable in higher-order gravity \cite{Berry2011}. So, we
should study the gauge fixing in higher-order gravity with greater
care. This will be dealt with in the next section.

For the time being, we take the trace of the linearized field equation:
\begin{equation}
\square\bar{h}+2\partial^{\nu}\partial^{\rho}\bar{h}_{\nu\rho}-\frac{6}{\kappa_{0}}\square\left[\partial^{\rho}\partial^{\sigma}\left(\bar{h}_{\rho\sigma}+\frac{1}{2}\eta_{\rho\sigma}\bar{h}\right)\right]+\frac{6\beta_{0}}{\kappa_{0}^{2}}\square\left[\square\partial^{\rho}\partial^{\sigma}\left(\bar{h}_{\rho\sigma}+\frac{1}{2}\eta_{\rho\sigma}\bar{h}\right)\right]=-2\chi T\,.\label{eq:FieldEq_WF_Tr}
\end{equation}
The solutions to the sixth-order differential equation (\ref{eq:FieldEq_WF_Box})
will be found after decomposing the trace-reverse tensor $\bar{h}_{\mu\nu}$
into a scalar part and a tensorial sector.

\section{SCALAR-TENSOR DECOMPOSITION IN THE WEAK-FIELD REGIME AND THE MODIFIED
HARMONIC GAUGE \label{sec:Scalar-tensor-decomposition}}

We define a new dimensionless scalar field\textbf{ $\Phi$} proportional
to the curvature scalar $R$: 
\begin{equation}
\kappa_{0}\Phi\equiv R=\partial^{\mu}\partial^{\nu}\left(\bar{h}_{\mu\nu}+\frac{1}{2}\eta_{\mu\nu}\bar{h}\right)\,,\label{eq:Phi}
\end{equation}
and we decompose the linearized metric into a form depending on both $\Phi$
and $\square\Phi$: 
\begin{equation}
\bar{h}_{\mu\nu}=\tilde{h}_{\mu\nu}+\eta_{\mu\nu}\left(\Phi-\frac{\beta_{0}}{\kappa_{0}}\square\Phi\right)\,,\label{eq:hbar_decomp}
\end{equation}
so that the trace is 
\begin{equation}
\bar{h}=\tilde{h}+4\left(\Phi-\frac{\beta_{0}}{\kappa_{0}}\square\Phi\right)\,.\label{eq:hbar_trace_decomp}
\end{equation}
Our trace-reverse tensor decomposition proposal [Eq. (\ref{eq:hbar_decomp})]
was inspired by the content in Ref. \cite{Accioly2016}.

We choose a gauge condition such that 
\[
\partial_{\mu}\bar{h}^{\mu\nu}=\partial^{\nu}\left(\Phi-\frac{\beta_{0}}{\kappa_{0}}\square\Phi\right)\,,
\]
which leads to the harmonic gauge on\textbf{ $\tilde{h}^{\mu\nu}$}:
\begin{equation}
\partial_{\mu}\tilde{h}^{\mu\nu}=0\qquad\left(\text{harmonic gauge}\right)\,.\label{eq:HarmGauge_h_tilde}
\end{equation}

Using Eqs. (\ref{eq:hbar_decomp}), (\ref{eq:hbar_trace_decomp}), and (\ref{eq:HarmGauge_h_tilde}),
the linearized field equation (\ref{eq:FieldEq_WF_Box}) reads 
\begin{equation}
\square\tilde{h}_{\mu\nu}=-2\chi T_{\mu\nu},\label{eq:FieldEq_h_tilde}
\end{equation}
and the trace of the linearized field equation (\ref{eq:FieldEq_WF_Tr})
reduces to 
\begin{equation}
\frac{\beta_{0}}{\kappa_{0}^{2}}\square^{2}\Phi-\frac{1}{\kappa_{0}}\square\Phi+\frac{1}{3}\Phi=-\frac{\chi}{3\kappa_{0}}T\,.\label{eq:FieldEq_Phi(T)}
\end{equation}
This is a fourth-order differential equation and, as such, is a delicate
matter to deal with. We will reduce its order below.

We claim that Eq. (\ref{eq:FieldEq_Phi(T)}) can the cast into the
form 
\begin{equation}
\left(\frac{\square}{\kappa_{0}}-K_{-}\right)\left(\beta_{0}\frac{\square}{\kappa_{0}}-K_{+}\right)\Phi=-\frac{\chi}{3\kappa_{0}}T\,.\label{eq:DiffEqKK}
\end{equation}
In order to determine the constants $K_{\pm}$ that allow for this
possibility, we open up the above expression and compare with Eq. (\ref{eq:FieldEq_Phi(T)}).
The result is 
\begin{align}
K_{-} & =\frac{1}{2\beta_{0}}\left[1-\sqrt{1-\frac{4}{3}\beta_{0}}\right]\,,\label{eq:Kminus}\\
K_{+} & =\frac{1}{2}\left[1+\sqrt{1-\frac{4}{3}\beta_{0}}\right].\label{eq:Kplus}
\end{align}

Let us further define 
\begin{equation}
\Phi\equiv\Phi_{+}\,,\label{eq:Phi_plus}
\end{equation}
and, motivated by a sector on the lhs of Eq. (\ref{eq:DiffEqKK}),
the new field 
\begin{equation}
\left(\beta_{0}\frac{\square}{\kappa_{0}}-K_{+}\right)\Phi_{+}\equiv\Phi_{-}\,.\label{eq:Phi_minus}
\end{equation}

Equation (\ref{eq:DiffEqKK}) is then written as 
\begin{equation}
\left(\frac{\square}{\kappa_{0}}-K_{-}\right)\Phi_{-}=-\frac{\chi}{3\kappa_{0}}T\,.\label{eq:DiffEq_Phi_minus}
\end{equation}
Now, Eqs. (\ref{eq:Phi_minus}), and (\ref{eq:DiffEq_Phi_minus}) may
be written in a more compelling form: 
\begin{equation}
\left(\square-m_{+}^{2}\right)\Phi_{+}=\frac{\kappa_{0}}{\beta_{0}}\Phi_{-}\,,\label{eq:FieldEq_Phi_plus}
\end{equation}
and 
\begin{equation}
\left(\square-m_{-}^{2}\right)\Phi_{-}=-\frac{\chi}{3}T\,,\label{eq:FieldEq_Phi_minus}
\end{equation}
where 
\begin{equation}
m_{\pm}^{2}=\frac{\kappa_{0}}{2\beta_{0}}\left[1\pm\sqrt{1-\frac{4}{3}\beta_{0}}\right].\label{eq:m_plus_minus}
\end{equation}
Notice that $\left[m_{\pm}^{2}\right]=\left(\text{length}\right)^{-2}$;
these parameters could be interpreted as masses of the fields $\Phi_{\pm}$.
The original differential equation (\ref{eq:FieldEq_Phi(T)}) was
thus reduced from a fourth-order equation to a pair of second-order
differential equations, viz. Equations (\ref{eq:FieldEq_Phi_plus}) and
(\ref{eq:FieldEq_Phi_minus}).

By using the definitions of the new scalar fields $\Phi_{\pm}$ in
Eq. (\ref{eq:hbar_decomp}) for $\bar{h}_{\mu\nu}$, we conclude that
the spacetime wiggles are given by 
\begin{equation}
\bar{h}_{\mu\nu}=\tilde{h}_{\mu\nu}+\eta_{\mu\nu}\left[\frac{\beta_{0}}{\kappa_{0}}m_{-}^{2}\Phi_{+}-\Phi_{-}\right]\,.\label{eq:h_bar(h_tilde,Phi_plus,Phi_minus)}
\end{equation}
This is the scalar-tensor decomposition of the gravitational waves
in our higher-order $R^{2}$ gravity.

The script to determine the scalar degree of freedom $\Phi$ in our
model---see Eq. (\ref{eq:Phi})---is the following: Given the source
type characterized by $T$, the first step is to integrate Eq. (\ref{eq:FieldEq_Phi_minus})
and find $\Phi_{-}$. The next step is to use $\Phi_{-}$ as the source
of $\Phi_{+}$, which is computed through Eq. (\ref{eq:FieldEq_Phi_plus}).
The field $\Phi_{-}$ is determined univocally given the adequate
boundary conditions (such as the requirement for a radiative solution);
$\Phi_{+}$ is then also determined univocally in terms of $\Phi_{-}$,
as will be explicitly shown in Sec. \ref{subsec:FieldEq}. The
conclusion is that there is a single scalar degree of freedom corresponding
to $\Phi$; the higher-order quadratic scalar curvature model exhibits
three degrees of freedom total: two encapsulated in $\tilde{h}_{\mu\nu}$,
and one embodied by $\Phi$.


\textit{Remark: $R^{2}$ gravity limit.}---If $\beta_{0}\ll1$, the mass parameters $m_{\pm}$ in Eq. (\ref{eq:m_plus_minus}) reduce to 
\begin{equation}
m_{+}^{2}\approx\frac{\kappa_{0}}{\beta_{0}}\left(1-\frac{1}{3}\beta_{0}\right)\text{ \ and \ }m_{-}^{2}\approx\frac{\kappa_{0}}{3}\left(1+\frac{1}{3}\beta_{0}\right)\,.\label{eq:m_pm_betasmall}
\end{equation}
Therefore, in the limit as $\beta_{0}\rightarrow0$, the decomposed
field equations (\ref{eq:Phi_minus}) and (\ref{eq:FieldEq_Phi_minus})
simplify to 
\[
\left(\square-\frac{\kappa_{0}}{3}\right)\Phi_{-}=-\frac{\chi}{3}T\text{ \ and \ }\left(\frac{\beta_{0}}{\kappa_{0}}\square-1\right)\Phi_{+}=\Phi_{-}\Rightarrow\Phi=\Phi_{+}\simeq-\Phi_{-}\,.
\]
This leads to the structure 
\[
\left(\square-\frac{\kappa_{0}}{3}\right)\Phi=\frac{\chi}{3}T\text{ }\,,
\]
which is typical of a pure $R^{2}$-gravity model \cite{Berry2011}.
$\blacksquare$


In the next section, we work on determining solutions to the field
equations (\ref{eq:FieldEq_h_tilde}), (\ref{eq:FieldEq_Phi_plus}),
and (\ref{eq:FieldEq_Phi_minus}) in order to characterize the gravitational
waves $\bar{h}_{\mu\nu}$.

\section{SOLUTIONS TO THE LINEARIZED FIELD EQUATIONS\label{sec:Solutions}}

\subsection{The sourceless equations \label{subsec:SourcelessEqs}}

The sourceless version of Eq. (\ref{eq:FieldEq_h_tilde}) is solved
by $\tilde{h}_{\mu\nu}=\epsilon_{\mu\nu}e^{iq_{\rho}x^{\rho}}+\epsilon_{\mu\nu}^{*}e^{-iq_{\rho}x^{\rho}}$.
This is the plane wave solution, $\tilde{h}_{\mu\nu}$ is a massless
spin-2 field, and the dispersion relation is 
\begin{equation}
\left|\mathbf{q}\right|=\frac{\omega}{c}\,,\label{eq:q}
\end{equation}
where $q^{0}=\omega/c$ and $\mathbf{q}=\left(q^{i}\right)$.

By setting $T=0$ in Eq. (\ref{eq:FieldEq_Phi_minus}), we get a differential
equation whose solution is $\Phi_{-}=\phi_{-}e^{ik_{\mu}x^{\mu}}+\phi_{-}^{*}e^{-ik_{\mu}x^{\mu}}$,
where 
\begin{equation}
\left|\mathbf{k}\right|=\sqrt{\frac{\omega^{2}}{c^{2}}-m_{-}^{2}}\,,\label{eq:k}
\end{equation}
with $k^{0}=\omega/c$ and $\mathbf{k}=\left(k^{i}\right)$. Then,
we are faced with two different possibilities: 
\begin{itemize}
\item [{(I.)}] $\omega\geqslant m_{-}c$. In this case, $\mathbf{k}$ is
a real number, the arguments of the exponentials in $\Phi_{-}$ are
complex, and the solution is oscillatory. 
\item [{(II.)}] $\omega<m_{-}c$. Here, $\mathbf{k}$ is a complex number,
the arguments of the exponentials in $\Phi_{-}$ are real, and the
solution is either decreasing (damped) or increasing with $x^{\mu}$.
The increasing solution blows up as $x\rightarrow\infty$ and should
be neglected as unphysical. 
\end{itemize}
The $\Phi_{+}$ field equation does not have a sourceless version:
$\Phi_{-}$ is the very source of $\Phi_{+}$, as indicated by Eq. (\ref{eq:FieldEq_Phi_plus}).

\subsection{The complete field equations \label{subsec:FieldEq}}

In the presence of a source, Eq. (\ref{eq:FieldEq_h_tilde}) is solved by 
\begin{equation}
\tilde{h}_{\mu\nu}=\frac{4G}{c^{4}}\int d^{3}x^{\prime}\frac{1}{\left|\mathbf{x}-\mathbf{x}^{\prime}\right|}T_{\mu\nu}\left(t_{r},\mathbf{x}^{\prime}\right)\,,\label{eq:h_tilde(T)}
\end{equation}
where $t_{r}=t-\left|\mathbf{x}-\mathbf{x}^{\prime}\right|/c$ is
the retarded time, and the vector $\mathbf{x}$ (the vector $\mathbf{x}^{\prime}$)
points from the origin of the coordinate system to the observer (to
the source). Equation (\ref{eq:h_tilde(T)}) is obtained through the retarded
Green's function. This is standard in GR \cite{Sabbata1985,Carroll2004},
except that in our case, $\tilde{h}_{\mu\nu}$ appears instead of $\bar{h}_{\mu\nu}$.

The formal solution of Eq. (\ref{eq:FieldEq_Phi_minus}) for $\Phi_{-}$
is in terms of the Green's function $G_{\Phi}\left(x^{\mu};x^{\prime\mu}\right)$:
\begin{equation}
\Phi_{-}\left(x^{\mu}\right)=-\frac{\chi}{3}\int G_{\Phi}\left(x^{\mu};x^{\prime\mu}\right)T\left(x^{\prime\mu}\right)d^{4}x^{\prime}\,.\label{eq:Phi_minus(G,T)}
\end{equation}
Following the detailed procedure in Appendix \ref{Appendix:Phi_minusFieldEq},
the explicit form of the retarded Green's function is [see Eq. (\ref{eq:G_Phi_Massive})]
\begin{equation}
G_{\Phi}\left(x^{\nu};x^{\prime\nu}\right)=-\frac{1}{4\pi}\frac{1}{c}\frac{1}{s}\delta\left(\tau-\frac{s}{c}\right)+\frac{1}{4\pi}\Theta\left(\tau-\frac{s}{c}\right)\frac{1}{c}\frac{m_{-}}{\sqrt{\tau^{2}-\left(\frac{s}{c}\right)^{2}}}J_{1}\left(m_{-}c\sqrt{\tau^{2}-\left(\frac{s}{c}\right)^{2}}\right)\,,\label{eq:G_Phi}
\end{equation}
where $\Theta\left(\xi\right)$ is the Heaviside step function, $\tau=\left(t-t^{\prime}\right)$,
and $s=\left|\mathbf{s}\right|=\left|\mathbf{x}-\mathbf{x}^{\prime}\right|$.
$J_{1}$ is the Bessel function of the first kind. Plugging Eq. (\ref{eq:G_Phi})
into Eq. (\ref{eq:Phi_minus(G,T)}) and manipulating with respect to $t^{\prime}$
leads to \cite{Naf2011} 
\begin{equation}
\Phi_{-}\left(\mathbf{x},t\right)=\frac{2}{3}\frac{G}{c^{4}}\int\frac{1}{s}T\left(\mathbf{x}^{\prime},t_{r}\right)d^{3}\mathbf{x}^{\prime}-m_{-}\frac{2}{3}\frac{G}{c^{4}}\int d^{3}\mathbf{x}^{\prime}\int_{\frac{s}{c}}^{\infty}\frac{J_{1}\left(m_{-}c\sqrt{\tau^{2}-\left(\frac{s}{c}\right)^{2}}\right)}{\sqrt{\tau^{2}-\left(\frac{s}{c}\right)^{2}}}T\left(\mathbf{x}^{\prime},t-\tau\right)d\tau\,.\label{eq:Phi_minus(T)}
\end{equation}
This equation appears in Ref. \cite{Cano2019} in the limit of great
distances from the source, where $s\simeq\left|\mathbf{x}\right|$.


\textit{Remark: The limits $m_{-}\rightarrow\infty$ and $m_{-}\rightarrow0$.}---
Let us comment on the form of Eq. (\ref{eq:Phi_minus(T)}) in relation
to the possible limits of the mass parameter $m_{-}$. For that, we
note that the change of variable $\xi=m_{-}c\sqrt{2\frac{s}{c}\left(\tau-\frac{s}{c}\right)}$
recasts $\Phi_{-}$ into the form 
\begin{align*}
\Phi_{-}\left(\mathbf{x},t\right)= & \frac{2}{3}\frac{G}{c^{4}}\int\frac{1}{s}T\left(\mathbf{x}^{\prime},t_{r}\right)d^{3}\mathbf{x}^{\prime}\\
 & -\frac{2}{3}\frac{G}{c^{4}}\int d^{3}\mathbf{x}^{\prime}\int_{0}^{\infty}\frac{J_{1}\left(\xi\sqrt{1+\left(\frac{\xi}{2m_{-}s}\right)^{2}}\right)}{s\sqrt{1+\left(\frac{\xi}{2m_{-}s}\right)^{2}}}T\left(\mathbf{x}^{\prime},t_{r}-2\frac{s}{c}\left(\frac{\xi}{2m_{-}s}\right)^{2}\right)d\xi\,.
\end{align*}
In the limit $m_{-}\rightarrow\infty$, this reduces to 
\begin{equation}
\Phi_{-}\left(\mathbf{x},t\right)\approx\frac{2}{3}\frac{G}{c^{4}}\int\frac{1}{s}T\left(\mathbf{x}^{\prime},t_{r}\right)d^{3}\mathbf{x}^{\prime}-\frac{2}{3}\frac{G}{c^{4}}\int\frac{1}{s}T\left(\mathbf{x}^{\prime},t_{r}\right)d^{3}\mathbf{x}^{\prime}\int_{0}^{\infty}J_{1}\left(\xi\right)d\xi=0\qquad\left(m_{-}\rightarrow\infty\right)\,,\label{eq:Phi_minus_large_m_minus}
\end{equation}
because $\int_{0}^{\infty}J_{1}\left(\xi\right)d\xi=1$. This was
expected: $m_{\ensuremath{-}}\rightarrow\infty$ amounts to taking $\kappa_{0}\rightarrow\infty$.
This means that the quadratic-curvature scalar term $\propto\kappa_{0}^{-1}R^{2}$
vanishes from the action, and we are left with the standard GR case.
This is the same as stating that the scalar mode vanishes.

On the other hand, for $m_{-}=0$, Eq. (\ref{eq:Phi_minus(T)}) reduces
to 
\begin{equation}
\Phi_{-}\left(\mathbf{x},t\right)=\frac{2}{3}\frac{G}{c^{4}}\int\frac{1}{\left|\mathbf{x}-\mathbf{x}^{\prime}\right|}T\left(\mathbf{x}^{\prime},t_{r}\right)d^{3}\mathbf{x}^{\prime}\qquad\left(m_{-}=0\right)\,.\label{eq:Phi_minus_small_m_minus}
\end{equation}
In this way, we see that the scalar mode loses its contribution from
the mass and exhibits a similar behavior to the massless spin-2 mode
of GR. $\blacksquare$


The form of the differential equation for $\Phi_{+}$ [Eq. (\ref{eq:FieldEq_Phi_plus})]
is the same as the one for $\Phi_{-}$ [Eq. (\ref{eq:FieldEq_Phi_minus})]:
it suffices to perform the mapping $\Phi_{-}\mapsto\Phi_{+}$, $m_{-}\mapsto m_{+}$,
and $\left(-\chi/3\right)T\mapsto\left(\kappa_{0}/\beta_{0}\right)\Phi_{-}$
to obtain the former from the later. Therefore, the solution to Eq. (\ref{eq:FieldEq_Phi_plus})
should be formally the same as the solution to Eq. (\ref{eq:FieldEq_Phi_minus}).
In fact, 
\begin{align}
\Phi_{+}\left(\mathbf{r},t\right)= & -\frac{1}{4\pi}\frac{\kappa_{0}}{\beta_{0}}\int\frac{1}{s}\Phi_{-}\left(\mathbf{x}^{\prime},t_{r}\right)d^{3}\mathbf{x}^{\prime}\nonumber \\
 & +\frac{m_{+}}{4\pi}\frac{\kappa_{0}}{\beta_{0}}\int d^{3}\mathbf{x}^{\prime}\int_{\frac{s}{c}}^{\infty}\frac{J_{1}\left(m_{+}c\sqrt{\tau^{2}-\left(\frac{s}{c}\right)^{2}}\right)}{\sqrt{\tau^{2}-\left(\frac{s}{c}\right)^{2}}}\Phi_{-}\left(\mathbf{x}^{\prime},t-\tau\right)d\tau\,.\label{eq:Phi_plus(T)}
\end{align}
As a consistency check, it pays to consider the limit $\beta_{0}\rightarrow0$:
This corresponds to the vanishing of the higher-order contribution
$R\square R$ in the action integral [Eq. (\ref{eq:Action})]. According
to Eq. (\ref{eq:m_pm_betasmall}), $\beta_{0}\rightarrow0$ leads to $m_{+}\rightarrow\infty$.
Then, $\Phi_{+}\rightarrow0$ by the same token below Eq. (\ref{eq:Phi_minus(T)}).

Equation (\ref{eq:Phi_minus(T)}) for $\Phi_{-}\left(\mathbf{x},t\right)$
is relatively easy to solve for pointlike sources, because the energy-momentum
tensor and its trace are given in terms of a Dirac delta. This shall
be done in a subsequent section. Notice that $\Phi_{-}$ thus calculated
is a source term for $\Phi_{+}$ in Eq. (\ref{eq:Phi_plus(T)}). This
introduces nonlocality in the computation of $\Phi_{+}$, since $\Phi_{-}$
is spread throughout the spacetime. Nonlocality is a subtle feature
to handle. However, we can bypass this difficulty by considering the
special case where $\beta_{0}\ll1$, which corresponds to regarding the
$R\square R$ contribution as a small perturbation when compared to
the $R+R^{2}$ part of the model. In this case, we invoke Eq. (\ref{eq:m_pm_betasmall})
to write Eq. (\ref{eq:FieldEq_Phi_plus}) as 
\[
\Phi_{+}\simeq-\Phi_{-}+\frac{\beta_{0}}{\kappa_{0}}\left(\square+\frac{\kappa_{0}}{3}\right)\Phi_{+}\qquad\left(\beta_{0}\ll1\right)\,.
\]
As long as $\beta_{0}$ is small, $\Phi_{+}=-\Phi_{-}+\left(\text{correction of order }\beta_{0}\right)$.
In this case, the higher-order terms are subdominant with respect
to the $R^{2}$ contribution. Therefore, we are allowed to replace
$\Phi_{+}=-\Phi_{-}$ in the last term of the above equation as a
first approximation: 
\begin{equation}
\Phi_{+}\simeq-\Phi_{-}-\frac{\beta_{0}}{\kappa_{0}}\left(\square+\frac{\kappa_{0}}{3}\right)\Phi_{-}\qquad\left(\beta_{0}\ll1\right)\,.\label{eq:Phi_plus(Phi_minus,beta0)}
\end{equation}
Now, the term $\square\Phi_{-}$ can be eliminated by utilizing the
field equation for $\Phi_{-}$, [Eq. (\ref{eq:FieldEq_Phi_minus})]
and the approximation for $m_{\pm}$ [Eq. (\ref{eq:m_pm_betasmall})].
Indeed, 
\[
\square\Phi_{-}\simeq-\frac{\chi}{3}T+\frac{\kappa_{0}}{3}\Phi_{-}\qquad\left(\beta_{0}\ll1\right)\,,
\]
so that (\ref{eq:Phi_plus(Phi_minus,beta0)}) reads 
\begin{equation}
\Phi_{+}\simeq-\left(1+\frac{2}{3}\beta_{0}\right)\Phi_{-}+\frac{\beta_{0}}{\kappa_{0}}\frac{\chi}{3}T\qquad\left(\beta_{0}\ll1\right)\,.\label{eq:Phi_plus(Phi-minus,T)}
\end{equation}
We then see that $\Phi_{+}$ can be written in terms of $\Phi_{-}$
and $T$. Consequently, we can account for the higher-order contribution
as a small correction to $R^{2}$ gravity thus avoiding the problem
of nonlocality.

\subsection{Multipole expansion of the scalar field \label{subsec:Multipole-expansion}}

The multipole expansion for the spin-2 mode $\bar{h}_{\mu\nu}$---in
our case, $\tilde{h}_{\mu\nu}$---is a standard procedure readily
found in textbooks \cite{Maggiore2007} as part of the itinerary to
describe gravitational waves. Below, we will develop the multipole
expansion for the spin-0 mode $\Phi_{-}$, Eq.~(\ref{eq:Phi_minus(T)}).
This will automatically apply to the mode $\Phi_{+}$, since it can
be written in terms of $\Phi_{-}$ under the conditions presented
at the end of the previous section.

We use the large-distances approximation: 
\begin{equation}
\frac{s}{c}=\frac{\left\vert \mathbf{x}-\mathbf{x}^{\prime}\right\vert }{c}\simeq\frac{r}{c}-\frac{1}{c}\left(\hat{\mathbf{n}}\cdot\mathbf{x}^{\prime}\right)\,,\label{eq:LargeDistApprox}
\end{equation}
where $r=\left\vert \mathbf{x}\right\vert $, and $\hat{\mathbf{n}}$
is the unit vector pointing along the direction of $\mathbf{x}$.
Moreover, we perform the following change of variables: 
\begin{equation}
\tau_{r}=\tau-\frac{s}{c}\simeq\tau-\frac{r}{c}+\frac{1}{c}\left(\hat{\mathbf{n}}\cdot\mathbf{x}^{\prime}\right)\,.\label{eq:tau_r}
\end{equation}
Hence, the integral with respect to $\tau$ in Eq. (\ref{eq:Phi_minus(T)})
reads: 
\begin{equation}
I_{\tau}=\int_{0}^{\infty}F\left(\tau_{r}\right)T\left(\mathbf{x}^{\prime},\left(t-\frac{r}{c}\right)-\tau_{r}+\frac{1}{c}\left(\hat{\mathbf{n}}\cdot\mathbf{x}^{\prime}\right)\right)d\tau_{r}\,,\label{eq:I_tau}
\end{equation}
with 
\begin{equation}
F\left(\tau_{r}\right)=\frac{J_{1}\left(m_{-}c\sqrt{2\tau_{r}}\sqrt{\frac{\tau_{r}}{2}+\frac{r}{c}}\right)}{\sqrt{2\tau_{r}}\sqrt{\frac{\tau_{r}}{2}+\frac{r}{c}}}\,.\label{eq:F(tau_r)}
\end{equation}
We have neglected the term $-\frac{1}{c}\left(\hat{\mathbf{n}}\cdot\mathbf{x}^{\prime}\right)$
in $F\left(\tau_{r}\right)$ due to the long-distance approximation.
The next step is to expand the trace of the energy-momentum tensor
in Eq. (\ref{eq:I_tau}) about 
\begin{equation}
\zeta\equiv\left(t-\frac{r}{c}\right)-\tau_{r}=t_{r}-\tau_{r}\,.\label{eq:zeta}
\end{equation}
We have 
\begin{equation}
T\left(\mathbf{x}^{\prime},\zeta+\frac{1}{c}\left(\hat{\mathbf{n}}\cdot\mathbf{x}^{\prime}\right)\right)\simeq T\left(\mathbf{x}^{\prime},\zeta\right)+\frac{x^{\prime i}n^{i}}{c}\left.\frac{\partial T}{\partial t}\right|_{\zeta}+\frac{1}{2c^{2}}x^{\prime i}x^{\prime j}n^{i}n^{j}\left.\frac{\partial^{2}T}{\partial t^{2}}\right|_{\zeta}+\cdots\,.\label{eq:T(zeta)}
\end{equation}

There is a $T\left(\mathbf{x}^{\prime},t_{r}\right)$ in the first
term of $\Phi_{-}$, cf. Eq. (\ref{eq:Phi_minus(T)}). It should be
expanded, too: 
\begin{align}
T\left(\mathbf{x}^{\prime},t_{r}\right) & \simeq T\left(\mathbf{x}^{\prime},t_{r}\right)+\frac{x^{\prime i}n^{i}}{c}\left.\frac{\partial T}{\partial t}\right|_{t_{r}}+\frac{1}{2c^{2}}x^{\prime i}x^{\prime j}n^{i}n^{j}\left.\frac{\partial^{2}T}{\partial t^{2}}\right|_{t_{r}}+\cdots\label{eq:T(t_r)}
\end{align}
The two expansions above can be substituted into Eq. (\ref{eq:Phi_minus(T)})
for $\Phi_{-}$. The result is 
\begin{equation}
\Phi_{-}\left(\mathbf{x},t\right)=\Phi_{-}^{M}\left(\mathbf{x},t\right)+\Phi_{-}^{D}\left(\mathbf{x},t\right)+\Phi_{-}^{Q}\left(\mathbf{x},t\right)+\cdots\,,\label{eq:Phi_minusMultipole}
\end{equation}
where 
\begin{align}
\Phi_{-}^{M}\left(\mathbf{x},t\right) & =\frac{2}{3}\frac{G}{c^{4}}\left\{ \frac{1}{r}\left[c^{2}\mathcal{M}\left(t_{r}\right)\right]-m_{-}\int_{0}^{\infty}d\tau_{r}F\left(\tau_{r}\right)\left[c^{2}\mathcal{M}\left(\zeta\right)\right]\right\} \,,\label{eq:Phi_minusM}
\end{align}
\begin{align}
\Phi_{-}^{D}\left(\mathbf{x},t\right) & =\frac{2}{3}\frac{G}{c^{4}}\left\{ \frac{1}{r}\left[cn^{i}\left.\frac{\partial\mathcal{M}^{i}}{\partial t}\right|_{t_{r}}\right]-m_{-}\int_{0}^{\infty}d\tau_{r}F\left(\tau_{r}\right)\left[cn^{i}\left.\frac{\partial\mathcal{M}^{i}}{\partial t}\right|_{\zeta}\right]\right\} \,,\label{eq:Phi_minusD}
\end{align}
\begin{align}
\Phi_{-}^{Q}\left(\mathbf{x},t\right) & =\frac{2}{3}\frac{G}{c^{4}}\left\{ \frac{1}{r}\left[\frac{1}{2}n^{i}n^{j}\left.\frac{\partial^{2}\mathcal{M}^{ij}}{\partial t^{2}}\right|_{t_{r}}\right]-m_{-}\int_{0}^{\infty}d\tau_{r}F\left(\tau_{r}\right)\left[\frac{1}{2}n^{i}n^{j}\left.\frac{\partial^{2}\mathcal{M}^{ij}}{\partial t^{2}}\right|_{\zeta}\right]\right\} \,,\label{eq:Phi_minusQ}
\end{align}
and $\mathcal{M}^{ijk\dots}$ are the mass moments built with the
trace of the energy-momentum tensor: 
\begin{equation}
\begin{cases}
\mathcal{M}\left(t\right)=\frac{1}{c^{2}}\int d^{3}\mathbf{x}^{\prime}T\left(\mathbf{x}^{\prime},t\right)\\
\mathcal{M}^{i}\left(t\right)=\frac{1}{c^{2}}\int d^{3}\mathbf{x}^{\prime}T\left(\mathbf{x}^{\prime},t\right)x^{\prime i}\\
\mathcal{M}^{ij}\left(t\right)=\frac{1}{c^{2}}\int d^{3}\mathbf{x}^{\prime}T\left(\mathbf{x}^{\prime},t\right)x^{\prime i}x^{\prime j}
\end{cases}\,.\label{eq:MassMultipoles}
\end{equation}
Equation (\ref{eq:Phi_minusMultipole}) is the multipole expansion for
the scalar mode $\Phi_{-}$. Therein, $\Phi_{-}^{M}$ is the monopole
term, $\Phi_{-}^{D}$ is the dipole contribution, and $\Phi_{-}^{Q}$
is the quadrupole moment. Equation (\ref{eq:Phi_minusMultipole}) will
be used to estimate the energy carried away by GW via the scalar modes.

Let us remark that the coordinate transformations---such as Eq. (\ref{eq:zeta})---and
the multipole expansion of the form (\ref{eq:Phi_minusMultipole})
are analogous to those in Ref. \cite{Naf2011}.

The fundamental equations for the quantities entering the spacetime
wiggles in our modified gravity model are now laid down. In order
to proceed any further, we should specify the matter-energy content
functioning as the source of gravitational waves. Our choice is to approach
binary systems of astrophysical compact objects.

\section{GW EMITTED BY A BINARY SYSTEM IN HIGHER-ORDER $R^{2}$ GRAVITY \label{sec:GW-BS}}

Our goal is to describe the dominant-order effects of the modified
gravity with respect to the standard case of GR. Accordingly, we will
consider the first terms of the multipole expansion of the previous
section.

\subsection{Energy-momentum tensor and the multipole moments \label{subsec:M-moments}}

We begin by writing the energy-momentum tensor for a (relativistic)
binary system in the center-of-mass frame \cite{Weinberg1972}: 
\begin{equation}
T^{\mu\nu}\left(t,\mathbf{x}\right)=\sum_{A}\frac{p_{A}^{\mu}p_{A}^{\nu}}{\gamma_{A}m_{A}}\delta^{\left(3\right)}\left(\mathbf{x}-\mathbf{x}_{A}\left(t\right)\right)\label{eq:T_munu}
\end{equation}
where $A=1,2$ once the two masses are considered as pointlike particles.
In this notation, $\mathbf{x}_{A}\left(t\right)$ is the vector representing
the trajectory of particle $A$. In the nonrelativistic case, $\gamma_{A}=1$
and $\left|p_{A}^{0}\right|\gg\left|p_{A}^{i}\right|$. Then \cite{Kim2019}\textbf{,
\[
T^{\mu\nu}\left(t,\mathbf{x}\right)\simeq\sum_{A}m_{A}c^{2}\delta_{0}^{\mu}\delta_{0}^{\nu}\delta^{\left(3\right)}\left(\mathbf{x}-\mathbf{x}_{A}\left(t\right)\right)\,,
\]
}whose trace is 
\begin{equation}
T\left(t,\mathbf{x}\right)=\eta_{\mu\nu}T^{\mu\nu}\left(t,\mathbf{x}\right)\simeq-\sum_{A}m_{A}c^{2}\delta^{\left(3\right)}\left(\mathbf{x}-\mathbf{x}_{A}\left(t\right)\right)\,.\label{eq:T_tot}
\end{equation}

We are now interested in building the mass moments $\mathcal{M}$,
$\mathcal{M}^{i}$ and $\mathcal{M}^{ij}$. The spin-2 contributions
are well known from the literature \cite{Maggiore2007}, and we shall
not repeat their calculation here. In the following, we will only
be concerned with the mass multipoles for the spin-0 mode. Substituting
Eq. (\ref{eq:T_tot}) into Eq. (\ref{eq:MassMultipoles}) results in 
\begin{equation}
\mathcal{M}\left(t\right)=-\left(m_{1}+m_{2}\right)\equiv-m\,.\label{eq:M}
\end{equation}
Therein, $m$ is the total mass of the binary system. Also, 
\begin{equation}
\mathcal{M}^{i}\left(t\right)=-\left(m_{1}x_{1}^{i}+m_{2}x_{2}^{i}\right)\,,\label{eq:Mi}
\end{equation}
where $x_{1}^{i}=x_{1}^{i}\left(t\right)$ is the $i$ component of
$\mathbf{x}_{1}$. Additionally, 
\begin{equation}
\mathcal{M}^{ij}\left(t\right)=-\left(m_{1}x_{1}^{i}x_{1}^{j}+m_{2}x_{2}^{i}x_{2}^{j}\right)\,.\label{eq:Mij}
\end{equation}

Next, we define the reduced mass $\mu=m_{1}m_{2}/m$, the center-of-mass
coordinate $m\mathbf{x}_{\text{CM}}=m_{1}\mathbf{x}_{1}+m_{2}\mathbf{x}_{2}$,
and the relative coordinate $\mathbf{x}_{0}=\mathbf{x}_{1}-\mathbf{x}_{2}$.
Thus, in the center-of-mass coordinate system, Eqs. (\ref{eq:M})--(\ref{eq:Mij})
read 
\begin{equation}
\begin{cases}
\mathcal{M}\left(t\right)=-m\\
\mathcal{M}^{i}\left(t\right)=-mx_{\text{CM}}^{i}=0\\
\mathcal{M}^{ij}\left(t\right)=-mx_{\text{CM}}^{i}x_{\text{CM}}^{j}-\mu x_{0}^{i}x_{0}^{j}=-\mu x_{0}^{i}x_{0}^{j}
\end{cases}\,.\label{eq:MassMomentsCM}
\end{equation}
The last step in above equation considers a reference frame fixed
at the center of mass: $\mathbf{x}_{\text{CM}}=0$. With that, the
binary system is described as a single particle of mass $\mu$ moving
around the center-of-mass point. The trajectory of this particle is
given by $\mathbf{x}_{0}\left(t\right)$.

These results will used to calculate the fields produced by the binary
system in the case of circular orbits. Before that, we close this
section with a few comments. We know from GR that the first contribution
of the spin-2 mode is quadrupolar in nature. In principle, the scalar
mode could be different in this regard---i.e., the monopole and dipole
moments could contribute to the radiated energy. However, Eq. (\ref{eq:MassMomentsCM})
shows that the monopole moment is constant and that the dipole moment
is null. Thus, the temporal derivatives of these quantities vanish,
and the conclusion is that the first contribution coming from the
scalar mode is also quadrupolar in nature. This is an expected result:
it follows as a consequence of the mass conservation and the linear
momentum conservation of the nonrelativistic binary system \cite{Shao2020}.
We note that the vanishing of both the monopole and dipole contributions
is due to the nonrelativistic approximation leading to a Newtonian
limit. Dipole emissions are present in the approach by Ref. \cite{Kim2019},
where PN corrections to the $R^{2}$ model are accounted for.

\subsection{The gravitational wave solution\label{subsec:Wave-BS}}

The trajectory $\mathbf{x}_{0}\left(t\right)$ of the reduced mass
$\mu$ representing the binary system is ultimately determined as
the solution of the central force problem within the gravitational
potential resulting from the weak-field limit of our modified gravity
model \cite{Medeiros2020}. The general trajectory may be very complicated,
but we adopt the approach in which the modifications lead to small
corrections to the nonrelativistic predictions. Accordingly, the trajectories
could be approximated as ellipses or, in an even more simplistic standpoint,
circles. In this paper, we choose to work with the circular orbit
approximation.

We fix the coordinate system at the center of mass, the circular orbit
lying in the $\left(x,y\right)$ plane. Then, the coordinates of the
reduced mass particle are 
\begin{equation}
x_{0}^{i}\left(t\right)=\left(R\cos\left(\omega_{s}t+\frac{\pi}{2}\right),R\sin\left(\omega_{s}t+\frac{\pi}{2}\right),0\right)\label{eq:x0}
\end{equation}
where $\pi/2$ is a convenient phase, $R$ is the orbital radius,
and $\omega_{s}$ is the angular frequency of the orbit. This equation
is necessary to evaluate the mass moments $\mathcal{M}$, $\mathcal{M}^{i}$,
$\mathcal{M}^{ij}$ appearing in Eqs. (\ref{eq:Phi_minusM})--(\ref{eq:Phi_minusQ}).
The multipole decomposition of the GW tensor and scalar modes depends
on those quantities, and it also depends on $\hat{\mathbf{n}}$---see Eq. (\ref{eq:LargeDistApprox}).

We can decompose the unit vector\footnote{For figures representing the unit vector $\hat{\mathbf{n}}=\left(n^{i}\right)$, please see Ref. \cite{Maggiore2007}, Figs. 3.2 and 3.6.} $n_{i}$ in terms of the polar angle
$\theta$ and azimuth angle $\phi$: 
\begin{equation}
n_{i}=\left(\sin\theta\sin\phi,\sin\theta\cos\phi,\cos\theta\right)\label{eq:ni}
\end{equation}
The angle $\phi$ is a phase. We recall that $n^{i}$ points out orthogonally
from the plane of the orbit and forms a angle $\theta$ with the line of sight.

Substituting Eq.~(\ref{eq:T_munu}) into Eq.~(\ref{eq:h_tilde(T)}) and considering
circular orbits, one gets the two polarizations $\tilde{h}_{+}$ and
$\tilde{h}_{\times}$ of the spin-2 mode that are typical of the transverse-traceless
gauge \cite{Maggiore2007}: 
\begin{equation}
\tilde{h}_{+}\left(\mathbf{x},t\right)=\frac{1}{r}\frac{4G}{c^{4}}\mu\omega_{s}^{2}R^{2}\left(\frac{1+\cos^{2}\theta}{2}\right)\cos\left[2\omega_{s}\left(t-\frac{r}{c}\right)+2\phi\right]\label{eq:h_tilde_plus}
\end{equation}
and 
\begin{equation}
\tilde{h}_{\times}\left(\mathbf{x},t\right)=\frac{1}{r}\frac{4G}{c^{4}}\mu\omega_{s}^{2}R^{2}\cos\theta\sin\left[2\omega_{s}\left(t-\frac{r}{c}\right)+2\phi\right]\,.\label{eq:h_tilde_cross}
\end{equation}
The frequency of the gravitational wave is $\omega_{\text{gw}}=2\omega_{s}$
and $r$ is the distance from the source's center of mass to the observer.

The solution to the spin-0 mode is obtained by substituting (\ref{eq:x0})
and (\ref{eq:ni}) into (\ref{eq:MassMomentsCM}) and (\ref{eq:Phi_minusQ}).
After a long calculation (covered in Appendix \ref{Appendix:Phi_minus}),
we get 
\begin{equation}
\Phi_{-}\left(\mathbf{x},t\right)=\begin{cases}
\frac{1}{6}\frac{1}{r}\frac{4G}{c^{4}}\mu\omega_{s}^{2}R^{2}\sin^{2}\theta\cos\left(2\omega_{s}t+2\phi\right)\left[\exp\left(-m_{-}r\sqrt{1-\left(\frac{2\omega_{s}}{m_{-}c}\right)^{2}}\right)\right]\,, & \left(2\omega_{s}<m_{-}c\right)\\
\frac{1}{6}\frac{1}{r}\frac{4G}{c^{4}}\mu\omega_{s}^{2}R^{2}\sin^{2}\theta\cos\left[2\omega_{s}\left(t-\frac{r}{c}\sqrt{1-\left(\frac{m_{-}c}{2\omega_{s}}\right)^{2}}\right)+2\phi\right]\,, & \left(2\omega_{s}>m_{-}c\right)
\end{cases}\label{eq:Phi_minus(x,t)}
\end{equation}
Equations (\ref{eq:h_tilde_plus}), (\ref{eq:h_tilde_cross}), and (\ref{eq:Phi_minus(x,t)})
contain the complete information needed to specify the form of gravitational
waves emitted by binary systems in circular orbits within the context
of higher-order $R^{2}$ gravity.

We emphasize that $\Phi_{-}\left(\mathbf{x},t\right)=\Phi_{-}^{Q}\left(\mathbf{x},t\right)$---i.e., 
the monopole and dipole moments do not radiate. It is clear that
the scalar mode displays a damping regime---the first line in Eq.
(\ref{eq:Phi_minus(x,t)})---and an oscillatory regime---the second
line in Eq. (\ref{eq:Phi_minus(x,t)}). This feature was already anticipated
in Sec. \ref{subsec:SourcelessEqs}; here we present the full functional
form of $\Phi_{-}$, taking into account the binary system as a source
of the gravitational field. The damping mode of $\Phi_{-}$ is a time-dependent
metric deformation featuring an exponential decay, and, for this reason,
it is only effective closer to the source; the oscillatory regime
of $\Phi_{-}$ constitutes the gravitational wave carrying the energy
away from the binary system.

By comparing Eqs. (\ref{eq:h_tilde_plus}), (\ref{eq:h_tilde_cross})
and the second line of Eq. (\ref{eq:Phi_minus(x,t)}), one concludes
that the differences between the tensor mode and the scalar oscillatory
mode occur in the amplitude, in the function of the angle $\theta$ related
to the line of sight, and in the wave number.

The amplitude of $\Phi_{-}$ is one sixth of the magnitude of both
$\tilde{h}_{+}$ and $\tilde{h}_{\times}$ (disregarding the $\theta$ dependence).

As for the angle $\theta$, at the value $\theta=0$, the tensor modes
are maximized and the scalar mode vanishes altogether; the scalar
mode is maximized at $\theta=\pi/2$, while $\tilde{h}_{\times}$
vanishes at this value of $\theta$, and $\tilde{h}_{+}$ reduces
its amplitude by half.

The nature of the difference between the tensor mode and the scalar
oscillatory mode regarding the wave number is more complex; it depends
on the ratio of the mass parameter $m_{-}c$ to the frequency $2\omega_{s}$.
In terms of the wavelength $\lambda$, 
\[
\lambda_{\tilde{h}_{+,\times}}=\frac{\pi c}{\omega_{s}}\qquad\text{and}\qquad\lambda_{\Phi_{-}}=\frac{\pi c}{\omega_{s}}\frac{1}{\sqrt{1-\left(\frac{m_{-}c}{2\omega_{s}}\right)^{2}}}\,.
\]
The argument of the square root is less than 1, leading to $\lambda_{\tilde{h}_{+,\times}}<\lambda_{\Phi_{-}}$.
Moreover, an increase of the mass parameter $m_{-}c$ (for a fixed
value of the frequency) implies an increase of the wavelength $\lambda_{\Phi_{-}}$;
in the limit $m_{-}c\rightarrow2\omega_{s}$ we ultimately get $\lambda_{\Phi_{-}}\rightarrow\infty$.
In this limit, the scalar mode of the gravitational wave ceases to
exist and $\Phi_{-}$ transits from the oscillatory regime to the
damping regime.

There are further differences between the tensor mode and the scalar
oscillatory mode besides those mentioned above. In effect, the scalar
mode displays two distinctive features: it bears a longitudinal polarization
component, and its propagation velocity depends on the wave frequency.
For more details on this regard, we refer the reader to Refs. \cite{Corda2008,Corda2007}.

Our next task will be to determine the power radiated by the binary
system in the form of the GW described by Eqs. (\ref{eq:h_tilde_plus})--(\ref{eq:Phi_minus(x,t)}).

\subsection{The power radiated \label{subsec:Power-BS}}

The energy per unit time per unit area emitted by the gravitational
source is given in terms of the energy-momentum pseudotensor $t_{\mu\nu}$
\cite{Landau1975}. It encompasses the energy stored in the gravitational
field solely, while the energy-momentum tensor $T_{\mu\nu}$ accounts
for the energy within the source of gravity. The energy-momentum tensor
$T_{\mu\nu}$ is only covariantly conserved, while the sum $\left(T_{\mu\nu}+t_{\mu\nu}\right)$
is strictly conserved \cite{Sabbata1985} in the sense that $\partial_{\nu}\left(T^{\mu\nu}+t^{\mu\nu}\right)=0$.
The energy-momentum pseudotensor $t_{\mu\nu}$ is the gravitational
analogue of the Poynting vector in electrodynamics; as such, the
determination of $t_{\mu\nu}$ for our modified gravity model is key
to the study of the associated radiation theory.

We refer to the classic papers by Isaacson \cite{Isaacson1968a,Isaacson1968b}
for a treatment of $t_{\mu\nu}$ in the context of GR. The energy-momentum
pseudotensor in our higher-order $R^{2}$ gravity is calculated in
Appendix \ref{Appendix:PseudoEnergyTensor} in great detail. For quadratic scalar curvature
gravity $\propto\left(R+R^{2}\right)$ with small higher-order corrections
(of the type $\propto R\square R$), it is 
\begin{equation}
t_{\mu\nu}=-\frac{1}{\chi}\left\langle \mathcal{G}_{\mu\nu}^{\left(2\right)}\right\rangle =-\frac{c^{4}}{8\pi G}\left\langle \mathcal{G}_{\mu\nu}^{\left(2\right)}\right\rangle \,,\label{eq:tmunu(Gmunu)}
\end{equation}
where 
\begin{equation}
\left\langle \mathcal{G}_{\mu\nu}^{\left(2\right)}\right\rangle \approx-\frac{1}{4}\left\langle \partial_{\mu}\tilde{h}_{\alpha\beta}\partial_{\nu}\tilde{h}^{\alpha\beta}\right\rangle -\frac{3}{2}\left(1+\frac{\beta_{0}}{3}\right)\left\langle \partial_{\mu}\Phi_{-}\partial_{\nu}\Phi_{-}\right\rangle +\frac{2\chi}{3}\frac{\beta_{0}}{\kappa_{0}}\left\langle \partial_{\mu}\Phi_{-}\partial_{\nu}T\right\rangle \qquad\left(\beta_{0}\ll1\right)\,.\label{eq:Gmunu(2)averaged}
\end{equation}
The brackets denote a spacetime average over several periods.

The GW power per unit solid angle $dP/d\Omega$ radiated by the source
is \cite{Maggiore2007} 
\begin{equation}
\frac{dP}{d\Omega}=cr^{2}t^{01}\,,\label{eq:dP_dOmega(t01)}
\end{equation}
where 
\begin{equation}
t^{01}=-\frac{c^{4}}{16\pi G}\left[\left\langle \partial_{0}\tilde{h}_{+}\partial_{1}\tilde{h}_{+}\right\rangle +\left\langle \partial_{0}\tilde{h}_{\times}\partial_{1}\tilde{h}_{\times}\right\rangle \right]-\frac{3c^{4}}{16\pi G}\left(1+\frac{\beta_{0}}{3}\right)\left\langle \partial_{0}\Phi_{-}\partial_{1}\Phi_{-}\right\rangle +\frac{2}{3}\frac{\beta_{0}}{\kappa_{0}}\left\langle \partial_{0}\Phi_{-}\partial_{1}T\right\rangle \,.\label{eq:t01}
\end{equation}

Due to the functional forms of $\tilde{h}_{+}$ and $\tilde{h}_{\times}$
in Eqs. (\ref{eq:h_tilde_plus}) and (\ref{eq:h_tilde_cross}), their
radial derivatives $\partial_{1}$ can be switched to time derivatives
$\partial_{0}$ up to order $\left(1/r\right)$: 
\begin{equation}
\partial_{1}\tilde{h}_{+,\times}=-\partial_{0}\tilde{h}_{+,\times}+\mathcal{O}\left(1/r^{2}\right)\,.\label{eq:d1htildepluscross}
\end{equation}

Analogously, for the spin-0 field $\Phi_{-}\left(\mathbf{x},t\right)$,
\begin{equation}
\partial_{1}\Phi_{-}\left(\mathbf{x},t\right)=\begin{cases}
\sqrt{\left(\frac{m_{-}c}{2\omega_{s}}\right)^{2}-1}\cot\left(2\omega_{s}t+2\phi\right)\partial_{0}\Phi_{-}\left(\mathbf{x},t\right)+\mathcal{O}\left(1/r^{2}\right)\,, & \left(2\omega_{s}<m_{-}c\right)\\
-\sqrt{1-\left(\frac{m_{-}c}{2\omega_{s}}\right)^{2}}\partial_{0}\Phi_{-}\left(\mathbf{x},t\right)+\mathcal{O}\left(1/r^{2}\right)\,, & \left(2\omega_{s}>m_{-}c\right)
\end{cases}\,,\label{eq:d1Phi_minus(x,t)}
\end{equation}
cf. Eq. (\ref{eq:Phi_minus(x,t)}).

By utilizing (\ref{eq:t01}), (\ref{eq:d1htildepluscross}) and (\ref{eq:d1Phi_minus(x,t)}),
Eq. (\ref{eq:dP_dOmega(t01)}) reads 
\begin{equation}
\frac{dP}{d\Omega}=\frac{r^{2}c^{3}}{16\pi G}\left[\left\langle \left(\partial_{t}\tilde{h}_{+}\right)^{2}\right\rangle +\left\langle \left(\partial_{t}\tilde{h}_{\times}\right)^{2}\right\rangle \right]+\frac{3r^{2}c^{3}}{16\pi G}\left(1+\frac{\beta_{0}}{3}\right)\left\langle \mathcal{F}\left(\mathbf{x},t\right)\right\rangle +\frac{2r^{2}c}{3}\frac{\beta_{0}}{\kappa_{0}}\left\langle \partial_{0}\Phi_{-}\partial_{1}T\right\rangle \label{eq:dP/dOmega(h,Phi,T)}
\end{equation}
with the definition 
\begin{equation}
\mathcal{F}\left(\mathbf{x},t\right)\equiv\begin{cases}
-\sqrt{\left(\frac{m_{-}c}{2\omega_{s}}\right)^{2}-1}\left(\partial_{t}\Phi_{-}\right)^{2}\cot\left(2\omega_{s}t+2\phi\right)\,, & \left(2\omega_{s}<m_{-}c\right)\\
\sqrt{1-\left(\frac{m_{-}c}{2\omega_{s}}\right)^{2}}\left(\partial_{t}\Phi_{-}\right)^{2}\,, & \left(2\omega_{s}>m_{-}c\right)
\end{cases}\,.\label{eq:F(x,t)}
\end{equation}
The first three terms in Eq. (\ref{eq:dP/dOmega(h,Phi,T)}) are calculated
via Eqs. (\ref{eq:h_tilde_plus}), (\ref{eq:h_tilde_cross}), and (\ref{eq:Phi_minus(x,t)}):
\begin{equation}
\left\langle \left(\partial_{t}\tilde{h}_{+}\right)^{2}\right\rangle =\frac{1}{2}\left(2\omega_{s}\right)^{2}\left[\frac{1}{r}\frac{4G}{c^{4}}\mu\omega_{s}^{2}R^{2}\left(\frac{1+\cos^{2}\theta}{2}\right)\right]^{2}\,,\label{eq:dh_plusdt}
\end{equation}
\begin{equation}
\left\langle \left(\partial_{t}\tilde{h}_{\times}\right)^{2}\right\rangle =\frac{1}{2}\left(2\omega_{s}\right)^{2}\left[\frac{1}{r}\frac{4G}{c^{4}}\mu\omega_{s}^{2}R^{2}\cos\theta\right]^{2}\,,\label{eq:dh_crossdt}
\end{equation}
\begin{equation}
\left\langle \left(\partial_{t}\Phi_{-}\right)^{2}\cot\left(2\omega_{s}t+2\phi\right)\right\rangle =0\qquad\left(2\omega_{s}<m_{-}c\right)\,,\label{eq:dPhi_minusdt_cot}
\end{equation}
\begin{equation}
\left\langle \left(\partial_{t}\Phi_{-}\right)^{2}\right\rangle =\frac{1}{2}\left(2\omega_{s}\right)^{2}\left[\frac{1}{r}\frac{2G}{3c^{4}}\mu\omega_{s}^{2}R^{2}\sin^{2}\theta\right]^{2}\qquad\left(2\omega_{s}>m_{-}c\right)\,.\label{eq:dPhi_minusdt}
\end{equation}
We have used $\left\langle \cos^{2}\xi\right\rangle =\left\langle \sin^{2}\xi\right\rangle =1/2$
and $\left\langle \sin\xi\cos\xi\right\rangle =0$. The fourth term
on the right-hand side of (\ref{eq:dP/dOmega(h,Phi,T)}) is not so
trivial. Appendix \ref{Appendix:Vanishing} shows that it vanishes
altogether: 
\begin{equation}
\left\langle \partial_{0}\Phi_{-}\partial_{1}T\right\rangle =0\,.\label{eq:d0Phid1T}
\end{equation}
All the above amounts to the conclusion that the power radiated per
unit solid angle is 
\begin{align}
\frac{dP}{d\Omega}= & \frac{2G\mu^{2}R^{4}\omega_{s}^{6}}{\pi c^{5}}\left\{ \left[\left(\frac{1+\cos^{2}\theta}{2}\right)^{2}+\cos^{2}\theta\right]\right.\nonumber \\
 & \left.+\Theta\left(2\omega_{s}-m_{-}c\right)\sqrt{1-\left(\frac{m_{-}c}{2\omega_{s}}\right)^{2}}\left(1+\frac{\beta_{0}}{3}\right)\left(\frac{1}{12}\sin^{4}\theta\right)\right\} \,.\label{eq:dPdOmega}
\end{align}
The first term recovers the result from GR.

One additional integration gives the expression for the total power
radiated: 
\begin{equation}
P=2\pi{\displaystyle \int\limits _{0}^{\pi}}\frac{dP}{d\Omega}\sin\theta d\theta=\frac{32}{5}\frac{G\mu^{2}}{c^{5}}R^{4}\omega_{s}^{6}\left\{ 1+\frac{\Theta\left(2\omega_{s}-m_{-}c\right)}{18}\sqrt{1-\left(\frac{m_{-}c}{2\omega_{s}}\right)^{2}}\left(1+\frac{\beta_{0}}{3}\right)\right\} .\label{eq:P}
\end{equation}
This completes our investigation of the power radiated by a generic
circular orbit binary system in the context of the higher-order $R^{2}$ model.
In the following sections, we will select two specific types of binary
systems, each of which presents its own features and links to observations.
Black-hole binaries are addressed in the next section. Then, in Sec.
\ref{sec:BP-Inspirals}, we study binary pulsars.

\section{INSPIRAL PHASE OF BINARY BLACK HOLES \label{sec:BH-Inspirals}}

In this section, we study the peculiarities of the radiation emitted
by black-hole binaries. Our main goal is to highlight the changes
in the gravitational waveforms coming from our modified gravity model.
Specifically, we wish to check in which way the GW pattern is affected
by the presence of the spin-0 mode alongside the regular tensorial
modes. Accordingly, we shall perform a comparison of the modified
GW strain from our higher-order quadratic-curvature scalar model with
the GW strain from GR.

We start by considering the black-hole binaries for simplicity reasons.
In fact, this type of source does not alter the orbital behavior expected
from the Newtonian limit of GR, even in the context of our modified
gravity \cite{Medeiros2020}. Binary black holes respect Kepler's
third law as it is known from classical mechanics, without changes.
This is not the case when the binary system is composed of two neutron
stars, for instance---see Sec. \ref{sec:BP-Inspirals}.

The emission of gravitational radiation by the black-hole binary equals
the loss of mechanical energy by the system. This is expressed as
an equation of energy balance. The power loss ultimately leads to
a reduction of the orbital radius; the orbital frequency increases,
and so does the gravitational wave frequency. This modifies the argument
of the functions $\tilde{h}_{+}$, $\tilde{h}_{\times}$, and $\Phi_{-}$.
The resulting modified strain can be compared to the one predicted
by GR. This is the general road map to what is done in this section.

\subsection{Gravitational potential, Kepler's third law, and balance equation\label{subsec:Kepler-BH}}

In order to determine the gravitational potential to which the binary
system will be subjected, we need to distinguish between the possible
types of sources for the gravitational field. One possibility is a
binary system composed of pulsars; this case will be explored in Sec.
\ref{sec:BP-Inspirals}. In the present section we address the possibility
of a binary system constituted by two black holes.\footnote{A third possibility would be a binary system formed by one black hole
and one pulsar. This will not be considered in the present paper.} In this case, the classical interpretation is that the masses are pointlike and there are event horizons associated with them. As pointed
out in Ref. \cite{Medeiros2020}, this fact indicates that the solution
describing a static black hole in the modified gravity context is
still Schwarzschild-like.\footnote{This was proved for the pure Starobinsky gravity in Refs. \cite{Nelson2010}
and \cite{Lu2015}.} In this situation, the gravitational potential reduces to the Newtonian
potential in the weak-field regime: 
\begin{equation}
V\left(r\right)=-\frac{Gm}{r}\,.\label{eq:V_BH}
\end{equation}
(Here, $r$ is the distance from the source to the point of where
the potential is evaluated.) Even though the binary system may start
off in a highly elliptical orbit, the gravitational wave emission
rapidly circularizes it \cite{Maggiore2007}. In the circular orbit
phase, the orbital angular frequency $\omega_{s}$ is related to the
orbital radius $r=R$ by Kepler's law: 
\begin{equation}
\omega_{s}^{2}=\frac{Gm}{R^{3}}\label{eq:Kepler-BH}
\end{equation}
In the case of binary pulsars, both the potential and the angular
frequency are changed due to the higher-order terms in our modified
gravity model.

Our objective is to determine the gravitational waveform in the nonrelativistic
limit of our model. For that, it is necessary to write the balance
equation by equating the power radiated in the form of gravitational
waves $P$ to (minus) the time variation of the orbital energy \cite{Maggiore2007,Sabbata1985},
\begin{equation}
\frac{d}{dt}E_{\text{orbit}}=\frac{Gm\mu}{2R^{2}}\dot{R}\,.\label{eq:dEdt-BH}
\end{equation}
Notice that $\dot{R}=dR/dt$ assumes that the orbital radius varies
over time. In fact, its reduction corresponds to the inspiral phase
of the orbital motion leading to the coalescence.

From Eqs. (\ref{eq:P}), (\ref{eq:Kepler-BH}), and (\ref{eq:dEdt-BH}),
$P=-dE_{\text{orbit}}/dt$ leads to 
\begin{equation}
\dot{\omega}_{s}=\frac{96}{5}\left(\frac{GM_{c}}{c^{3}}\right)^{5/3}\omega_{s}^{11/3}\left\{ 1+\frac{\Theta\left(2\omega_{s}-m_{-}c\right)}{18}\sqrt{1-\left(\frac{m_{-}c}{2\omega_{s}}\right)^{2}}\left(1+\frac{\beta_{0}}{3}\right)\right\} \,,\label{eq:balance-BH}
\end{equation}
where $M_{c}=m^{2/5}\mu^{3/5}$ is the chirp mass. Equation (\ref{eq:balance-BH})
is a differential equation for the orbital frequency as a function
of time. The term within curly brackets is the additional contribution
coming from the scalar mode. It will only be effective when $\Theta\left(2\omega_{s}-m_{-}c\right)=1$---i.e., 
for the oscillatory mode of the spin-0 contribution. In the damping
regime, $2\omega_{s}<m_{-}c$, the curly brackets reduces to 1,
and we recover the GR prediction.

The next natural step is to solve Eq. (\ref{eq:balance-BH}) for both
the oscillatory and the damping regimes. First, we address the oscillatory
case, for which $2\omega_{s}>m_{-}c$. By changing the variables to 
\begin{equation}
u=\left(\frac{m_{-}c}{2\omega_{s}}\right)^{8/3}\qquad\text{and}\qquad\tau=\left(t_{\text{coal}}-t\right)\,,\label{eq:u_tau}
\end{equation}
where $t_{\text{coal}}$ is the coalescence time value, it follows that
\begin{equation}
\frac{du}{d\tau}=a\left\{ 1+\frac{1}{18}\sqrt{1-u^{3/4}}\left(1+\frac{\beta_{0}}{3}\right)\right\} \,,\label{eq:dudtau}
\end{equation}
with 
\begin{equation}
a\equiv\frac{1}{5}\left(\frac{GM_{c}}{c^{3}}\right)^{5/3}\left(4m_{-}c\right)^{8/3}\,.\label{eq:a}
\end{equation}
The second term in the curly brackets of Eq. (\ref{eq:dudtau}) is
a small correction and can be treated perturbatively. Thus, 
\begin{equation}
u=u^{(0)}+u^{(1)}\,,\label{eq:u_pert}
\end{equation}
so that 
\begin{equation}
\frac{du^{(0)}}{d\tau}=a\Rightarrow u^{(0)}=a\tau\,,\label{eq:u(0)}
\end{equation}
and 
\begin{equation}
\frac{du^{(1)}}{d\left(a\tau\right)}=\frac{1}{18}\sqrt{1-\left(a\tau\right)^{3/4}}\left(1+\frac{\beta_{0}}{3}\right)\Rightarrow u^{(1)}\simeq\frac{1}{18}\left(1+\frac{\beta_{0}}{3}\right)\left[a\tau-\frac{2}{7}\left(a\tau\right)^{7/4}+...\right]\,,\label{eq:u(1)}
\end{equation}
since $a\tau<1$.\footnote{The constraint $a\tau<1$ is equivalent to the oscillatory regime
condition---see Eqs. (\ref{eq:u_tau}) and (\ref{eq:u(0)}) in the
face of $2\omega_{s}>m_{-}c$.} The initial condition $u^{\left(0\right)}\left(0\right)=0$ and $u^{\left(1\right)}\left(0\right)=0$
is assumed. Returning to the variable $\omega_{s}$ via Eq. (\ref{eq:u_tau})
leads to 
\begin{equation}
\omega_{s}\left(\tau\right)\simeq\left(\frac{5}{256}\frac{1}{\tau}\right)^{3/8}\left(\frac{GM_{c}}{c^{3}}\right)^{-5/8}\left\{ 1-\frac{1}{48}\left(1+\frac{\beta_{0}}{3}\right)\left[1-\frac{2}{7}\left(a\tau\right)^{3/4}\right]\right\} \qquad\left(a\tau<1\right)\,,\label{eq:omega_s_osc_BH}
\end{equation}
where we have kept only the first two terms of the series in the square
brackets of Eq. (\ref{eq:u(1)}).

Notice that Eq. (\ref{eq:omega_s_osc_BH}) leads to a divergent $\omega_{s}$
as $\tau$ approaches zero. This is the coalescence event taken as
initial condition in the solution of the differential equation (\ref{eq:dudtau}).
As $\tau$ increases, we are further away from the coalescence---see 
Eq. (\ref{eq:u_tau}). The variable $\tau$ may increase up to
values $\sim\left(1/a\right)$; above this value, the oscillatory regime
transits to a damping mode and the validity condition for the above
solution breaks down.

Close to $\tau\sim0$, the binary system leaves the nonrelativistic
regime, and our above estimate for $\omega_{s}\left(\tau\right)$ fails;
however, the functional behavior of $\omega_{s}\left(\tau\right)$
should serve as a semiquantitative description of the system dynamics.
This is analogous to the standard treatment of the inspiral phase
of binaries in the context of GR.

The damping regime is characterized by $2\omega_{s}<m_{-}c$. In this
case, Eq. (\ref{eq:balance-BH}) reduces to 
\[
\dot{\omega}_{s}=\frac{96}{5}\left(\frac{GM_{c}}{c^{3}}\right)^{5/3}\omega_{s}^{11/3}\,,
\]
whose solution is 
\begin{equation}
\omega_{s}=\left(\frac{5}{256}\frac{1}{\left(K-t\right)}\right)^{3/8}\left(\frac{GM_{c}}{c^{3}}\right)^{-5/8}\,.\label{eq:omega_s(K)_damp_BH}
\end{equation}
The integration constant $K$ is set by demanding continuity with
the oscillatory case at $a\tau=1$. Therefore, equating (\ref{eq:omega_s_osc_BH})
and (\ref{eq:omega_s(K)_damp_BH}) at $\tau=\left(1/a\right)$, we
have 
\[
K=t_{\text{coal}}-\frac{1}{a}\left\{ 1-\left[1-\frac{1}{48}\left(1+\frac{\beta_{0}}{3}\right)\frac{5}{7}\right]^{-8/3}\right\} \,,
\]
leading to 
\begin{equation}
\omega_{s}=\left(\frac{5}{256}\right)^{3/8}\left(\frac{GM_{c}}{c^{3}}\right)^{-5/8}\frac{1}{\left(\tau-\frac{1}{a}\left\{ 1-\left[1-\frac{1}{48}\left(1+\frac{\beta_{0}}{3}\right)\frac{5}{7}\right]^{-8/3}\right\} \right)^{3/8}}\qquad\left(a\tau>1\right)\,.\label{eq:omega_s_damp_BH}
\end{equation}
Gathering (\ref{eq:omega_s_osc_BH}) and (\ref{eq:omega_s_damp_BH}),
\begin{equation}
\omega_{s}\left(\tau\right)\simeq\frac{5^{3/8}}{2^{3}}\left(\frac{GM_{c}}{c^{3}}\right)^{-5/8}\times\begin{cases}
\left(\tau-\frac{1}{a}\left\{ 1-\left[1-\frac{1}{48}\left(1+\frac{\beta_{0}}{3}\right)\frac{5}{7}\right]^{-8/3}\right\} \right)^{-3/8}\,, & a\tau>1\\
\tau^{-3/8}\left\{ 1-\frac{1}{48}\left(1+\frac{\beta_{0}}{3}\right)\left[1-\frac{2}{7}\left(a\tau\right)^{3/4}\right]\right\} \,, & a\tau<1
\end{cases}\,.\label{eq:omega_s_BH}
\end{equation}
This expression is useful for determining the waveform of the gravitational
wave in the binary inspiral phase.

\subsection{Gravitational waveform}

The fact that $\omega_{s}$ is now a function of time, as per Eq.
(\ref{eq:omega_s_BH}), modifies the waveform of both the tensor modes
$\left(\tilde{h}_{+},\tilde{h}_{\times}\right)$ and the scalar mode
$\Phi_{-}$. In this section, we will investigate how that happens.

In the quasicircular motion, the time variation of the orbital radius
$\left|\dot{R}\right|$ should be much smaller than the tangential
velocity $\omega_{s}R$---i.e., $\dot{R}/\left(\omega_{s}R\right)\ll1$.
This requirement is mapped to $\dot{\omega}_{s}/\omega_{s}^{2}\ll1$
due to Kepler's law [Eq. (\ref{eq:Kepler-BH})]. Under these circumstances,
we are allowed to write the waveforms (\ref{eq:h_tilde_plus}), (\ref{eq:h_tilde_cross}),
and (\ref{eq:Phi_minus(x,t)}) as 
\begin{equation}
\tilde{h}_{+}\left(\mathbf{x},t\right)=\mathcal{A}\left(r\right)\left(\frac{1+\cos^{2}\theta}{2}\right)\,\left[\omega_{s}\left(t\right)\right]^{2/3}\cos\left(2\int\omega_{s}\left(t\right)dt\right)\,,\label{eq:h_tilde_plus(omega_s)}
\end{equation}
\begin{equation}
\tilde{h}_{\times}\left(\mathbf{x},t\right)=\mathcal{A}\left(r\right)\cos\theta\,\left[\omega_{s}\left(t\right)\right]^{2/3}\cos\left(2\int\omega_{s}\left(t\right)dt-\frac{\pi}{2}\right)\,,\label{eq:h_tilde_cross(omega_s)}
\end{equation}
\begin{equation}
\Phi_{-}\left(\mathbf{x},t\right)=\begin{cases}
\frac{1}{6}\mathcal{A}(r)\sin^{2}\theta\left[\omega_{s}\left(t\right)\right]^{2/3}\cos\left(2\int\omega_{s}\left(t\right)dt+\mathcal{B}\right)\exp\left(-\frac{r}{r_{d}}\right)\,, & a\tau>1\\
\frac{1}{6}\mathcal{A}(r)\sin^{2}\theta\left[\omega_{s}\left(t\right)\right]^{2/3}\cos\left(2\int\omega_{s}\left(t\right)dt+\mathcal{C}\right)\,, & a\tau<1
\end{cases}\,,\label{eq:Phi_minus(omega_s)}
\end{equation}
where 
\begin{equation}
\mathcal{A}\left(r\right)\equiv\frac{1}{r}\frac{4\left(GM_{c}\right)^{5/3}}{c^{4}}\,,\label{eq:A(r)}
\end{equation}
the quantities $\mathcal{B}$ and $\mathcal{C}$ represent constant
phases, and $r_{d}$ is a constant distance scale associated with the
damping regime. The integral in Eqs. (\ref{eq:h_tilde_plus(omega_s)})--(\ref{eq:Phi_minus(omega_s)})
is solved by using Eq. (\ref{eq:omega_s_BH}). Notice that there are
two different regimes for the function $\omega_{s}\left(t\right)$;
this fact leads to 
\begin{equation}
\int\omega_{s}\left(t\right)dt=-\left(\frac{5GM_{c}}{c^{3}}\right)^{-5/8}\times\begin{cases}
\left[\tau-\frac{1}{a}\left\{ 1-\left[1-\frac{1}{48}\left(1+\frac{\beta_{0}}{3}\right)\frac{5}{7}\right]^{-8/3}\right\} \right]^{5/8}-\mathcal{K}\,, & a\tau>1\\
\tau^{5/8}\left\{ 1-\frac{1}{48}\left(1+\frac{\beta_{0}}{3}\right)\left[1-\frac{10}{77}\left(a\tau\right)^{3/4}\right]\right\} \,, & a\tau<1
\end{cases}\,.\label{eq:int_omega_s}
\end{equation}
We determine the integration constant $\mathcal{K}$ by equating both
regimes of Eq. (\ref{eq:int_omega_s}) at $\tau=1/a$: 
\begin{equation}
\mathcal{K}=\left[1-\frac{1}{48}\left(1+\frac{\beta_{0}}{3}\right)\frac{5}{7}\right]^{-5/3}-\left[1-\frac{1}{48}\left(1+\frac{\beta_{0}}{3}\right)\left(1-\frac{10}{77}\right)\right]\,.\label{eq:K_damping}
\end{equation}

\begin{figure}[h]
\begin{centering}
\includegraphics[scale=0.4]{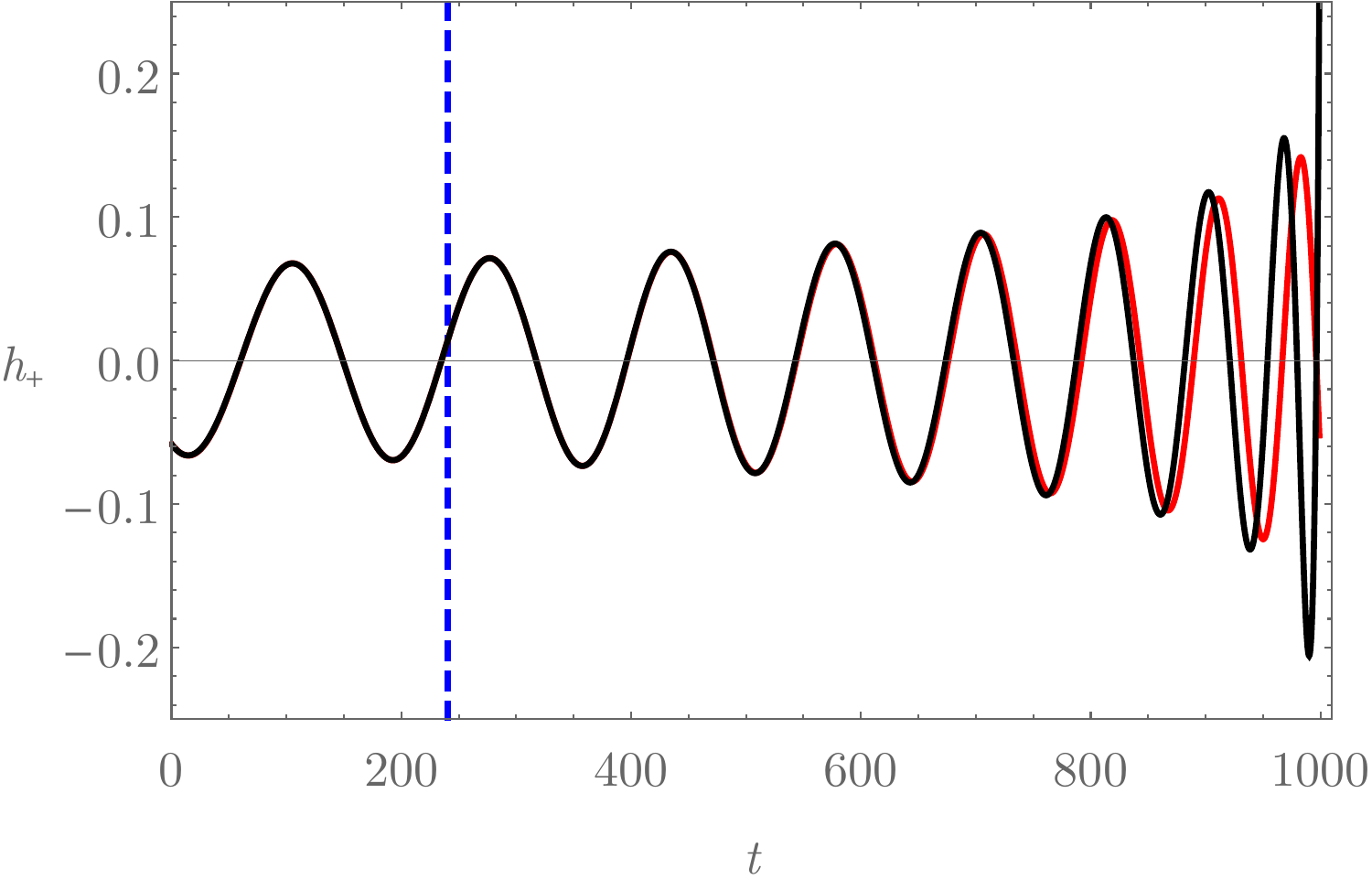} 
\par\end{centering}
\caption{Gravitational waveforms for the spin-2 mode emitted by a BH binary
inspiral. The $x$ axis gives the time $t$ (in arbitrary units);
the $y$ axis measures the strain $h_{+}$ in units of $\mathcal{A}(r)\left(1+\cos^{2}\theta\right)/2$.
The red curve corresponds to the prediction coming from GR. The black
curve shows the spin-2 mode wave pattern for our higher-order gravity.
The initial condition is such that the two strain curves coincide at
$t=0$. The vertical dashed line marks the transition from the damping
regime (on the left) to the oscillating regime, and the scalar mode is emitted 
from that point on. The transition depends on the factor $m_{-}c=3.43\times10^{-2}$
corresponding to the time $t=240$. In our units, $\left(GM_{c}/c^{3}\right)=1$;
for simplicity, we have set $\beta_{0}=0$. }
\label{fig:wave_form} 
\end{figure}

The waveform time dependences of the modes $\tilde{h}_{+}$, $\tilde{h}_{\times}$,
and $\Phi_{-}$ are the same up to a constant phase factor. However,
these modes differ from their GR counterparts. This is illustrated
in Fig. \ref{fig:wave_form}, where the wave profile of the modified
gravity spin-2 mode $\tilde{h}_{+}$ is superimposed on GR's spin-2
mode $\bar{h}_{+}$.

The curve for the strain in our modified gravity model coincides with
the curve for the strain from GR in the region to the left of the
vertical dashed line of Fig. \ref{fig:wave_form}; in this region,
the binary system does not emit the scalar-mode radiation. To the
right of the vertical dashed line, the system gives off the extra
scalar polarization; in this circumstance, the frequency of the strain
in our modified gravity increases more rapidly than the frequency
of the strain in GR. This increase is subtle, and it is noticeable
only after a few oscillation periods (Fig. \ref{fig:wave_form}).
The larger frequency increase leads to an earlier coalescence in our
modified gravity.

We emphasize that the more the binary system approaches the merger,
the weaker are the hypotheses used in our modelling, namely pointlike
masses and nonrelativistic dynamics. In this sense, the region of
the plot in Fig. \ref{fig:wave_form} close to the merger should be
taken as a qualitative description.

\section{INSPIRAL PHASE OF BINARY PULSARS \label{sec:BP-Inspirals}}

The previous section dealt with black-hole binaries. The second possibility
that we are going to contemplate here is a binary system composed
of a pair of pulsars. Our focus is to constrain the coupling constants
of our higher-order $R^{2}$ model utilizing observational data. In
particular, we draw information from the observations of the Hulse-Taylor
binary pulsar \cite{Hulse1975,Taylor1982,Weisberg2004}.

The motivation to treat binary pulsars in a separate section stems
from the fact that their orbital dynamics is distinct from the one
followed by binary black holes. In effect, the gravitational potential
determining the orbits of the pair of neutron stars in the modified
gravity context is not the Newtonian potential alone, as in Eq. (\ref{eq:V_BH}),
it but carries Yukawa-type terms, as we will see in Eq. (\ref{eq:V_StaPod})
below.

We shall study two regimes: (i) the regime where the scalar mode propagates
and the orbital dynamics modifications with respect to GR are maximum,
and (ii) the regime where the scalar mode dies off and the spin-2
mode alone propagates; in this regime, we approach only small corrections
to the ordinary gravitational potential. It is this second regime
that allows us to observationally constrain the parameters in our modified
gravity model.

\subsection{Gravitational potential and modified Kepler's law \label{subsec:Kepler-BP}}

The complete relativistic solution in the context of modified gravity
describing an extended static spherically symmetric mass has not been
determined analytically yet. However, the gravitational potential
for this type of source in the weak-field regime far from the source
is already known\footnote{The potential of an extended spherically symmetric source in $R^{2}$-gravity
was calculated in Ref. \cite{Berry2011}. It is clear that their result
reduces to the point-like one for distances much larger that the typical
size of the source.}: it bears corrections of the Yukawa type besides the conventional
$\left(1/r\right)$ term. Accordingly, the modified potential of higher-order
$R^{2}$ gravity in the nonrelativistic regime is \cite{Medeiros2020}:
\begin{equation}
V\left(r\right)=-\frac{Gm}{r}\left(1+a_{+}e^{-m_{-}r}+a_{-}e^{-m_{+}r}\right)\,,\label{eq:V_StaPod}
\end{equation}
where the parameters $m_{\pm}$ are given by Eq. (\ref{eq:m_plus_minus})
and 
\begin{equation}
a_{\pm}=\pm\frac{1}{6}\left(\frac{1\pm\sqrt{1-\frac{4\beta_{0}}{3}}}{\sqrt{1-\frac{4\beta_{0}}{3}}}\right)\,.\label{eq:apm(beta0)}
\end{equation}

In this paper, we will consider small corrections to the predictions
by $R^{2}$ gravity coming from our higher-order $R^{2}$ gravity;
this means that we will be taking $\beta_{0}\ll1$ to achieve our
main results. Within this condition, the masses $m_{\pm}$ are given
by Eq.~(\ref{eq:m_pm_betasmall}) and 
\begin{equation}
a_{+}\simeq\frac{1}{3}\left(1+\frac{\beta_{0}}{3}\right)\qquad\text{and}\qquad a_{-}\simeq-\frac{\beta_{0}}{9}\qquad\left(\beta_{0}\ll1\right)\,.\label{eq:a_pm_betasmall}
\end{equation}

The potential (\ref{eq:V_StaPod}) can be used to build the nonrelativistic
equation of motion for the binary system. Assuming a circular orbit,
where $r=R=\text{constant}$, this procedure leads to 
\begin{equation}
\omega_{s}^{2}=\frac{Gm}{R^{3}}\left[1+a_{+}\left(1+m_{-}R\right)e^{-m_{-}R}+a_{-}\left(1+m_{+}R\right)e^{-m_{+}R}\right]\,,\label{eq:Kepler}
\end{equation}
which is an extended version of Kepler's third law.

Equations (\ref{eq:m_pm_betasmall}) and (\ref{eq:a_pm_betasmall}) are
then used in (\ref{eq:Kepler}) to write down Kepler's law in terms
of the coupling constants $\kappa_{0}$ and $\beta_{0}$: 
\begin{align}
\omega_{s}^{2} & \simeq\frac{Gm}{R^{3}}\left[1+\frac{1}{3}\left(1+\frac{\beta_{0}}{3}+\left(1+\frac{\beta_{0}}{2}\right)\Delta\right)\left(1-\frac{\beta_{0}}{6}\right)^{\Delta}e^{-\Delta}\right.\nonumber \\
 & \left.-\frac{1}{3}\left(\frac{\beta_{0}}{3}+\sqrt{\frac{\beta_{0}}{3}}\left(1-\frac{\beta_{0}}{6}\right)\Delta\right)\left(1-\frac{\beta_{0}}{6}\right)^{-\sqrt{\frac{3}{\beta_{0}}}\Delta}e^{-\sqrt{\frac{3}{\beta_{0}}}\Delta}\right]\qquad\left(\beta_{0}\ll1\right)\,,\label{eq:Kepler(Delta)}
\end{align}
where 
\begin{equation}
\Delta\equiv\sqrt{\frac{\kappa_{0}}{3}}R\,.\label{eq:Delta}
\end{equation}
The dimensionless quantity $\Delta$ is a measure of the importance
of $R^{2}$ term relative to the Einstein-Hilbert term on an orbital
scale. The greater (smaller) the $\Delta$, the smaller (greater)
the influence of the modified gravity term at this scale. \bigskip{}

\textit{Remark: The $R^{2}$-limit.}---
If $\beta_{0}\rightarrow0$, Eq. (\ref{eq:Kepler(Delta)}) reads
\begin{equation}
\omega_{s}^{2}\simeq\frac{Gm}{R^{3}}\left[1+\frac{1}{3}\left(1+\Delta\right)e^{-\Delta}\right]\,.\label{eq:Kepler_Sta}
\end{equation}
This result is consistent with Eq. (40) in Ref. \cite{Naf2011}. $\blacksquare$


\subsubsection{The oscillatory limit}

Let us study the limit of the generalized Kepler law [Eq. (\ref{eq:Kepler(Delta)})]
as $\Delta\ll1$. This is interesting because we have seen in Sec.
\ref{subsec:Wave-BS} that the $\Phi_{-}$ solution displays two distinct
regimes: the damped regime and the oscillatory regime. To select the
oscillatory solution for $\Phi_{-}$ ultimately corresponds to taking
$\Delta\ll1$. In fact, the oscillatory regime occurs as 
\[
\frac{2\omega_{s}}{m_{-}c}>1\Rightarrow\frac{2\omega_{s}}{c}>\sqrt{\frac{\kappa_{0}}{3}}\left(1+\frac{\beta_{0}}{3}\right)^{1/2}\,,
\]
where $2\omega_{s}$ is the frequency of the gravitational wave. For
a binary system in a circular orbit, the tangential scalar velocity
is $v=\omega_{s}R$. Then, the condition above is the same as 
\begin{equation}
\frac{2}{R}\frac{v}{c}\gtrsim\sqrt{\frac{\kappa_{0}}{3}}\left(1+\frac{\beta_{0}}{6}\right)\overset{\frac{v}{c}\ll1}{\Rightarrow}\Delta\left(1+\frac{\beta_{0}}{6}\right)\simeq\Delta\ll1\,,\label{eq:v_over_c}
\end{equation}
since $\beta_{0}\ll1$.

In the case where $\Delta\ll1$, we also have $e^{-\Delta}\simeq1-\Delta\approx1$,
and terms scaling as $\beta_{0}\Delta$ may be neglected, since both
$\beta_{0}$ and $\Delta$ are small. Hence, Eq. (\ref{eq:Kepler(Delta)})
reduces to 
\[
\omega_{s}^{2}\simeq\frac{Gm}{R^{3}}\left[\frac{4}{3}+\frac{\beta_{0}}{9}\epsilon\left(\Delta,\beta_{0}\right)\right]\,,
\]
where 
\[
\epsilon\left(\Delta,\beta_{0}\right)\equiv\left[1-\left(1+\sqrt{\frac{3}{\beta_{0}}}\Delta\right)\left[\left(1-\frac{\beta_{0}}{6}\right)e\right]^{-\sqrt{\frac{3}{\beta_{0}}}\Delta}\right]\,.
\]
A numerical analysis reveals that $0<\epsilon\left(\Delta,\beta_{0}\right)\leqslant1$
by keeping in mind that $\beta_{0}\ll1$. At most, 
\begin{equation}
\omega_{s}^{2}\simeq\frac{Gm}{R^{3}}\left(\frac{4}{3}+\frac{\beta_{0}}{9}\right)\qquad\left(\Delta\ll1\right)\,.\label{eq:Kepler_Delta_small}
\end{equation}
This expression is troublesome, since the factor $\left(4/3\right)$
enhances the orbital frequency in Kepler's law in a way that should
have been detected in orbital dynamics observations. This modification
is significant enough to render our prediction incompatible with observations.
This implies that our model in unphysical in the oscillatory regime
for binary pulsars.\textcolor{blue}{{} }The regime $\Delta\ll1$ corresponds
to the dominance of the $R^{2}$ term over the Einstein-Hilbert term
in the action [Eq. (\ref{eq:Action})]. Reference \cite{Pechlaner1966} has shown
a long time ago that this regime leads to the appearance of the factor
4/3 as a correction to the gravitational potential and Kepler's third
law. Moreover, the same reference proves that the weak-field regime
of a theory built exclusively from the term $R^{2}$ is inconsistent.
These arguments indicate that the physical regime is the one for $\Delta\gg1$,
which will be studied next.

\subsubsection{The damping limit}

Now, the case $\Delta\gg1$ will be considered. Because $\beta_{0}\ll1$,
we have $\Delta/\sqrt{\beta_{0}}\gg1$, and Eq. (\ref{eq:Kepler(Delta)})
reduces to 
\begin{equation}
\omega_{s}^{2}\simeq\frac{Gm}{R^{3}}\left[1+\frac{1}{3}\Delta e^{-\Delta}\left(1+\frac{\beta_{0}}{2}\right)\left(1-\frac{\beta_{0}}{6}\right)^{\Delta}\right]\qquad\left(\Delta\gg1\right)\,,\label{eq:Kepler_Delta_large}
\end{equation}
which is very close to the Newtonian dynamics. This is the regime
that will lead to constraints upon the additional coupling constants
of our modified gravity model.

\subsection{The balance equation in the damping regime \label{subsec:BalanceEq-BP}}

We have already mentioned that the energy balance expression equates
the power radiated to the time derivative of the orbital energy of
the system. The power equation for a generic binary system in our modified
gravity model was deduced in Sec. \ref{subsec:Power-BS}. Here
we begin by constructing the orbital energy $E_{\text{orbit}}=E_{\text{kin}}+\mu V\left(r\right)$
of the binary system in the context of our modified gravity. From
Eqs. (\ref{eq:V_StaPod}), (\ref{eq:a_pm_betasmall}), and (\ref{eq:Delta}),
\begin{align}
E_{\text{orbit}} & =\frac{1}{2}\mu\omega_{s}^{2}R^{2}\nonumber \\
 & -\frac{Gm\mu}{R}\left\{ 1+\frac{1}{3}\left(1+\frac{\beta_{0}}{3}\right)\exp\left[-\left(1+\frac{\beta_{0}}{6}\right)\Delta\right]-\frac{\beta_{0}}{9}\exp\left[-\sqrt{\frac{3}{\beta_{0}}}\left(1-\frac{\beta_{0}}{6}\right)\Delta\right]\right\} \,.\label{eq:E_orbit(w_s)}
\end{align}

Equation (\ref{eq:E_orbit(w_s)}) assumes a circular orbit.

Using the modified Kepler law (\ref{eq:Kepler(Delta)}), the above
equation turns to 
\begin{align}
E_{\text{orbit}} & =-\frac{1}{2}\frac{Gm\mu}{R}\left\{ 1+\frac{1}{3}\left[\left(1+\frac{\beta_{0}}{3}\right)-\left(1+\frac{\beta_{0}}{2}\right)\Delta\right]e^{-\Delta\left(1+\frac{\beta_{0}}{6}\right)}\right.\nonumber \\
 & \left.-\frac{\beta_{0}}{9}\left[1-\left(1-\frac{\beta_{0}}{6}\right)\sqrt{\frac{3}{\beta_{0}}}\Delta\right]e^{-\sqrt{\frac{3}{\beta_{0}}}\Delta\left(1-\frac{\beta_{0}}{6}\right)}\right\} \,,\label{eq:E_orbit}
\end{align}
where we have utilized the approximation 
\begin{equation}
\left(1-\frac{\beta_{0}}{6}\right)^{\Delta}=e^{\Delta\ln\left(1-\frac{\beta_{0}}{6}\right)}\simeq e^{\Delta\left(-\frac{\beta_{0}}{6}\right)}\,.\label{eq:beta0_risen_Delta}
\end{equation}
We notice that $\dot{\Delta}=\Delta\left(\dot{R}/R\right)$ and obtain
the time derivative of the orbital energy as 
\begin{align}
\frac{d}{dt}E_{\text{orbit}} & =\frac{Gm\mu}{2R^{2}}\dot{R}\left\{ 1+\frac{1}{3}\left[\left(1+\frac{\beta_{0}}{3}\right)+\left(1+\frac{\beta_{0}}{2}\right)\Delta-\left(1+2\frac{\beta_{0}}{3}\right)\Delta^{2}\right]e^{-\Delta\left(1+\frac{\beta_{0}}{6}\right)}\right.\nonumber \\
 & \left.-\frac{\beta_{0}}{9}\left[1+\left(1-\frac{\beta_{0}}{6}\right)\sqrt{\frac{3}{\beta_{0}}}\Delta-\left(1-\frac{\beta_{0}}{3}\right)\left(\sqrt{\frac{3}{\beta_{0}}}\Delta\right)^{2}\right]e^{-\sqrt{\frac{3}{\beta_{0}}}\Delta\left(1-\frac{\beta_{0}}{6}\right)}\right\} \,.\label{eq:d(Eorbit)dt}
\end{align}

We equate Eq. (\ref{eq:d(Eorbit)dt}) to Eq. (\ref{eq:P}), to get 
\begin{align}
\frac{64}{5}\frac{\mu}{m}\frac{\left(\omega_{s}R\right)^{6}}{c^{5}}\left\{ 1+\frac{\Theta\left(2\omega_{s}-m_{-}c\right)}{18}\sqrt{1-\left(\frac{m_{-}c}{2\omega_{s}}\right)^{2}}\left(1+\frac{\beta_{0}}{3}\right)\right\} =\nonumber \\
=-\dot{R}\left\{ 1+\frac{1}{3}\left[\left(1+\frac{\beta_{0}}{3}\right)+\left(1+\frac{\beta_{0}}{2}\right)\Delta-\left(1+2\frac{\beta_{0}}{3}\right)\Delta^{2}\right]e^{-\Delta\left(1+\frac{\beta_{0}}{6}\right)}\right.\nonumber \\
\left.-\frac{\beta_{0}}{9}\left[1+\left(1-\frac{\beta_{0}}{6}\right)\sqrt{\frac{3}{\beta_{0}}}\Delta-\left(1-\frac{\beta_{0}}{3}\right)\left(\sqrt{\frac{3}{\beta_{0}}}\Delta\right)^{2}\right]e^{-\sqrt{\frac{3}{\beta_{0}}}\Delta\left(1-\frac{\beta_{0}}{6}\right)}\right\} \,,\label{eq:DiffEqR}
\end{align}
where $\omega_{s}=\omega_{s}\left(\Delta\right)$ is obtained via
Eq. (\ref{eq:Kepler(Delta)}). Equation (\ref{eq:DiffEqR}) expresses the energy
balance in the regime of small higher-order corrections: $\beta_{0}\ll1$.
The left-hand side of this equation contains the spin-2 contribution
(first term) and the contribution from the spin-0 mode (carrying the
Heaviside function). The right-hand side of the equation accounts
for the orbital energy loss, and the associated orbital radius decreasing,
in the context of our higher-order $R^{2}$ model.

In principle, Eq. (\ref{eq:DiffEqR}) could be explored in both the
oscillatory regime and the damping regime. However, we have seen above
that the oscillatory regime is unphysical for binary pulsars. Consequently,
our focus will be on the damping regime for which $2\omega_{s}<m_{-}c\Rightarrow\Theta\left(2\omega_{s}-m_{-}c\right)=0$.

\subsubsection{Corrections to the time variation of the orbital period due to $R^{2}$ gravity\label{subsec:Pbdot-Sta}}

\begin{figure}[h]
\begin{centering}
\includegraphics[scale=0.4]{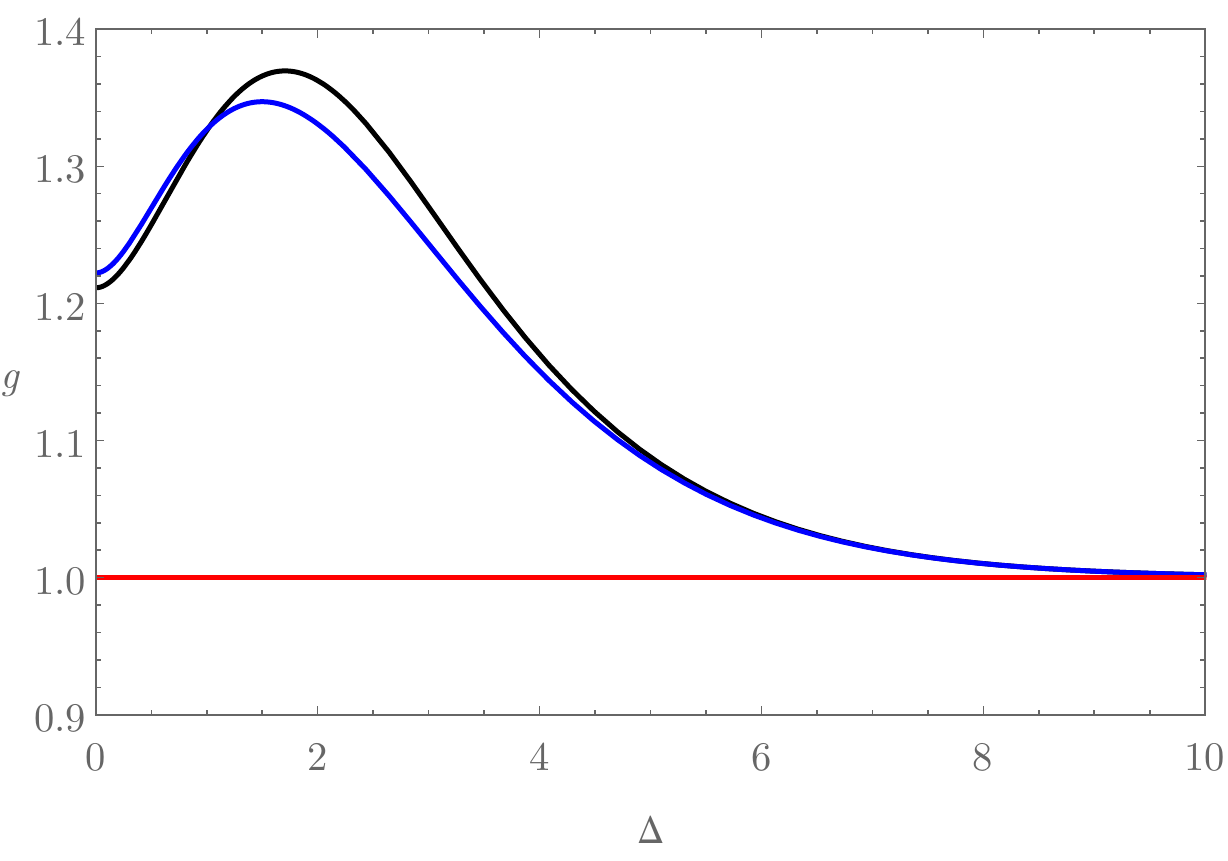} 
\par\end{centering}
\caption{Plot of the function $g\left(\Delta\right)$ (black line) and
its approximation for large values of $\Delta$ (blue line). The
red line marks the result expected from GR: $g\left(\Delta\right)=1$.}
\label{fig:g(Delta)} 
\end{figure}

For the sake of simplicity, we shall consider the pure $R^{2}$-gravity
case first. This is done by taking $\beta_{0}\rightarrow0$ in Eqs.
(\ref{eq:DiffEqR}) and (\ref{eq:Kepler(Delta)}). Later on, we shall
reintroduce the contribution of the higher-order terms through a small
$\beta_{0}$. Accordingly, the energy balance equation reduces to
\begin{equation}
\frac{64}{5}\frac{\mu}{m}\frac{\left(R\omega_{s}\right)^{6}}{c^{5}}=-\dot{R}\left\{ 1+\frac{1}{3}\left[1+\Delta\left(1-\Delta\right)\right]e^{-\Delta}\right\} \,,\label{eq:BalanceEq_Sta}
\end{equation}
with 
\begin{align}
\omega_{s}^{2} & \simeq\frac{Gm}{R^{3}}\left[1+\frac{1}{3}\left(1+\Delta\right)e^{-\Delta}\right]\,,\label{eq:omega_s_Sta}
\end{align}
where the frequency $\omega_{s}$ can be expressed in terms of the
orbital period $T$ as $\omega_{s}=2\pi/T$. By taking the time derivative
of Eq. (\ref{eq:omega_s_Sta}) and recalling Eq. (\ref{eq:Delta}),
we get 
\begin{equation}
\frac{2}{3}\frac{\dot{T}}{T}=\frac{\dot{R}}{R}\left\{ 1+\frac{1}{3}\frac{\Delta^{2}}{\left(1+\Delta\right)}\left[1-\left(\frac{T}{2\pi}\right)^{2}\frac{Gm}{R^{3}}\right]\right\} \,.\label{eq:TdotT}
\end{equation}
In GR, we miss the second term in the curly brackets. Our next step
is to use Eq. (\ref{eq:BalanceEq_Sta}) to replace $\dot{R}$ in Eq. (\ref{eq:TdotT});
the result is 
\begin{equation}
\dot{T}=-\frac{192\pi}{5}\left(\frac{GM_{c}}{c^{3}}\right)^{5/3}\left(\frac{T}{2\pi}\right)^{-5/3}g\left(\Delta\right)\label{eq:Tdot(T,Delta)_damping}
\end{equation}
where 
\begin{equation}
g\left(\Delta\right)=\frac{\left\{ 1+\frac{1}{3}\left[1+\Delta+\frac{1}{3}\Delta^{2}\right]e^{-\Delta}\right\} }{\left\{ 1+\frac{1}{3}\left[1+\Delta-\Delta^{2}\right]e^{-\Delta}\right\} }\left[1+\frac{1}{3}\left(1+\Delta\right)e^{-\Delta}\right]^{2/3}\,.\label{eq:g(Delta)}
\end{equation}
Equation (\ref{eq:g(Delta)}) is obtained after we make use of the modified
Kepler's law [Eq. (\ref{eq:omega_s_Sta})] multiple times. All changes to
GR predictions due to our model are encapsulated in the function $g\left(\Delta\right)$.
In fact, the $R^{2}$-gravity model tends to GR in the limit as $\kappa_{0}\rightarrow\infty$,
thus $\Delta\rightarrow\infty$ and $\lim_{\Delta\rightarrow\infty}g\left(\Delta\right)=1$.
Figure \ref{fig:g(Delta)} shows the plot of $g=g\left(\Delta\right)$.
The same figure displays the $g\left(\Delta\right)$ curve in the regime
of large values of the argument, in which case Eq. (\ref{eq:g(Delta)})
reduces to 
\begin{equation}
g\left(\Delta\right)\approx1+\frac{2}{9}\left(1+\Delta+2\Delta^{2}\right)e^{-\Delta}\,,\qquad\left(\Delta\gg1\right)\,.\label{eq:g(Delta)_Delta_large}
\end{equation}
It is clear from Fig. \ref{fig:g(Delta)} that the smaller the value
of $\Delta$, the larger the differences with respect to GR. The parameter
$\Delta$ can be estimated by using Eq. (\ref{eq:Tdot(T,Delta)_damping})
and observational data. Our goal will be to determine the smallest
value of $\Delta$ admissible by observations.

In order to use Eq. (\ref{eq:Tdot(T,Delta)_damping}) for comparison
with observations, we need to account for the eccentricity of the binary
system. For this reason, we will introduce (in a nonrigorous way)
the eccentricity factor \cite{Weisberg2004,Maggiore2007,Peters1963}
\begin{equation}
f\left(e\right)=\frac{1}{\left(1-e^{2}\right)^{7/2}}\left(1+\frac{73}{24}e^{2}+\frac{37}{96}e^{4}\right)\label{eq:f(e)}
\end{equation}
as a multiplicative term\footnote{The inclusion of $f\left(e\right)$ as a multiplicative factor in
Eq. (\ref{eq:Pb_dot_Sta}) assumes that the corrections due to the
modified gravity are independent of (or very little dependent on)
the eccentricity.} in the expression for the time variation of the orbital period [Eq.
(\ref{eq:Tdot(T,Delta)_damping})]:
\begin{equation}
\dot{P}_{b,R^{2}\text{-gravity}}=\dot{T}\times f\left(e\right)\,.\label{eq:Pb_dot_Sta}
\end{equation}

Our idea is to constrain parameter $\Delta$ from the Hulse and Taylor
binary pulsar PSR B1913+16 \cite{Hulse1975,Taylor1982} studied in
detail in Ref. \cite{Weisberg2004}. The latter reference provides
the parameter values in Table \ref{table:PSR}. 
\begin{table}
\begin{centering}
\begin{tabular}{|c|c|}
\hline 
Parameter  & Value\tabularnewline
\hline 
\hline 
$e$  & $0.6171338(4)$\tabularnewline
\hline 
$P_{b}\left(\text{days}\right)$  & $0.322997448930(4)$\tabularnewline
\hline 
$m_{1}\left(M_{\astrosun}\right)$  & $1.4414\pm0.0002$\tabularnewline
\hline 
$m_{2}\left(M_{\astrosun}\right)$  & $1.3867\pm0.0002$\tabularnewline
\hline 
$\dot{P}_{b,\text{corrected}}\left(\text{s/s}\right)$  & $-\left(2.4056\pm0.0051\right)\times10^{-12}$\tabularnewline
\hline 
\end{tabular}
\par\end{centering}
\caption{Parameters for the binary pulsar PSR B1913+16 as given by \cite{Weisberg2004}.
The quantity $\dot{P}_{b,\text{corrected}}$ takes into account corrections
to the observed period variation $\dot{P}_{b}$ due to the relative
acceleration between the solar system and the binary pulsar.}
\label{table:PSR} 
\end{table}

There are four parameters in Table \ref{table:PSR} affecting the
evaluation of $\dot{P}_{b,R^{2}\text{-gravity}}$: namely, $e$, $P_{b}$,
$m_{1}$, and $m_{2}$. When the uncertainties of these parameters
are accounted for via error propagation, one concludes that the uncertainty
in $\left(\dot{P}_{b,R^{2}\text{-gravity}}\times10^{-12}\right)$
is of order $10^{-5}$. On the other hand, the uncertainty in $\left(\dot{P}_{b,\text{corrected}}\times10^{-12}\right)$
is of order $10^{-3}$. Hence, we conclude that $\dot{P}_{b,\text{corrected}}$
dominates the uncertainties, the others being negligible by comparison.
We will constrain the parameter $\Delta$ by considering that $\dot{P}_{b,R^{2}\text{-gravity}}$
should be within the interval 
\[
\dot{P}_{b,\text{min}}<\dot{P}_{b,R^{2}\text{-gravity}}<\dot{P}_{b,\text{max}}\,,
\]
where 
\[
\dot{P}_{b,\text{min}}=\dot{P}_{b,\text{corrected}}-3\sigma=-2.42\times10^{-12}\,\text{s/s}
\]
and 
\[
\dot{P}_{b,\text{max}}=\dot{P}_{b,\text{corrected}}+3\sigma=-2.39\times10^{-12}\,\text{s/s}\,.
\]

\begin{figure}[h]
\begin{centering}
\includegraphics[scale=0.5]{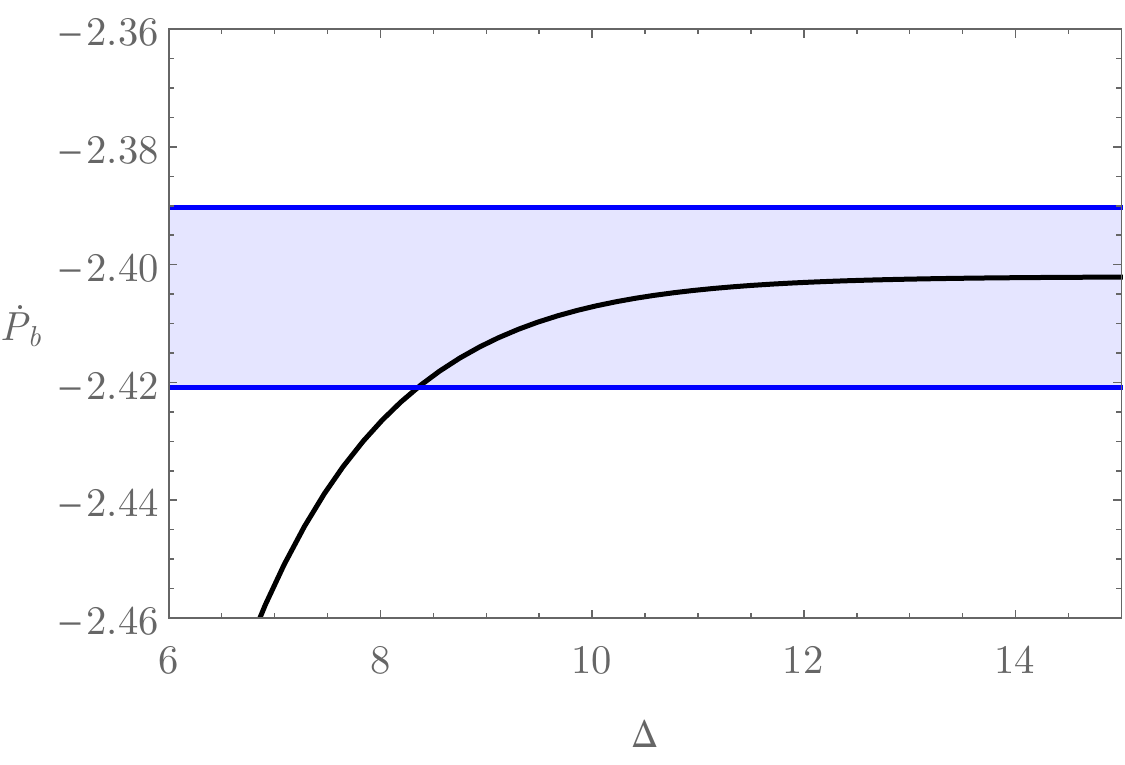} 
\par\end{centering}
\caption{Plot of the predicted $\dot{P}_{b,R^{2}\text{-gravity}}$ as a function
of $\Delta$. The lower (upper) straight line corresponds to $\dot{P}_{b,\text{min}}$
($\dot{P}_{b,\text{max}}$). The region between the straight lines
containing the curve $\dot{P}_{b,R^{2}\text{-gravity}}\left(\Delta\right)$
determines the domain of possible values for $\Delta$, the minimum
of which corresponds to the intersection of the lower straight line
with the curve $\dot{P}_{b,R^{2}\text{-gravity}}\left(\Delta\right)$
at $\Delta_{\text{min}}=8.3$.}
\label{fig:DeltaConstraint} 
\end{figure}

The plot of $\dot{P}_{b,R^{2}\text{-gravity}}\left(\Delta\right)$
in Fig. \ref{fig:DeltaConstraint} intercepts the horizontal line
$\dot{P}_{b,\text{min}}$ at $\Delta_{\text{min}}=8.3$. The conclusion
is that $\Delta$ should be greater than $8.3$ in order for $\dot{P}_{b,R^{2}\text{-gravity}}$
to be compatible with the observed time variation of the period within
the $3\sigma$ uncertainty interval.

In the following, we shall estimate the observational constraint upon
$\kappa_{0}=3\left(\Delta/R\right)^{2}$ and check if the damping
condition $2\omega_{s}<m_{-}c$ is satisfied. In order to do that,
we have to estimate the orbital distance $R$. This value will lie
in the interval set by the semimajor axis $a$ and the semiminor axis
$b$ of the elliptical orbit of the Hulse-Taylor binary pulsar---see
Ref. \cite{Taylor1982}. According to this reference, $R_{\text{max}}\equiv a=1.9490\times10^{9}\,\text{m}$,
and $R_{\text{min}}\equiv b=\sqrt{1-e^{2}}a=1.5336\times10^{9}$, where
we have used the eccentricity value in Table \ref{table:PSR}.

We know from Fig. \ref{fig:DeltaConstraint} that $\Delta\geqslant8.3$.
Then, $\Delta_{\text{min}}=8.3$, and the more restrictive estimate
upon $R^{2}$-term coupling is achieved by taking $R=R_{\text{min}}$:
\begin{equation}
\kappa_{0}>3\left(\frac{\Delta_{\text{min}}}{R_{\text{min}}}\right)^{2}\simeq8.8\times10^{-17}\,\text{m}^{-2}\qquad\text{or}\qquad\frac{1}{\kappa_{0}}\lesssim1.1\times10^{16}\,\text{m}^{2}\,.\label{eq:kappa_0_constraint}
\end{equation}

It is interesting to compare this constraint to other results available
in the literature. The estimates related to GW are $a_{2}<1.2\times10^{18}\,\text{m}^{2}$
in Ref. \cite{Berry2011} and $a<1.7\times10^{17}\,\text{m}^{2}$
in Ref. \cite{Naf2011}, with the notational correspondences $\left(1/\kappa_{0}\right)\leftrightarrow a_{2}\leftrightarrow a$,
respectively. Notice that we obtain an order-of-magnitude improvement
in the constraint of $\left(1/\kappa_{0}\right)$; however, this should
be taken with a grain of salt, since we have introduced the eccentricity
factor in an \emph{ad hoc} fashion. In a future work, we intend to
refine this estimate by computing the effect of eccentricity in a
deductive, rigorous way in our analysis.

Now, we will check if the damping constraint ($2\omega_{s}<m_{-}c$)
is satisfied by our estimate of $\kappa_{0}$. First, we use the lowest
value of $\kappa_{0}$ in Eq. (\ref{eq:kappa_0_constraint}) to obtain
\begin{equation}
m_{-}=\sqrt{\frac{\kappa_{0}}{3}}\Rightarrow m_{-}c>1.6\,\text{s}^{-1}\,.\label{eq:(m_c)_constraint}
\end{equation}
On the other hand, utilizing the observed period in Table \ref{table:PSR},
\begin{equation}
\omega_{s}=\frac{2\pi}{T}\Rightarrow2\omega_{s}=\frac{4\pi}{P_{b}\left(\text{s}\right)}=4.5\times10^{-4}\,\text{s}^{-1}\,.\label{eq:2omega_s}
\end{equation}
By comparing Eqs. (\ref{eq:(m_c)_constraint}) and (\ref{eq:2omega_s}),
we conclude that $2\omega_{s}<m_{-}c$, consequently checking the
validity of our analysis.

\subsubsection{Corrections to the time variation of the orbital period due to higher-order
$R^{2}$ gravity\label{subsec:Pbdot-StaPod}}

The development above on the pure quadratic-curvature scalar case
shows that $\Delta\gtrsim8$. This will also be true when one adds
to the $R^{2}$-gravity model the higher-order contributions carrying
the coupling constant $\beta_{0}$. In fact, the terms scaling with
$\beta_{0}$ in the action of our higher-order quadratic-curvature scalar
model are considered as a small correction to those accompanied by
$R^{2}$-gravity coupling $\kappa_{0}$. Accordingly, by considering
$\beta_{0}\ll1$ and $\Delta\gtrsim8$ in Eqs. (\ref{eq:DiffEqR})
and (\ref{eq:Kepler(Delta)}), and following a procedure similar to
the one in the previous subsection, we get 
\begin{align}
\dot{T}\simeq & -\frac{192\pi}{5}\left(\frac{GM_{c}}{c^{3}}\right)^{5/3}\left(\frac{T}{2\pi}\right)^{-5/3}\nonumber \\
 & \times\left\{ 1+\frac{2}{9}\left[\left(1+\frac{\beta_{0}}{3}\right)+\left(1+\frac{\beta_{0}}{2}\right)\Delta+2\left(1+2\frac{\beta_{0}}{3}\right)\Delta^{2}\right]e^{-\Delta\left(1+\frac{\beta_{0}}{6}\right)}\right\} \,.\label{eq:Tdot(T,Delta)}
\end{align}
Note that Eq. (\ref{eq:Tdot(T,Delta)}) obediently reduces to Eq. (\ref{eq:Tdot(T,Delta)_damping}),
with $g\left(\Delta\right)$ given by Eq. (\ref{eq:g(Delta)_Delta_large})
in the limit as $\beta_{0}\rightarrow0$.

The interval of values assumed by the pair $\left(\beta_{0},\Delta\right)$
can be estimated by using Eq. (\ref{eq:Tdot(T,Delta)}) and observational
data in much the same way as is done in the case of pure $R^{2}$-type
corrections to GR (Sec. \ref{subsec:Pbdot-Sta}). We allow $\beta_{0}$
to vary in the interval\footnote{The interval $0<\beta_{0}<3/4$ is demanded by the requirement of
stability of the gravitational potential \cite{Medeiros2020}.} $\left[0,3/4\right]$ and estimate the values admitted by $\Delta$
that satisfy $\dot{P}_{b,\text{min}}<\dot{P}_{b,\text{higher-order}}<\dot{P}_{b,\text{max}}$,
where $\dot{P}_{b,\text{higher-order}}=\dot{T}\times f\left(e\right)$.
Thereafter, we can transfer the constraint upon $\Delta$ onto the
coupling $\kappa_{0}$. This is done by estimating $R=R_{\text{min}}$
and using $\kappa_{0}=3\left(\Delta_{\text{min}}/R_{\text{min}}\right)^{2}$,
with the understanding that now $\Delta_{\text{min}}=\Delta_{\text{min}}\left(\beta_{0}\right)$.
The plot of $\left(1/\kappa_{0}\right)$ as a function of $\beta_{0}$
is shown in Fig. \ref{fig-(beta0,kappa0inverse)}. 

\begin{figure}[h]
\begin{centering}
\includegraphics[scale=0.5]{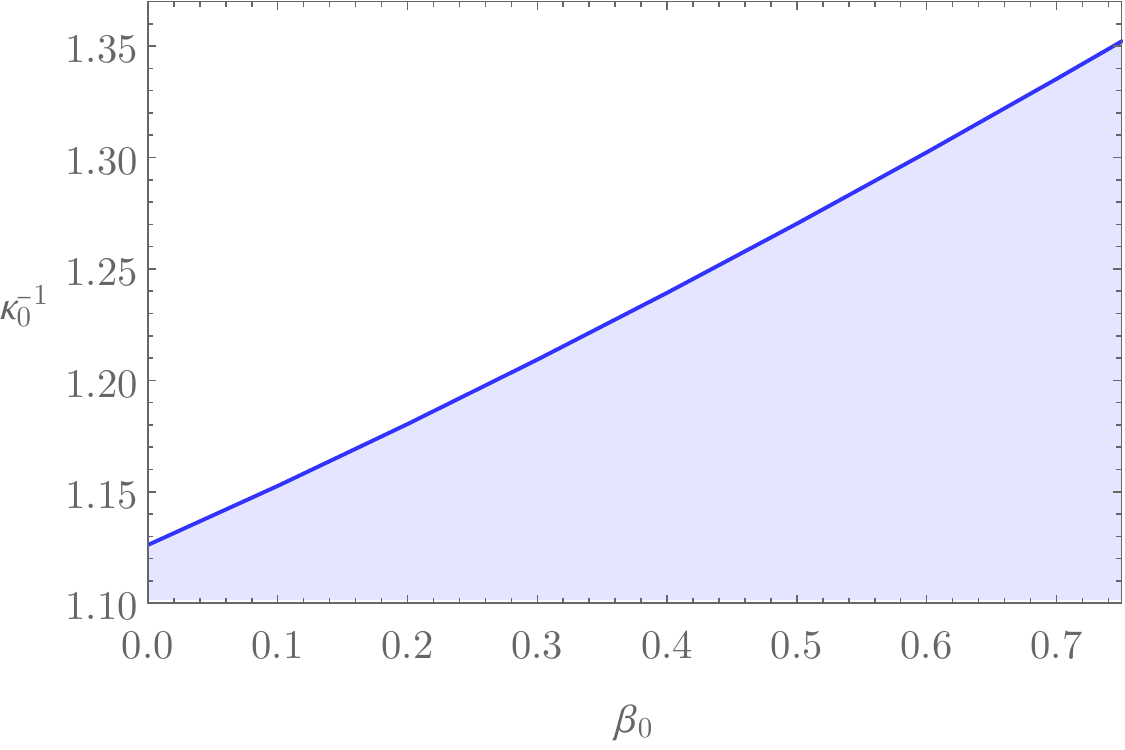} 
\par\end{centering}
\caption{ Plot of $\beta_{0}\times\kappa_{0}^{-1}$. The $y$ axis contains the
values of $\kappa_{0}^{-1}$ in units of $\left(10^{16}\,\text{m}^{2}\right)$.
The allowed region for the parameter space is the shadowed region
below the curve. }
\label{fig-(beta0,kappa0inverse)} 
\end{figure}

Even though $\beta_{0}$ can take values in the interval $\left[0,3/4\right]$,
we have assumed a small $\beta_{0}$ in our approximations along the
computations. For this reason, the most reliable region of the parameter
space in Fig. \ref{fig-(beta0,kappa0inverse)} is close to the origin.

\section{FINAL COMMENTS\label{sec:Final-comments}}

In this paper, we have studied the gravitational waves emitted by binary
systems in the context of the higher-order quadratic-curvature scalar
model built from an action integral involving the Einstein-Hilbert
term plus $R^{2}$ and $R\square R$ contributions. We have assumed
nonrelativistic motion of the pair of point particles in the binary system
and a circular orbit.

The gravitational wave solution was computed by considering the dominant
terms in the multipole expansion. It exhibits two modes, $\Phi_{-}$
and $\Phi_{+}$, for the decomposed massive scalar field $\Phi$ in
addition to the expected spin-2 mode $\tilde{h}_{\mu\nu}$.\footnote{We have seen that the monopole and dipole contributions for the scalar
modes do not contribute to the GW solution in the center-of-mass reference
frame. This is true as long as we admit a nonrelativistic approximation
for the orbital motion.} The spin-0 modes feature two different behaviors: the oscillatory
regime, which plays the actual GW part, and the damped regime, which
is exponentially suppressed.

In order to study the power radiated in the form of gravitational
waves, it was necessary to build the higher-order energy-momentum
pseudotensor, after which we established the balance equation.

We have determined the wave pattern given off by black-hole binaries
during the inspiral phase. The strain displays an increase in amplitude
and frequency that is a bit more pronounced than in the relativistic
case; the chirp occurs before the ordinary GR prediction. This effect
is a result of the extra channeling through the scalar mode: besides
the power radiated by $\tilde{h}_{\mu\nu}$-type emission, there is
the energy loss via the radiation of the spin-0 field $\Phi$.

Our modification in the description of gravity also affects the orbital
dynamics of a binary pulsar system. The change in the orbital dynamics
alters the balance equation and, as a consequence, the differential
equation of the orbital period. The time variation in the orbital
period of the Hulse-Taylor binary pulsar was used to constrain the
coupling constants $\kappa_{0}$ and $\beta_{0}$ related to the $R^{2}$
and $R\square R$ terms.

In fact, the coupling constant $\left(1/\kappa_{0}\right)$ of the
$R^{2}$ term was restricted to $\kappa_{0}^{-1}\lesssim1.1\times10^{16}\,\text{m}^{2}$.
This constraint is several orders of magnitude less restrictive than
the one offered by the E\"ot-Wash torsion balance experiment \cite{Kapner2007}, namely $a \lesssim {10}^{-9}\,\text{m}^{2}$. This estimation is based on the result $\sqrt{3a} = \lambda \gtrsim 56 \, \mu\text{m}$ is excluded within 95 \% confidence level according to Ref. \cite{Kapner2007}. The difference is rooted
in the fact that the data related to GW are astrophysical in nature,
while torsion balance experiments probe earthly gravity effects. Even
though our constraint is much less restrictive, it is of fundamental
importance to test alternative models of gravity in as many different
scales as possible. This is a sensitive aspect indeed, given that our
model (or extensions of it) might feature a screening mechanism \cite{Jana2019,Khoury2004};
this possibility remains to be checked and explored in a future work.
In any case, our constraint on $\kappa_{0}^{-1}$ is about 1 order
of magnitude better than the ones given in Refs. \cite{Berry2011,Naf2011},
which use similar approaches to ours.

In this work, we assume that the contribution of the term $R\square R$
to the modified gravity model is incremental. Accordingly, the associated
coupling constant $\beta_{0}$ should be small. We have shown that
the small-valued $\beta_{0}$ modifies very little the results obtained
in the analysis of the gravitational waves in the pure $\left(R+R^{2}\right)$ case.

It should be mentioned that the constraint over $\kappa_{0}$ lacks
rigor, given the fact that we have introduced the eccentricity function
$f\left(e\right)$ in an \emph{ad hoc} fashion. This weak point in
our treatment automatically motivates a future work in which the noncircular
orbits of the pair of astrophysical bodies in the binary system are
considered from the start. It is well known from the standard GR treatment
that taking into account elliptical orbits leads to the emission of
GW with a multitude of frequencies. As a result of considering eccentric
orbits in our modified gravity model, we might conclude that the gravitational
waves present a spin-0 oscillatory mode with frequencies high enough
to meaningfully contribute to the power radiated within the observational
bounds. This remains to be checked.

\begin{acknowledgments}
S. G. V. thanks Universidade Federal de Alfenas for the hospitality and
CAPES-Brazil for financial support. L. G. M. acknowledges CNPq-Brazil (Grant
No 308380/2019-3) for partial financial support. R. R. C. is grateful to CNPq-Brazil
(Grant No 309984/2020-3) for financial support. The authors thank the referee for helpful comments.
\end{acknowledgments}

\appendix

\section{RESOLVING $\Phi_{-}$ FIELD EQUATION \label{Appendix:Phi_minusFieldEq}}

Let us consider the scalar equation for $\Phi_{-}$, Eq. (\ref{eq:FieldEq_Phi_minus})
from Sec. \ref{sec:Scalar-tensor-decomposition}: 
\begin{equation}
\left(\square-m_{-}^{2}\right)\Phi_{-}=Q\,,\label{eq:FieldEq_Phi_Minus(Q)}
\end{equation}
where $Q\equiv-\frac{\chi}{3}T$. The solution is of the form 
\begin{equation}
\Phi_{-}\left(x^{\mu}\right)=\int G_{\Phi}\left(x^{\mu};x^{\prime\mu}\right)Q\left(x^{\prime\mu}\right)d^{4}x^{\prime}\,,\label{eq:Phi_Minus_Int_G}
\end{equation}
where the differential equation for the Green's function $G_{\Phi}\left(x^{\mu};x^{\prime\mu}\right)$,
\begin{equation}
\left(\square-m_{-}^{2}\right)G_{\Phi}\left(x^{\nu};x^{\prime\nu}\right)=\delta^{4}\left(x^{\nu}-x^{\prime\nu}\right)\,,\label{eq:G_DiffEq}
\end{equation}
is solved via the Fourier transform 
\begin{equation}
G_{\Phi}\left(x^{\nu};x^{\prime\nu}\right)=\frac{1}{\left(2\pi\right)^{2}}\int e^{iq_{\mu}\left(x^{\mu}-x^{\prime\mu}\right)}g_{\Phi}\left(q_{\nu}\right)d^{4}q\,.\label{eq:G_Fourier}
\end{equation}
Plugging Eq. (\ref{eq:G_Fourier}) into Eq. (\ref{eq:G_DiffEq}) and utilizing
the integral form of the delta function, we obtain 
\begin{equation}
g_{\Phi}\left(q_{\nu}\right)=\frac{1}{\left(2\pi\right)^{2}\left[-q_{\mu}q^{\mu}-m_{-}^{2}\right]}\,.\label{eq:g_Phi(q)}
\end{equation}
Substituting Eq. (\ref{eq:g_Phi(q)}) into Eq. (\ref{eq:G_Fourier}), we find 
\begin{equation}
G_{\Phi}\left(x^{\nu};x^{\prime\nu}\right)=\frac{1}{\left(2\pi\right)^{4}}\int d^{3}\mathbf{q}e^{i\mathbf{q}.\left(\mathbf{x}-\mathbf{x}^{\prime}\right)}\left\{ \int dq_{0}e^{iq_{0}\left(x^{0}-x^{\prime0}\right)}\frac{1}{\left[\left(q_{0}\right)^{2}-\mathbf{q}^{2}-m_{-}^{2}\right]}\right\} \,,\label{eq:G(I_0)}
\end{equation}
which depends on the integral 
\begin{equation}
I_{0}=\int_{-\infty}^{+\infty}dq_{0}e^{iq_{0}\left(x^{0}-x^{\prime0}\right)}\frac{1}{\left[\left(q_{0}\right)^{2}-p^{2}\right]}\,,\label{eq:I_0}
\end{equation}
where 
\begin{equation}
p^{2}\equiv\mathbf{q}^{2}+m_{-}^{2}\,.\label{eq:p(q,m)}
\end{equation}
The integral $I_{0}$ is solved by analytical extension and the use
of the Cauchy integral formula---see Ref. \cite{Jackson1999}. Physical
arguments lead us to choose the retarded Green's function, and to the
result 
\begin{equation}
I_{0}=\begin{cases}
0\,, & x^{0}<x^{\prime0}\\
-\frac{2\pi}{p}\sin\left[p\left(x^{0}-x^{\prime0}\right)\right]\,, & x^{0}>x^{\prime0}
\end{cases}\,.\label{eq:I_0(p,x_0)}
\end{equation}

Inserting Eq. (\ref{eq:I_0(p,x_0)}) into the expression (\ref{eq:G(I_0)})
for $G_{\Phi}\left(x^{\nu};x^{\prime\nu}\right)$ gives 
\begin{equation}
G_{\Phi}\left(x^{\nu};x^{\prime\nu}\right)=-\frac{1}{\left(2\pi\right)^{3}}\int q^{2}\sin\theta dqd\theta d\phi\frac{e^{iqs\cos\theta}}{\sqrt{\mathbf{q}^{2}+m_{-}^{2}}}\sin\left[\sqrt{\mathbf{q}^{2}+m_{-}^{2}}c\tau\right]\,,\label{eq:G(tau,s)}
\end{equation}
where 
\begin{equation}
c\tau\equiv\left(x^{0}-x^{\prime0}\right)=\left(ct-ct^{\prime}\right)\,,\label{eq:tau}
\end{equation}
\begin{equation}
s=\left|\mathbf{s}\right|=\left|\mathbf{x}-\mathbf{x}^{\prime}\right|\,,\label{eq:s}
\end{equation}
and we take $\mathbf{q}$ in spherical coordinates. Moreover, we have
set the $z$ axis in the direction of vector $\mathbf{s}$, such that
$\mathbf{q}\cdot\left(\mathbf{x}-\mathbf{x}^{\prime}\right)\equiv\mathbf{q}\cdot\mathbf{s}=qs\cos\theta$.

By integrating over the angular coordinates, Eq. (\ref{eq:G(tau,s)})
takes on the form 
\begin{equation}
G_{\Phi}\left(x^{\nu};x^{\prime\nu}\right)=-\frac{2}{\left(2\pi\right)^{2}}\frac{1}{s}\int_{0}^{\infty}qdq\sin\left(qs\right)\frac{\sin\left(E_{q}c\tau\right)}{E_{q}}\,,\label{eq:G_Phi(E_q)}
\end{equation}
where 
\begin{equation}
E_{q}\equiv\sqrt{q^{2}+m_{-}^{2}}\,.\label{eq:E_q}
\end{equation}

The massless case, $m_{-}=0$, is well known from the literature. The
result consistent with $x^{0}>x^{\prime0}$ is 
\begin{equation}
G_{\Phi}\left(x^{\nu};x^{\prime\nu}\right)=-\frac{1}{4\pi}\frac{1}{s}\delta\left(s-c\tau\right)\,,\qquad\left(m_{-}=0\right)\,.\label{eq:G_Phi_Massless}
\end{equation}

For the complete massive case, it is useful to write the sines in
Eq. (\ref{eq:G_Phi(E_q)}) in terms of exponentials and use the identity
\[
\int_{-\infty}^{\infty}\frac{q}{E_{q}}dq\exp\left[-i\left(E_{q}c\tau-qs\right)\right]=-i\frac{\partial}{\partial s}\int_{-\infty}^{\infty}dq\frac{\exp\left[-i\left(E_{q}c\tau-qs\right)\right]}{E_{q}}
\]
to write 
\begin{equation}
G_{\Phi}\left(x^{\nu};x^{\prime\nu}\right)=\frac{i}{8\pi^{2}}\frac{1}{s}\frac{\partial}{\partial s}\left[\int_{-\infty}^{\infty}dq\frac{\exp\left[-i\left(E_{q}c\tau-qs\right)\right]}{E_{q}}-\int_{-\infty}^{\infty}dq\frac{\exp\left[i\left(E_{q}c\tau-qs\right)\right]}{E_{q}}\right]\,.\label{eq:G_Phi_Massive_Int(Exp)}
\end{equation}
The change of variables \cite{Greiner2009} 
\begin{equation}
E_{q}=m_{-}\cosh\eta\,,\qquad\text{and}\qquad q=m_{-}\sinh\eta\label{eq:E_q(eta)}
\end{equation}
is suggestive, once it respects $E_{q}^{2}-q^{2}=m_{-}^{2}$ , cf. 
Eq. (\ref{eq:E_q}). Then, 
\begin{align}
G_{\Phi}\left(x^{\nu};x^{\prime\nu}\right) & =\frac{i}{8\pi^{2}}\frac{1}{s}\frac{\partial}{\partial s}\left[\int_{-\infty}^{\infty}d\eta\exp\left[-im_{-}\left(c\tau\cosh\eta-s\sinh\eta\right)\right]\right.\nonumber \\
 & \left.-\int_{-\infty}^{\infty}d\eta\exp\left[im_{-}\left(c\tau\cosh\eta-s\sinh\eta\right)\right]\right]\,.\label{eq:G_Phi_Massive_Int(Hyp)}
\end{align}
We should have $\tau>0$; otherwise, the Green's function is null.

There are three possible cases: 
\begin{enumerate}
\item \textbf{Timelike separation}: $c\tau>s$. In this case, it is always
true that $\tau>0$, because $s=\left|\mathbf{x}-\mathbf{x}^{\prime}\right|>0$.
It is convenient to define the variable $\theta$ such that 
\begin{equation}
c\tau=\sqrt{\left(c\tau\right)^{2}-s^{2}}\cosh\theta\,,\qquad\text{and}\qquad s=\sqrt{\left(c\tau\right)^{2}-s^{2}}\sinh\theta\,.\label{eq:tau(theta)_timelike}
\end{equation}
Hence, $\left(c\tau\cosh\eta-s\sinh\eta\right)=\sqrt{\left(c\tau\right)^{2}-s^{2}}\cosh\left(\eta-\theta\right)$.
Furthermore, by defining 
\begin{equation}
u=\left(\eta-\theta\right)\,,\label{eq:u}
\end{equation}
we cast Eq. (\ref{eq:G_Phi_Massive_Int(Hyp)}) into the form 
\begin{equation}
G_{\Phi}\left(x^{\nu};x^{\prime\nu}\right)=\frac{1}{4\pi^{2}}\frac{1}{s}\frac{\partial}{\partial s}\int_{-\infty}^{\infty}du\sin\left[m_{-}\sqrt{\left(c\tau\right)^{2}-s^{2}}\cosh u\right]\,.\label{eq:G_Phi(u)_timelike}
\end{equation}
Using the integral representation of the Bessel function \cite{Gradshteyn2007},
\[
J_{0}\left(z\right)=\frac{1}{\pi}\int_{-\infty}^{\infty}\sin\left(z\cosh u\right)du\qquad\left(\text{Re}[z]>0\right)\,,
\]
and the recurrency relation, 
\[
\frac{dJ_{0}\left(z\right)}{dz}=-J_{1}\left(z\right)\,,
\]
in Eq. (\ref{eq:G_Phi(u)_timelike}), we finally obtain 
\begin{equation}
G_{\Phi}\left(x^{\nu};x^{\prime\nu}\right)=\frac{1}{4\pi}\frac{m_{-}}{\sqrt{\left(c\tau\right)^{2}-s^{2}}}J_{1}\left(m_{-}\sqrt{\left(c\tau\right)^{2}-s^{2}}\right)\qquad\left(c\tau>s\right)\,.\label{eq:G_Phi_timelike}
\end{equation}
\item \textbf{Spacelike separation}: $c\tau<s$. In this case, the convenient
change of variables is 
\begin{equation}
c\tau=\sqrt{s^{2}-\left(c\tau\right)^{2}}\sinh\theta\,\qquad\text{and}\qquad s=\sqrt{s^{2}-\left(c\tau\right)^{2}}\cosh\theta\,.\label{eq:tau(theta)_spacelike}
\end{equation}
Following an analogous reasoning to the previous case, we have 
\begin{equation}
G_{\Phi}\left(x^{\nu};x^{\prime\nu}\right)=-\frac{1}{4\pi^{2}}\frac{1}{s}\frac{\partial}{\partial s}\left\{ \int_{-\infty}^{\infty}du\sin\left[m_{-}\sqrt{s^{2}-\left(c\tau\right)^{2}}\sinh u\right]\right\} =0\qquad\left(0<c\tau<s\right)\,,\label{eq:G_Phi_spacelike}
\end{equation}
where $u$ is defined in Eq. (\ref{eq:u}). 
\item \textbf{Lightlike case}: $c\tau=s$. The lightlike case corresponds
to a massless particle moving on the light cone. This case was previously
treated; the corresponding Green's function is Eq. (\ref{eq:G_Phi_Massless}). 
\end{enumerate}
All the possibilities can be unified by utilizing the Heaviside step
function 
\begin{equation}
\Theta\left(\xi\right)=\begin{cases}
0, & \xi<0\\
1, & \xi>0
\end{cases}\,.\label{eq:Heaviside}
\end{equation}
Therefore, 
\begin{equation}
G_{\Phi}\left(x^{\nu};x^{\prime\nu}\right)=-\frac{1}{4\pi}\frac{1}{c}\frac{1}{s}\delta\left(\tau-\frac{s}{c}\right)+\frac{1}{4\pi}\Theta\left(\tau-\frac{s}{c}\right)\frac{1}{c}\frac{m_{-}}{\sqrt{\tau^{2}-\left(\frac{s}{c}\right)^{2}}}J_{1}\left(m_{-}c\sqrt{\tau^{2}-\left(\frac{s}{c}\right)^{2}}\right)\,.\label{eq:G_Phi_Massive}
\end{equation}
This is the complete Green's function that solves Eqs. (\ref{eq:FieldEq_Phi_Minus(Q)})
through (\ref{eq:Phi_Minus_Int_G}).

\section{THE SPIN-0 FIELD $\Phi_{-}$ \label{Appendix:Phi_minus}}

In Eq. (\ref{eq:Phi_minusMultipole}) of Sec. \ref{subsec:Multipole-expansion},
we have written $\Phi_{-}$ as a multipole decomposition $\Phi_{-}=\Phi_{-}^{M}+\Phi_{-}^{D}+\Phi_{-}^{Q}+\cdots$.
It can be concluded from Eqs. (\ref{eq:Phi_minusM}), (\ref{eq:Phi_minusD}),
and (\ref{eq:MassMomentsCM}) that the monopole term $\Phi_{-}^{M}$
and dipole mode $\Phi_{-}^{D}$ do not contribute to the gravitational
wave emitted by a binary system in the center-of-mass reference frame:
$\Phi_{-}^{M}$ does not radiate, and $\Phi_{-}^{D}$ is zero.

The quadrupole mode will be the dominant contribution for the scalar
mode of the gravitational radiation, 
\begin{equation}
\Phi_{-}\left(\mathbf{x},t\right)=\Phi_{-}^{Q}\left(\mathbf{x},t\right)\equiv\Phi_{-}^{Q_{1}}\left(\mathbf{x},t\right)+\Phi_{-}^{Q_{2}}\left(\mathbf{x},t\right),\label{eq:Phi_minus(PhiQ1,PhiQ2)}
\end{equation}
where 
\begin{equation}
\Phi_{-}^{Q_{1}}\left(\mathbf{x},t\right)=\frac{2}{3}\frac{G}{c^{4}}\frac{1}{r}\left[\frac{1}{2}n^{i}n^{j}\left.\frac{\partial^{2}\mathcal{M}^{ij}}{\partial t^{2}}\right|_{t_{r}}\right]\label{eq:Phi_minusQ1}
\end{equation}
and 
\begin{equation}
\Phi_{-}^{Q_{2}}\left(\mathbf{x},t\right)=-m_{-}\frac{2}{3}\frac{G}{c^{4}}\int_{0}^{\infty}d\tau_{r}F\left(\tau_{r}\right)\left[\frac{1}{2}n^{i}n^{j}\left.\frac{\partial^{2}\mathcal{M}^{ij}}{\partial t^{2}}\right|_{\zeta}\right]\,,\label{eq:Phi_minusQ2}
\end{equation}
as per Eqs. (\ref{eq:F(tau_r)}) and (\ref{eq:Phi_minusQ}). We recall
that $\zeta=\left(t-\frac{r}{c}\right)-\tau_{r}$. The unit vector
$n_{i}$ points out orthogonally from the plane of the orbit, forming
an angle $\theta$ with the line of sight, cf. Eq. (\ref{eq:ni}).
From Eqs. (\ref{eq:x0}) and (\ref{eq:MassMomentsCM}), we calculate
the time derivatives of $\mathcal{M}^{ij}$: 
\begin{equation}
\ddot{\mathcal{M}}^{11}\left(t\right)=-2\mu R^{2}\omega_{s}^{2}\cos\left(2\omega_{s}t\right)=-\ddot{\mathcal{M}}^{22}\,,\qquad\ddot{\mathcal{M}}^{12}\left(t\right)=-2\mu R^{2}\omega_{s}^{2}\sin\left(2\omega_{s}t\right)\,.\label{eq:MassMomentsDeriv}
\end{equation}
Therefore,

\begin{equation}
\left[\frac{1}{2}n^{i}n^{j}\left.\frac{\partial^{2}\mathcal{M}^{ij}}{\partial t^{2}}\right|_{\xi}\right]=\left[\mu R^{2}\omega_{s}^{2}\sin^{2}\theta\cos\left(2\omega_{s}\xi+2\phi\right)\right]\,.\label{eq:KernelQ1Q2}
\end{equation}
From Eq. (\ref{eq:KernelQ1Q2}), we immediately obtain 
\begin{equation}
\Phi_{-}^{Q_{1}}\left(\mathbf{x},t\right)=\frac{2\mu R^{2}\omega_{s}^{2}}{3}\frac{G}{c^{4}}\frac{1}{r}\sin^{2}\theta\cos\left(2\omega_{s}t_{r}+2\phi\right)\label{eq:Phi_MinusQ1(x,t)}
\end{equation}
and 
\begin{align}
\Phi_{-}^{Q_{2}}\left(\mathbf{x},t\right) & =-m_{-}\frac{2}{3}\frac{G}{c^{4}}\mu R^{2}\omega_{s}^{2}\sin^{2}\theta\nonumber \\
 & \times\int_{0}^{\infty}d\tau_{r}\frac{J_{1}\left(m_{-}c\sqrt{2\tau_{r}}\sqrt{\frac{\tau_{r}}{2}+\frac{r}{c}}\right)}{\sqrt{2\tau_{r}}\sqrt{\frac{\tau_{r}}{2}+\frac{r}{c}}}\cos\left(-2\omega_{s}\left[\tau_{r}-\left(t-\frac{r}{c}\right)\right]+2\phi\right)\,.\label{eq:Phi_MinusQ2(intJ1cos)}
\end{align}
Hereafter, we will deal with Eq. (\ref{eq:Phi_MinusQ2(intJ1cos)}).
The integration variable can be changed to 
\begin{equation}
\tau_{r}=\frac{x}{m_{-}c}-\frac{r}{c}\,,\label{eq:T_bar_r(x)}
\end{equation}
so that 
\begin{equation}
\Phi_{-}^{Q_{2}}\left(\mathbf{x},t\right)=-m_{-}\frac{2}{3}\frac{G}{c^{4}}\mu R^{2}\omega_{s}^{2}\sin^{2}\theta\left\{ \cos\left[2\omega_{s}t+2\phi\right]I_{1}\left(r\right)+\sin\left[2\omega_{s}t+2\phi\right]I_{2}\left(r\right)\right\} \,,\label{eq:Phi_minusQ2(I1,I2)}
\end{equation}
where we denote 
\begin{equation}
I_{1}\left(r\right)=I_{1}\left(a,b\right)\equiv\int_{a}^{\infty}dx\frac{J_{1}\left(\sqrt{x^{2}-a^{2}}\right)}{\sqrt{x^{2}-a^{2}}}\cos\left(bx\right)\label{eq:I1}
\end{equation}
and 
\begin{equation}
I_{2}\left(r\right)=I_{2}\left(a,b\right)\equiv\int_{a}^{\infty}dx\frac{J_{1}\left(\sqrt{x^{2}-a^{2}}\right)}{\sqrt{x^{2}-a^{2}}}\sin\left(bx\right)\,,\label{eq:I2}
\end{equation}
with the definitions 
\begin{equation}
a\equiv m_{-}r\,,\quad\text{and}\quad b\equiv\frac{2\omega_{s}}{m_{-}c}\,.\label{eq:(a,b)}
\end{equation}

In order to solve $I_{1}\left(a,b\right)$ and $I_{2}\left(a,b\right)$,
consider the following integrals in Ref. \cite{Gradshteyn2007}: 
\begin{equation}
H_{1}\left(a,b\right)=b\int_{a}^{\infty}dxJ_{0}\left(\sqrt{x^{2}-a^{2}}\right)\cos\left(bx\right)=\begin{cases}
\frac{b}{\sqrt{1-b^{2}}}\exp\left[-a\sqrt{1-b^{2}}\right]\,, & 0<b<1\\
-b\frac{\sin\left(a\sqrt{b^{2}-1}\right)}{\sqrt{b^{2}-1}}\,, & b>1
\end{cases}\label{eq:H1}
\end{equation}
and 
\begin{equation}
H_{2}\left(a,b\right)=b\int_{a}^{\infty}dxJ_{0}\left(\sqrt{x^{2}-a^{2}}\right)\sin\left(bx\right)=\begin{cases}
0\,, & 0<b<1\\
b\frac{\cos\left(a\sqrt{b^{2}-1}\right)}{\sqrt{b^{2}-1}}\,, & b>1
\end{cases}\,.\label{eq:H2}
\end{equation}

Our strategy will be to solve $I_{1}\left(a,b\right)$ first by utilizing
Eq. (\ref{eq:H1}). Accordingly, let us work out the integral in 
\[
H_{1}\left(a,b\right)=b\int_{a}^{\infty}J_{0}\left(\sqrt{x^{2}-a^{2}}\right)\cos\left(bx\right)dx\,.
\]
We perform an integration by parts, recalling the following properties
of the Bessel function: 
\[
\frac{d}{dz}J_{0}\left(z\right)=-J_{1}\left(z\right)\,,\quad J_{0}\left(0\right)=1\quad\text{and}\quad\lim_{x\rightarrow\infty}J_{0}\left(x\right)=0\,.
\]
The result is 
\[
H_{1}\left(a,b\right)=-\sin\left(ab\right)+\int_{a}^{\infty}\frac{x}{\sqrt{x^{2}-a^{2}}}J_{1}\left(\sqrt{x^{2}-a^{2}}\right)\sin\left(bx\right)dx\,.
\]

We will recognize $I_{1}\left(a,b\right)$ in the expression above
after we integrate it with respect to $b$: 
\[
\int H_{1}db=\frac{1}{a}\cos\left(ab\right)-\int_{a}^{\infty}\frac{J_{1}\left(\sqrt{x^{2}-a^{2}}\right)}{\sqrt{x^{2}-a^{2}}}\cos\left(bx\right)dx\,,
\]
or, due to Eq. (\ref{eq:I1}), 
\begin{equation}
I_{1}\left(a,b\right)=-\int H_{1}db+\frac{1}{a}\cos\left(ab\right)\,.\label{eq:I1(H1)}
\end{equation}
The integral $\int H_{1}db$ appearing in Eq. (\ref{eq:I1(H1)}) is evaluated
with the help of the results on the right-hand side of Eq. (\ref{eq:H1}):
\[
\int H_{1}db=\begin{cases}
\frac{1}{a}\exp\left[-a\sqrt{1-b^{2}}\right]+K_{1}\,, & 0<b<1\\
\frac{1}{a}\cos\left[a\sqrt{b^{2}-1}\right]+C_{1}\,, & b>1
\end{cases}\,,
\]
where $K_{1}$ and $C_{1}$ are integration constants. Plugging back
in the definitions of $a$ and $b$ in terms of parameters of our model---Eq. (\ref{eq:(a,b)})---leads to 
\begin{equation}
I_{1}\left(r\right)=\begin{cases}
-\frac{1}{m_{-}r}\left[\exp\left(-m_{-}r\sqrt{1-\left(\frac{2\omega_{s}}{m_{-}c}\right)^{2}}\right)-\cos\left(\frac{2\omega_{s}r}{c}\right)\right]-K_{1}\,, & 2\omega_{s}<m_{-}c\\
-\frac{1}{m_{-}r}\left[\cos\left(m_{-}r\sqrt{\left(\frac{2\omega_{s}}{m_{-}c}\right)^{2}-1}\right)-\cos\left(\frac{2\omega_{s}r}{c}\right)\right]-C_{1}\,, & 2\omega_{s}>m_{-}c
\end{cases}\,.\label{eq:I1(r)}
\end{equation}

The integral $I_{2}\left(r\right)$ in Eq. (\ref{eq:I2}) is solved by
using Eq. (\ref{eq:H2}) for $H_{2}\left(a,b\right)$. In a procedure
completely analogous to what we have done above for $I_{1}\left(r\right)$,
we compute 
\begin{equation}
I_{2}\left(r\right)\begin{cases}
\frac{1}{m_{-}r}\sin\left(\frac{2\omega_{s}r}{c}\right)-K_{2}\,, & 2\omega_{s}<m_{-}c\\
-\frac{1}{m_{-}r}\left[\sin\left(m_{-}r\sqrt{\left(\frac{2\omega_{s}}{m_{-}c}\right)^{2}-1}\right)-\sin\left(\frac{2\omega_{s}r}{c}\right)\right]-C_{2}\,, & 2\omega_{s}>m_{-}c
\end{cases}\,,\label{eq:I2(r)}
\end{equation}
with $K_{2}$ and $C_{2}$ being integration constants.

Now, we plug Eqs. (\ref{eq:I1(r)}) and (\ref{eq:I2(r)}) into
Eq. (\ref{eq:Phi_minusQ2(I1,I2)}) for $\Phi_{-}^{Q_{2}}$ and separate
the result into two regimes: 
\begin{itemize}
\item[(1)] For $m_{-}c>2\omega_{s}$, 
\begin{align}
\Phi_{-}^{Q_{2}}\left(\mathbf{x},t\right) & =m_{-}\frac{2}{3}\frac{G}{c^{4}}\mu R^{2}\omega_{s}^{2}\sin^{2}\theta\cos\left[2\omega_{s}t+2\phi\right]\nonumber \\
 & \times\left\{ \frac{1}{m_{-}r}\left[\exp\left(-m_{-}r\sqrt{1-\left(\frac{2\omega_{s}}{m_{-}c}\right)^{2}}\right)-\cos\left(\frac{2\omega_{s}r}{c}\right)\right]+K_{1}\right\} \nonumber \\
 & +m_{-}\frac{2}{3}\frac{G}{c^{4}}\mu R^{2}\omega_{s}^{2}\sin^{2}\theta\sin\left[2\omega_{s}t+2\phi\right]\left\{ -\frac{1}{m_{-}r}\sin\left(\frac{2\omega_{s}r}{c}\right)+K_{2}\right\} \,.\label{eq:Phi_minusQ2(x,t)_damping}
\end{align}
\item[(2)] For $m_{-}c<$ $2\omega_{s}$, 
\begin{align}
\Phi_{-}^{Q_{2}}\left(\mathbf{x},t\right) & =m_{-}\frac{2}{3}\frac{G}{c^{4}}\mu R^{2}\omega_{s}^{2}\sin^{2}\theta\cos\left[2\omega_{s}t+2\phi\right]\nonumber \\
 & \times\left\{ \frac{1}{m_{-}r}\left[\cos\left(m_{-}r\sqrt{\left(\frac{2\omega_{s}}{m_{-}c}\right)^{2}-1}\right)-\cos\left(\frac{2\omega_{s}r}{c}\right)\right]+C_{1}\right\} \nonumber \\
 & +m_{-}\frac{2}{3}\frac{G}{c^{4}}\mu R^{2}\omega_{s}^{2}\sin^{2}\theta\sin\left[2\omega_{s}t+2\phi\right]\nonumber \\
 & \times\left\{ \frac{1}{m_{-}r}\left[\sin\left(m_{-}r\sqrt{\left(\frac{2\omega_{s}}{m_{-}c}\right)^{2}-1}\right)-\sin\left(\frac{2\omega_{s}r}{c}\right)\right]+C_{2}\right\} \,.\label{eq:Phi_minusQ2(x,t)_oscillating}
\end{align}
\end{itemize}
At this point, it is necessary to set the constants $K_{1,2}$ to
recover the suitable limits. As $m_{-}\rightarrow\infty\Rightarrow\kappa_{0}\rightarrow\infty$,
we must recover the GR results, which means that 
\[
\lim_{m_{-}\rightarrow\infty}\left(\Phi_{-}^{Q_{1}}+\Phi_{-}^{Q_{2}}\right)=0\Rightarrow\lim_{m_{-}\rightarrow\infty}\Phi_{-}^{Q_{2}}=-\lim_{m_{-}\rightarrow\infty}\Phi_{-}^{Q_{1}}
\]
This can be achieved if $K_{1}=K_{2}=0$. Moreover, by imposing continuity
of Eqs. (\ref{eq:Phi_minusQ2(x,t)_oscillating}) and (\ref{eq:Phi_minusQ2(x,t)_damping})
at $m_{-}c=2\omega_{s}$, we conclude that $C_{1}=C_{2}=0$. Under
these choices, and substituting Eqs. (\ref{eq:Phi_MinusQ1(x,t)}),
(\ref{eq:Phi_minusQ2(x,t)_oscillating}), and (\ref{eq:Phi_minusQ2(x,t)_damping})
into Eq. (\ref{eq:Phi_minus(PhiQ1,PhiQ2)}), it follows that 
\begin{equation}
\Phi_{-}\left(\mathbf{x},t\right)=\begin{cases}
\frac{1}{6}\frac{1}{r}\frac{4G}{c^{4}}\mu\omega_{s}^{2}R^{2}\sin^{2}\theta\cos\left(2\omega_{s}t+2\phi\right)\left[\exp\left(-m_{-}r\sqrt{1-\left(\frac{2\omega_{s}}{m_{-}c}\right)^{2}}\right)\right]\,, & \left(2\omega_{s}<m_{-}c\right)\\
\frac{1}{6}\frac{1}{r}\frac{4G}{c^{4}}\mu\omega_{s}^{2}R^{2}\sin^{2}\theta\cos\left[2\omega_{s}\left(t-\frac{r}{c}\sqrt{1-\left(\frac{m_{-}c}{2\omega_{s}}\right)^{2}}\right)+2\phi\right]\,, & \left(2\omega_{s}>m_{-}c\right)
\end{cases}\,.\label{eq:Phi_minusQ(x,t)}
\end{equation}
This is precisely Eq. (\ref{eq:Phi_minus(x,t)}) in the main text.

\section{THE ENERGY-MOMENTUM PSEUDOTENSOR \label{Appendix:PseudoEnergyTensor}}

The energy effectively carried out by the gravitational wave depends
on the energy-momentum pseudotensor $t_{\mu\nu}$. It is given by
\cite{Sabbata1985} 
\begin{equation}
t_{\mu\nu}=-\frac{c^{4}}{8\pi G}\left\langle \mathcal{G}_{\mu\nu}^{\left(2\right)}\right\rangle \,,\label{eq:tmunu(calGmunu)}
\end{equation}
where $\mathcal{G}_{\mu\nu}^{\left(2\right)}$ is the geometry sector
(left-hand side) of the gravitational field equations. The label $^{\left(2\right)}$
in Eq. (\ref{eq:tmunu(calGmunu)}) means that $\mathcal{G}_{\mu\nu}^{\left(2\right)}$
contains only terms of quadratic order in $h_{\mu\nu}$. Therefore,
we move on to the task of calculating $\mathcal{G}_{\mu\nu}^{\left(2\right)}$
in our modified theory of gravity from Eqs. (\ref{eq:FieldEq}), (\ref{eq:Hmunu}),
and (\ref{eq:Imunu}): 
\begin{equation}
\mathcal{G}_{\mu\nu}=G_{\mu\nu}+\frac{1}{2\kappa_{0}}H_{\mu\nu}+\frac{\beta_{0}}{2\kappa_{0}^{2}}I_{\mu\nu}\,.\label{eq:calGmunu}
\end{equation}
We will decompose this object in orders of contributions involving
the metric tensor. According to the weak-field decomposition discussed
in Sec. \ref{sec:HOSG-WeakField}, the metric reads $g_{\mu\nu}=g_{\mu\nu}^{\left(0\right)}+g_{\mu\nu}^{\left(1\right)}$,
where $g_{\mu\nu}^{\left(0\right)}=\eta_{\mu\nu}$ and $g_{\mu\nu}^{\left(1\right)}=h_{\mu\nu}$.
Then,

\begin{equation}
\mathcal{G}_{\mu\nu}=\mathcal{G}_{\mu\nu}^{(0)}+\mathcal{G}_{\mu\nu}^{(1)}+\mathcal{G}_{\mu\nu}^{(2)}\,.\label{eq:calGmunu(0,1,2)}
\end{equation}
The superscript $^{(0)}$ indicates quantities dependent on the background;
the label $^{(1)}$ denotes objects that depend linearly on $h_{\mu\nu}$;
and, we repeat, quantities with the label $^{(2)}$ depend quadratically
on $h_{\mu\nu}$.

The zeroth-order contribution is built with the flat background metric
$\eta_{\mu\nu}$, so: $R_{\mu\nu}^{(0)}=0\Rightarrow\mathcal{G}_{\mu\nu}^{(0)}=0$.

The first-order objects $G_{\mu\nu}^{(1)}$, $H_{\mu\nu}^{(1)}$, and
$I_{\mu\nu}^{(1)}$ are used to build the object $\mathcal{G}_{\mu\nu}^{(1)}$
according to Eq. (\ref{eq:calGmunu}).

The second-order term of $\mathcal{G}_{\mu\nu}$ assumes the form
\begin{align}
\mathcal{G}_{\mu\nu}^{\left(2\right)} & =G_{\mu\nu}^{\left(2\right)}+\frac{1}{2\kappa_{0}}H_{\mu\nu}^{\left(2\right)}+\frac{\beta_{0}}{2\kappa_{0}^{2}}I_{\mu\nu}^{\left(2\right)}\label{eq:CalGmunu(2)}
\end{align}
where 
\begin{equation}
G_{\mu\nu}^{\left(2\right)}=R_{\mu\nu}^{\left(2\right)}-\frac{1}{2}g_{\mu\nu}^{\left(1\right)}R^{\left(1\right)}-\frac{1}{2}g_{\mu\nu}^{\left(0\right)}R^{\left(2\right)}\,,\label{eq:Gmunu(2)(Curvature)}
\end{equation}
\begin{align}
H_{\mu\nu}^{\left(2\right)} & =2R_{\mu\nu}^{(1)}R^{(1)}-\frac{1}{2}g_{\mu\nu}^{\left(0\right)}\left[R^{\left(1\right)}\right]^{2}-2\partial_{\nu}\left[\partial_{\mu}R^{\left(2\right)}\right]+2\left.\Gamma^{\left(1\right)}\right._{\nu\mu}^{\lambda}\left[\partial_{\lambda}R^{\left(1\right)}\right]\nonumber \\
 & +2\left[g_{\mu\nu}^{\left(1\right)}\partial_{\alpha}\partial^{\alpha}R^{\left(1\right)}+g_{\mu\nu}^{\left(0\right)}\partial_{\alpha}\partial^{\alpha}R^{\left(2\right)}+g_{\mu\nu}^{\left(0\right)}\left.\Gamma^{\left(1\right)}\right._{\alpha\beta}^{\alpha}\partial^{\beta}R^{\left(1\right)}\right]\,,\label{eq:Hmunu(2)(Curvature)}
\end{align}
and 
\begin{align}
I_{\mu\nu}^{\left(2\right)} & =\partial_{\mu}R^{\left(1\right)}\partial_{\nu}R^{\left(1\right)}-2R_{\mu\nu}^{\left(1\right)}\partial_{\alpha}\partial^{\alpha}R^{\left(1\right)}-\frac{1}{2}g_{\mu\nu}^{\left(0\right)}\partial_{\rho}R^{\left(1\right)}\partial^{\rho}R^{\left(1\right)}\nonumber \\
 & +2\partial_{\nu}\partial_{\mu}\left[\partial_{\alpha}\partial^{\alpha}R^{\left(2\right)}+\left.\Gamma^{\left(1\right)}\right._{\alpha\lambda}^{\alpha}\partial^{\lambda}R^{\left(1\right)}\right]-2\left.\Gamma^{\left(1\right)}\right._{\nu\mu}^{\beta}\partial_{\beta}\left[\partial_{\alpha}\partial^{\alpha}R^{\left(1\right)}\right]\nonumber \\
 & -2g_{\mu\nu}^{\left(1\right)}\partial_{\beta}\partial^{\beta}\partial_{\alpha}\partial^{\alpha}R^{\left(1\right)}-2g_{\mu\nu}^{\left(0\right)}\partial_{\beta}\partial^{\beta}\partial_{\alpha}\partial^{\alpha}R^{\left(2\right)}-2g_{\mu\nu}^{\left(0\right)}\partial_{\beta}\partial^{\beta}\left(\left.\Gamma^{\left(1\right)}\right._{\alpha\lambda}^{\alpha}\partial^{\lambda}R^{\left(1\right)}\right)\nonumber \\
 & -2g_{\mu\nu}^{\left(0\right)}\left.\Gamma^{\left(1\right)}\right._{\alpha\beta}^{\alpha}\partial^{\beta}\partial_{\rho}\partial^{\rho}R^{\left(1\right)}\,.\label{eq:Imunu(2)(Curvature)}
\end{align}

The objects $G_{\mu\nu}^{(2)}$, $H_{\mu\nu}^{(2)}$, and $I_{\mu\nu}^{(2)}$
in Eqs. (\ref{eq:Gmunu(2)(Curvature)}), (\ref{eq:Hmunu(2)(Curvature)}),
and (\ref{eq:Imunu(2)(Curvature)}) depend on $g_{\mu\nu}^{\left(0\right)}$,
$g_{\mu\nu}^{\left(1\right)}$, $\left.\Gamma^{\left(1\right)}\right._{\nu\mu}^{\lambda}$,
$R_{\mu\nu}^{\left(1\right)}$, $R^{(1)}$, $R_{\mu\nu}^{\left(2\right)}$,
and $R^{(2)}$.

We know that $g_{\mu\nu}^{\left(0\right)}=\eta_{\mu\nu}$. Now, we
will be concerned with $g_{\mu\nu}^{\left(1\right)}$. Equations (\ref{eq:hbar})
and (\ref{eq:h_bar(h_tilde,Phi_plus,Phi_minus)}) allow us to cast
$g_{\mu\nu}^{\left(1\right)}$ into the form 
\begin{equation}
g_{\mu\nu}^{\left(1\right)}=h_{\mu\nu}=\tilde{h}_{\mu\nu}+\eta_{\mu\nu}\tilde{\Phi}\,,\label{eq:gmunu(1)(htilde,Phitilde)}
\end{equation}
where 
\begin{equation}
\tilde{\Phi}\equiv-\left[\frac{\beta_{0}}{\kappa_{0}}m_{-}^{2}\Phi_{+}-\Phi_{-}\right]\,.\label{eq:Phitilde}
\end{equation}
Moreover, from the field equations (\ref{eq:FieldEq_h_tilde}), (\ref{eq:FieldEq_Phi_plus}),
and (\ref{eq:FieldEq_Phi_minus}) in vacuum,\footnote{We are concerned with the propagating waves in free space, after it
leaves the source.} it follows that 
\begin{equation}
\square\tilde{h}_{\mu\nu}=0\qquad\text{and}\qquad\Box\tilde{\Phi}=-\frac{\kappa_{0}}{3}\Phi_{+}\label{eq:Box(htilde,Phitilde)}
\end{equation}
[where we have used the definitions in Eq. (\ref{eq:m_plus_minus}) of $m_{\pm}$].

Using Eqs. (\ref{eq:gmunu(1)(htilde,Phitilde)}), (\ref{eq:Box(htilde,Phitilde)}),
and the transverse-traceless character of $\tilde{h}_{\mu\nu}$, i.e.,
$\partial^{\nu}\tilde{h}_{\mu\nu}=\tilde{h}=0$, the first-order quantities
$\left.\Gamma^{\left(1\right)}\right._{\,\mu\nu}^{\sigma}$, $R_{\mu\nu}^{\left(1\right)}$,
and $R^{(1)}$ can be expressed as 
\begin{align}
\left.\Gamma^{\left(1\right)}\right._{\,\mu\nu}^{\sigma} & =\frac{1}{2}\eta^{\sigma\rho}\left(\partial_{\mu}h_{\nu\rho}+\partial_{\nu}h_{\rho\mu}-\partial_{\rho}h_{\mu\nu}\right)\nonumber \\
 & =\frac{1}{2}\left[\partial_{\mu}\tilde{h}_{\nu}^{\,\sigma}+\delta_{\nu}^{\,\sigma}\partial_{\mu}\tilde{\Phi}+\partial_{\nu}\tilde{h}_{\,\mu}^{\sigma}+\delta_{\mu}^{\,\sigma}\partial_{\nu}\tilde{\Phi}-\partial^{\sigma}\tilde{h}_{\mu\nu}-\eta_{\mu\nu}\partial^{\sigma}\tilde{\Phi}\right]\,,\label{eq:Gamma(1)}
\end{align}
\begin{equation}
R_{\mu\nu}^{(1)}=\frac{1}{2}\left[\partial_{\sigma}\partial_{\mu}h_{\nu}^{\,\sigma}-\eta^{\rho\sigma}\partial_{\rho}\partial_{\sigma}h_{\mu\nu}-\partial_{\mu}\partial_{\nu}h+\partial_{\nu}\partial_{\rho}h_{\mu}^{\,\rho}\right]=\frac{1}{2}\left[-2\partial_{\nu}\partial_{\mu}\tilde{\Phi}-\eta_{\mu\nu}\Box\tilde{\Phi}\right]\,,\label{eq:Ricci(1)}
\end{equation}
and 
\begin{equation}
R^{\left(1\right)}=\partial_{\mu}\partial_{\nu}h^{\mu\nu}-\eta^{\mu\nu}\partial_{\mu}\partial_{\nu}h=-3\Box\tilde{\Phi}\,.\label{eq:R(1)}
\end{equation}

The second-order objects $R_{\mu\nu}^{\left(2\right)}$ and $R^{(2)}$
in terms of $h_{\mu\nu}$ are \cite{Maggiore2007} 
\begin{align}
2R_{\mu\nu}^{\left(2\right)} & =\frac{1}{2}\partial_{\mu}h_{\alpha\beta}\partial_{\nu}h^{\alpha\beta}+h^{\alpha\beta}\partial_{\mu}\partial_{\nu}h_{\alpha\beta}-h^{\alpha\beta}\partial_{\nu}\partial_{\beta}h_{\alpha\mu}-h^{\alpha\beta}\partial_{\mu}\partial_{\beta}h_{\alpha\nu}+\nonumber \\
 & +h^{\alpha\beta}\partial_{\alpha}\partial_{\beta}h_{\mu\nu}+\partial^{\beta}h_{\nu}^{\alpha}\partial_{\beta}h_{\alpha\mu}-\partial^{\beta}h_{\nu}^{\alpha}\partial_{\alpha}h_{\beta\mu}-\partial_{\beta}h^{\alpha\beta}\partial_{\nu}h_{\alpha\mu}\nonumber \\
 & +\partial_{\beta}h^{\alpha\beta}\partial_{\alpha}h_{\mu\nu}-\partial_{\beta}h^{\alpha\beta}\partial_{\mu}h_{\alpha\nu}-\frac{1}{2}\partial^{\alpha}h\partial_{\alpha}h_{\mu\nu}+\frac{1}{2}\partial^{\alpha}h\partial_{\nu}h_{\alpha\mu}+\frac{1}{2}\partial^{\alpha}h\partial_{\mu}h_{\alpha\nu}\label{eq:Ricci(2)(h)}
\end{align}
and 
\begin{equation}
R^{\left(2\right)}=-h^{\mu\nu}R_{\mu\nu}^{\left(1\right)}+\eta^{\mu\nu}R_{\mu\nu}^{\left(2\right)}\,.\label{eq:R(2)(h)}
\end{equation}
Equations (\ref{eq:Ricci(2)(h)}) and (\ref{eq:R(2)(h)}) can be written
in terms of $\tilde{h}_{\mu\nu}$ and $\tilde{\Phi}$ via the replacement
$h_{\mu\nu}=\tilde{h}_{\mu\nu}+\eta_{\mu\nu}\tilde{\Phi}$.

There is a further possible simplification in the above expressions.
The energy-momentum pseudotensor in Eq. (\ref{eq:tmunu(calGmunu)}) depends
on the average of $\mathcal{G}_{\mu\nu}^{\left(2\right)}$. The averaging
process will affect each term in the sum (\ref{eq:CalGmunu(2)}),
which ultimately acts upon the several terms of Eqs. (\ref{eq:Gmunu(2)(Curvature)})--(\ref{eq:Imunu(2)(Curvature)})
and, consequently, upon those terms in Eqs. (\ref{eq:Gamma(1)})--(\ref{eq:R(2)(h)}).

The argument of wavelike solutions is of the D'Alembert type, which
means that we can interchange time derivatives by space derivatives
and conversely (up to a global sign), e.g., $\partial_{x}f\left(x-ct\right)=-c^{-1}\partial_{t}f\left(x-ct\right)$.
This enables us to perform integration by parts in the averaging
process and transfer any type of derivative from one term to the other
(up to irrelevant surface terms),\footnote{Surface terms are neglected because the spatial average is taken over
a length that is typically much larger than the solution's wavelength. } as in the example below: 
\begin{align*}
\left\langle h^{\alpha\beta}\partial_{\alpha}\partial_{\beta}h_{\mu\nu}\right\rangle  & =\left\langle \tilde{h}^{\alpha\beta}\partial_{\alpha}\partial_{\beta}h_{\mu\nu}\right\rangle +\left\langle \eta^{\alpha\beta}\tilde{\Phi}\partial_{\alpha}\partial_{\beta}h_{\mu\nu}\right\rangle \\
 & =\left\langle \partial_{\beta}\left(\tilde{h}^{\alpha\beta}\partial_{\alpha}h_{\mu\nu}\right)\right\rangle -\left\langle \partial_{\beta}\tilde{h}^{\alpha\beta}\partial_{\alpha}h_{\mu\nu}\right\rangle +\left\langle \tilde{\Phi}\square h_{\mu\nu}\right\rangle \\
 & =\left(\text{surface term}\right)-\left\langle \left(\partial_{\beta}\tilde{h}^{\alpha\beta}\right)\partial_{\alpha}h_{\mu\nu}\right\rangle +\left\langle \tilde{\Phi}\square\tilde{h}_{\mu\nu}\right\rangle +\left\langle \eta_{\mu\nu}\tilde{\Phi}\square\tilde{\Phi}\right\rangle =\left\langle \eta_{\mu\nu}\tilde{\Phi}\square\tilde{\Phi}\right\rangle \,,
\end{align*}
where we have neglected the surface term in the last step, used the
harmonic gauge $\partial_{\beta}\tilde{h}^{\alpha\beta}=0$ and invoked
the vacuum field equation $\square\tilde{h}_{\mu\nu}=0$.

By proceeding as described above, we are led to 
\begin{equation}
2\left\langle G_{\mu\nu}^{\left(2\right)}\right\rangle =-\frac{1}{2}\left\langle \partial_{\mu}\tilde{h}_{\alpha\beta}\partial_{\nu}\tilde{h}^{\alpha\beta}\right\rangle +\left\langle \partial_{\nu}\tilde{\Phi}\partial_{\mu}\tilde{\Phi}\right\rangle -\frac{1}{2}\eta_{\mu\nu}\left\langle \tilde{\Phi}\square\tilde{\Phi}\right\rangle \,;\label{eq:Gmunu(2)}
\end{equation}
\begin{align}
\left\langle H_{\mu\nu}^{\left(2\right)}\right\rangle  & =12\left\langle \left(\partial_{\mu}\partial_{\nu}\tilde{\Phi}\right)\left(\square\tilde{\Phi}\right)\right\rangle +\frac{3}{2}\eta_{\mu\nu}\left\langle \left(\square\tilde{\Phi}\right)^{2}\right\rangle \,,\label{eq:Hmunu(2)}
\end{align}
and 
\begin{equation}
\left\langle I_{\mu\nu}^{\left(2\right)}\right\rangle =21\left\langle \left(\partial_{\mu}\square\tilde{\Phi}\right)\left(\partial_{\nu}\square\tilde{\Phi}\right)\right\rangle -\frac{3}{2}\eta_{\mu\nu}\left\langle \tilde{\Phi}\square^{3}\tilde{\Phi}\right\rangle \,.\label{eq:Imunu(2)}
\end{equation}

Finally, we substitute Eqs. (\ref{eq:Gmunu(2)}), (\ref{eq:Hmunu(2)}),
and (\ref{eq:Imunu(2)}) into Eq. (\ref{eq:CalGmunu(2)}) to obtain the
expression of $\left\langle \mathcal{G}_{\mu\nu}^{\left(2\right)}\right\rangle $
in terms of $\tilde{h}_{\alpha\beta}$ and $\tilde{\Phi}$. Alternatively,
we can write $\left\langle \mathcal{G}_{\mu\nu}^{\left(2\right)}\right\rangle $
in terms of $\tilde{h}_{\alpha\beta}$, $\Phi_{+}$, and $\Phi_{-}$
by utilizing the definition of $\tilde{\Phi}$ in Eq. (\ref{eq:Phitilde}),
the $\Phi_{+}$ and $\Phi_{-}$ field equations (\ref{eq:FieldEq_Phi_plus})
and (\ref{eq:FieldEq_Phi_minus}), the definition [Eq. (\ref{eq:m_plus_minus})]
of $m_{\pm}$, and some integration by parts: 
\begin{align}
\left\langle \mathcal{G}_{\mu\nu}^{\left(2\right)}\right\rangle  & =-\frac{1}{4}\left\langle \partial_{\mu}\tilde{h}_{\alpha\beta}\partial_{\nu}\tilde{h}^{\alpha\beta}\right\rangle +\frac{3}{4}\sqrt{1-\frac{4}{3}\beta_{0}}\left[1-\sqrt{1-\frac{4}{3}\beta_{0}}\right]\left\langle \partial_{\mu}\Phi_{+}\partial_{\nu}\Phi_{+}\right\rangle \nonumber \\
 & +\left\{ -\frac{1}{2}\left[1-\sqrt{1-\frac{4}{3}\beta_{0}}\right]+2\right\} \left\langle \partial_{\mu}\Phi_{-}\partial_{\nu}\Phi_{+}\right\rangle +\frac{1}{2}\left\langle \partial_{\nu}\Phi_{-}\partial_{\mu}\Phi_{-}\right\rangle \,.\label{eq:CalGmunu(2)(htilde,Phi_pm)}
\end{align}
We emphasize that the term with $\left\langle \partial_{\mu}\Phi_{-}\partial_{\nu}\Phi_{+}\right\rangle $
is symmetric. This is true within the averaging processes.

Equation (\ref{eq:CalGmunu(2)(htilde,Phi_pm)}) determines the energy-momentum
pseudotensor. Its 00 component corresponds to the energy density of
the gravitational wave. This energy density, as given by the above
expression, is not necessarily positive-definite. In fact, while the
first term---the massless spin-2 contribution---is indeed positive
for $\mu=\nu=0$, the analogous terms related to $\Phi_{-}$ and $\Phi_{+}$
are not. This so happens due to the fact that the $R\square R$ term
in the action [Eq. (\ref{eq:Action})] introduces ghosts in the scalar sector
of our model. We will see below that this potential conumdrum is avoided
if we treat the higher-order term as a small correction to the $R^{2}$ term.


\textit{Remark: Higher-order term as a small correction
to $R^{2}$ case.}---Let us consider the expression for $\mathcal{G}_{\mu\nu}^{\left(2\right)}$,
Eq. (\ref{eq:CalGmunu(2)(htilde,Phi_pm)}), in the case $\beta_{0}\ll1$.
In this context, 
\[
\sqrt{1-\frac{4}{3}\beta_{0}}\simeq\left(1-\frac{2}{3}\beta_{0}\right)\,,
\]
and Eq. (\ref{eq:CalGmunu(2)(htilde,Phi_pm)}) reduces to 
\begin{align}
\left\langle \mathcal{G}_{\mu\nu}^{\left(2\right)}\right\rangle  & =-\frac{1}{4}\left\langle \partial_{\mu}\tilde{h}_{\alpha\beta}\partial_{\nu}\tilde{h}^{\alpha\beta}\right\rangle +2\left\langle \partial_{\mu}\Phi_{-}\partial_{\nu}\Phi_{+}\right\rangle +\frac{1}{2}\left\langle \partial_{\nu}\Phi_{-}\partial_{\mu}\Phi_{-}\right\rangle \nonumber \\
 & +\frac{1}{2}\beta_{0}\left(\left\langle \partial_{\mu}\Phi_{+}\partial_{\nu}\Phi_{+}\right\rangle -\frac{2}{3}\left\langle \partial_{\mu}\Phi_{-}\partial_{\nu}\Phi_{+}\right\rangle \right)\,.\label{eq:CalGmunu(2)(htilde,Phi_pm)betasmall}
\end{align}
Furthermore, Eq. (\ref{eq:Phi_plus(Phi-minus,T)}) can be used to
eliminate $\Phi_{+}$ of Eq. (\ref{eq:CalGmunu(2)(htilde,Phi_pm)betasmall})
in favor of $\Phi_{-}$ and $T$: 
\begin{equation}
\left\langle \mathcal{G}_{\mu\nu}^{\left(2\right)}\right\rangle \approx-\frac{1}{4}\left\langle \partial_{\mu}\tilde{h}_{\alpha\beta}\partial_{\nu}\tilde{h}^{\alpha\beta}\right\rangle -\frac{3}{2}\left(1+\frac{\beta_{0}}{3}\right)\left\langle \partial_{\mu}\Phi_{-}\partial_{\nu}\Phi_{-}\right\rangle +\frac{2\chi}{3}\frac{\beta_{0}}{\kappa_{0}}\left\langle \partial_{\mu}\Phi_{-}\partial_{\nu}T\right\rangle \,.\label{eq:CalGmunu(2)(htilde,Phi_m,T)betasmall}
\end{equation}
This is precisely Eq. (\ref{eq:Gmunu(2)averaged}) in the main text.

Equation (\ref{eq:CalGmunu(2)(htilde,Phi_m,T)betasmall}) reduces to the
corresponding result in Ref. \cite{Berry2011} when $\beta_{0}=0$
under the identification $\Phi_{-}\rightarrow-a_{2}R^{\left(1\right)}$.

We shall see in Appendix \ref{Appendix:Vanishing} that the last term
of Eq. (\ref{eq:CalGmunu(2)(htilde,Phi_m,T)betasmall}) vanishes.
Consequently, $\left\langle \mathcal{G}_{00}^{\left(2\right)}\right\rangle $
represents a positive energy density for $\beta_{0}\ll1$, which guarantees
a consistent physical interpretation to the emitted gravitational
waves predicted within our model. $\blacksquare$

\section{THE VANISHING OF $\left\langle \partial_{0}\Phi_{-}\partial_{1}T\right\rangle $
FOR THE BINARY SYSTEM \label{Appendix:Vanishing}}

In this appendix, we would like to calculate the quantity $\left\langle \partial_{0}\Phi_{-}\partial_{1}T\right\rangle $
in the last term in Eq. (\ref{eq:dP/dOmega(h,Phi,T)}) expressing
the power radiate per unit solid angle by the binary system. It reads
\begin{equation}
\left\langle \partial_{0}\Phi_{-}\partial_{1}T\right\rangle =\frac{1}{V}\intop_{0}^{V}\left[\frac{1}{\mathcal{T}}\intop_{0}^{\mathcal{T}}\partial_{0}\Phi_{-}\partial_{1}Tdt\right]d^{3}x=\frac{1}{c}\frac{1}{V}\frac{1}{\mathcal{T}}\intop_{0}^{V}\intop_{0}^{\mathcal{T}}\left[\partial_{r}\left[\left(\partial_{t}\Phi_{-}\right)T\right]-\left(\partial_{r}\partial_{t}\Phi_{-}\right)T\right]dtd^{3}x\,,\label{eq:d0Phi_minusd1T}
\end{equation}
where $\mathcal{T}$ is a timescale encompassing a multitude of periods
$2\pi/\left(2\omega_{s}\right)$; analogously, $V$ is a region several
times larger than the characteristic length $\left(2\pi c\right)/\left(2\omega_{s}\right)$.
The first term in the last step can be cast in the form of a surface
term. The surface integral then vanishes because it is taken outside
the source.

The remaining term is $\intop_{0}^{V}\left(\partial_{r}\partial_{t}\Phi_{-}^{Q}\right)Td^{3}x$.
In order to calculate this term, let us compute $\left(\partial_{r}\partial_{t}\Phi_{-}^{Q}\right)$
first. For the \textit{oscillatory mode}---see the second line in
Eq. (\ref{eq:Phi_minus(x,t)})--- 
\begin{equation}
\left(\partial_{r}\partial_{t}\Phi_{-}^{Q}\right)\simeq\mathcal{O}\left(1/r^{2}\right)+\mathcal{J}\left(\theta\right)\frac{1}{r}\cos\left[2\omega_{s}\left(t-\frac{r}{c}\sqrt{1-\left(\frac{m_{-}c}{2\omega_{s}}\right)^{2}}\right)+2\phi\right]\,,\label{eq:drdtPhi_minusQ}
\end{equation}
where 
\begin{equation}
\mathcal{J}\left(\theta\right)\equiv\left(2\omega_{s}\right)^{2}\frac{2\mu R^{2}\omega_{s}^{2}}{3}\frac{G}{c^{5}}\sqrt{1-\left(\frac{m_{-}c}{2\omega_{s}}\right)^{2}}\sin^{2}\theta\,.\label{eq:calJ}
\end{equation}
Now, 
\begin{equation}
\intop_{0}^{V}\left(\partial_{r}\partial_{t}\Phi_{-}^{Q}\right)Td^{3}x=\intop_{0}^{V}\mathcal{J}\left(\theta\right)\frac{1}{r}\cos\left[2\omega_{s}\left(t-\frac{r}{c}\sqrt{1-\left(\frac{m_{-}c}{2\omega_{s}}\right)^{2}}\right)+2\phi\right]Td^{3}x\,.\label{eq:Int_drdtPhi_minusQ_T}
\end{equation}
The trace of the energy-momentum tensor $T$ of a binary system of
nonrelativistic point particles is 
\begin{equation}
T\left(t,\mathbf{x}\right)=-m_{1}c^{2}\delta^{\left(3\right)}\left(\mathbf{x}-\mathbf{x}_{1}\left(t\right)\right)-m_{2}c^{2}\delta^{\left(3\right)}\left(\mathbf{x}-\mathbf{x}_{2}\left(t\right)\right)\,.\label{eq:T_point-like}
\end{equation}
For more details on $T$, we refer the reader to Sec. \ref{subsec:M-moments}.
In the reference frame at the center of mass, $\mathbf{x}_{\text{CM}}=0$,
and the position vectors of the point particles are simplified to
$\mathbf{x}_{1,2}=\left(m_{2,1}/m\right)\mathbf{x}_{0}$, with the
relative position vector given by $\mathbf{x}_{0}=\left(R\cos\left(\omega_{s}t+\frac{\pi}{2}\right),R\sin\left(\omega_{s}t+\frac{\pi}{2}\right),0\right)$.
Then, 
\begin{equation}
T\left(t,\mathbf{x}\right)=-\sum_{A=1,2}m_{A}c^{2}\delta^{\left(3\right)}\left(\mathbf{x}-\frac{m-m_{A}}{m}\mathbf{x}_{0}\right)\,.\label{eq:T_point-like_CM}
\end{equation}
In spherical coordinates, 
\begin{equation}
\delta^{\left(3\right)}\left(\mathbf{x}-\frac{m-m_{A}}{m}\mathbf{x}_{0}\right)=\frac{1}{\left(\frac{m-m_{A}}{m}R\right)^{2}}\delta\left(r-\frac{m-m_{A}}{m}R\right)\delta\left(\phi-\left(\omega_{s}t-\frac{\pi}{2}\right)\right)\delta\left(\theta-\frac{\pi}{2}\right)\,.\label{eq:delta3spherical}
\end{equation}
Plugging Eqs. (\ref{eq:T_point-like_CM}) and (\ref{eq:delta3spherical})
into Eq. (\ref{eq:Int_drdtPhi_minusQ_T}), 
\begin{equation}
\intop_{0}^{V}\left(\partial_{r}\partial_{t}\Phi_{-}^{Q}\right)Td^{3}x=\sum_{A}\frac{m_{A}c^{2}}{\left(\frac{m-m_{A}}{m}R\right)}\mathcal{J}\left(\frac{\pi}{2}\right)\cos\left[2\omega_{s}\left(2t-\frac{m-m_{A}}{m}\frac{R}{c}\sqrt{1-\left(\frac{m_{-}c}{2\omega_{s}}\right)^{2}}\right)\right]\,.\label{eq:Int_drdtPhi_minusQ_T_solved}
\end{equation}
The integral above appears in Eq. (\ref{eq:d0Phi_minusd1T}): 
\begin{equation}
\left\langle \partial_{0}\Phi_{-}\partial_{1}T\right\rangle =-\frac{1}{c}\frac{1}{V}\frac{1}{\mathcal{T}}\intop_{0}^{\mathcal{T}}\left[\intop_{0}^{V}\left(\partial_{r}\partial_{t}\Phi_{-}^{Q}\right)Td^{3}x\right]dt\,,\label{eq:d0Phi_minusd1T(Int)}
\end{equation}
which then reads 
\begin{align}
\left\langle \partial_{0}\Phi_{-}\partial_{1}T\right\rangle  & =-\frac{1}{c}\frac{1}{V}\frac{1}{\mathcal{T}}\sum_{A}\frac{m_{A}c^{2}}{\left(\frac{m-m_{A}}{m}R\right)}\mathcal{J}\left(\frac{\pi}{2}\right)\nonumber \\
 & \times\intop_{0}^{\mathcal{T}}\cos\left[2\omega_{s}\left(2t-\frac{m-m_{A}}{m}\frac{R}{c}\sqrt{1-\left(\frac{m_{-}c}{2\omega_{s}}\right)^{2}}\right)\right]dt=0\,,\label{eq:d0Phi_minusd1T_null}
\end{align}
because we are integrating the cosine over one period.

The study of the term $\left\langle \partial_{0}\Phi_{-}\partial_{1}T\right\rangle $
for the \textit{damping mode} is entirely analogous to the above case
and leads to the same result: $\left\langle \partial_{0}\Phi_{-}\partial_{1}T\right\rangle =0$.

\end{document}